\documentclass{cornellerratum}
\usepackage{graphicx}
\usepackage{amsmath}
\usepackage{latexsym}
\usepackage{amsfonts}
\usepackage{amssymb}
\usepackage{bbm}
\usepackage{verbatim}

\newcommand{\sgn}{\mathop{\rm sgn}}
\newcommand{\erf}{\mathop{\rm erf}}
\newcommand{\Ai}{\mathop{\rm Ai}}
\newcommand{\Bi}{\mathop{\rm Bi}}
\newcommand{\sun}{\odot}

\newcommand{\ddr}{\frac{\partial}{\partial r}}
\newcommand{\Domt}{\Delta \tilde \omega}

\renewcommand{\vec}{\mathbf}

\font\tenbg=cmmib10 at 10pt
\def \rvecphi{{\hbox{\tenbg\char'036}}}

\long\def\symbolfootnotetext[#1]#2{\begingroup%
\def\thefootnote{\fnsymbol{footnote}}\footnotetext[#1]{#2}\endgroup} 

\begin{document}

\title{Collisionless Beam-Radiation Processes in the Laboratory and Astrophysics}
\author{Bjoern Sebastian Schmekel}
\conferraldate{August}{2005}
\maketitle

\copyrightholder{Bjoern Sebastian Schmekel}
\copyrightyear{2005}
\makecopyright

\begin{abstract}
Plasma instabilities can be encountered in many branches of
physics. This work focuses on relativistic plasmas with
applications in theoretical astrophysics and particle
accelerator physics. Even though these fields seem to be
unrelated the underlying plasma physics processes are often
very similar. Two plasma instabilities - the beam-beam
instability and the coherent synchrotron radiation instability -
are analyzed. The former severely limits the achievable
luminosity in storage rings and is related to the
two-stream instability which has been proposed as a candidate
for the radiation mechanism of radio pulsars.  The main
emphasis is on coherent synchrotron radiation which can
lead to prohibitive energy losses in bunch compressors.
Coherent synchrotron radiation also makes up the intense
emission of radio waves by pulsars. Simple models based
on the linearized Vlasov equation and relativistic 
magnetohydrodynamics which allow to compute detailed spectra 
of the emitted radiation are developed.
\end{abstract}

\begin{erratum}
Some powers of $\gamma$ in (\ref{dvphi}) and subsequent equations
are incorrect. In Phys. Rev. ST Accel. Beams {\bf 9}, 114401 (2006)
(also available as preprint arXiv:astro-ph/0608566) those
mistakes have been fixed.
\end{erratum}

\begin{biosketch}
Bjoern S. Schmekel was born in West Berlin on June 14th, 1977.
With his parents he moved to Elmshorn (a suburb located 30km
north of Hamburg) where he graduated from high school in 1997.
As a high school student he was mainly interested in
optimizing photomultiplier based scintillation detectors and
designing high voltage power supplies for the operation thereof.
While still in high school he started to work at Fermi National
Accelerator Laboratory in Batavia, Illinois with Ryuji Yamada
on the decays of the top quark in the 6-jet channel. He continued
his work at Fermilab during his summer vacations in college.
In 1997 he had to work in a hospital in Bayreuth as a contentious
objector for a year. Not possessing any useful skills he had to serve
in a ``champagne unit'', so he could enroll at the Open University in Hagen 
in mathematics. One year later he transferred to the University
of Hamburg and received his "Vordiplom" in physics in late 1999.
In 2000 he came to Cornell as a graduate student in the department
of physics. He has been working on a variety of problems
in theoretical physics, most notably in plasma physics 
(with applications in astrophysics and particle accelerator physics)
and gravity. 
In 1998 the German National Academic Foundation
(Studienstiftung des deutschen Volkes) elected him a fellow.
He is also a member of the American Physical Society.
\end{biosketch}

\begin{dedication}
To the memory of Joe Rogers
\end{dedication}

\begin{acknowledgements}
This is a book about facts. If for some reasons you object to
facts I have to ask you to close this book immediately
and ask the librarian to reshelve it for you.
Some people may experience severe health and mental problems when
exposed to facts. I strongly recommend that those people
seek a fact free environment as soon as possible - and there
exist plenty.

\bigskip

Congratulations! You decided to keep reading.
Well then, I warned you, but let me begin by
thanking those who contributed to the success and to the
content of my research. First of all I would like to
thank the American taxpayers (represented by the
National Science Foundation and my longtime sponsor -
the U.S. Department of Energy) who understand the
importance and the potential of fundamental research
in theoretical physics. Without their financial
contributions this book would not have been written.
More specifically I would like to thank the Laboratory 
for Nuclear Studies (now called Laboratory for Elementary 
Particle Physics) and David Hammer from the Cornell 
Laboratory for Plasma Studies who graciously
allocated the necessary resources to me.
Thanks are also due to Ira M. Wasserman who agreed
to serve as the chair of my special committee even though
I had no formal qualifications as an astrophysicist whatsoever.
This must have made him very suspicious and it demonstrates
that there are still people who have patience and endless trust in me.
Maybe he started to regret his decision, but now it is too late.
Being a very stubborn character it is very hard to teach me anything, 
but hopefully I have learnt something. The same is of course true
for Richard V.E. Lovelace with whom I did most of the calculations
in parallel. His patience, knowledge and skills 
made working with him extremely enjoyable. Richard's vast amount of experience
always convinced and reassured me that everything I did was actually
worth doing. 

It is with regret that one of my teachers could not see the completion
of this work. Joseph T. Roger's untimely death came as a surprise to
many of us. In the obituary for Joe I wrote:

{ \it " ... I was reminded of his cheerful personality, his endless patience
and of course his knowledge of physics. The former was the reason
undergrads in our department used to call him "Happy Joe" (I doubt
he was actually aware of this, though) ... }

Unfortunately, there is not much I can do except express my sincere condolences
to Joe's family once again. 

I was rather fortunate to have met
Georg H. Hoffstaetter who helped me finish a paper on the beam-beam
interaction in storage rings which I started writing together with Joe
before he was diagnosed with terminal cancer. Georg's passion
and enthusiasm for particle accelerators can become contagious 
and it simply will not stop - even when rowing across a / the Lost Lake
somewhere in Oregon during heavy snowfall.

There is a huge number of support staff that struggled with my problems
and they deserve being mentioned:
Lori Beyea-Powers (accounting), Cora Jackson (travel arrangements), 
Joyce Oliver (accounting), Rosemary French (support for teaching), Tom Shannon (IT),
Chuck Jessop (licensing) and many more. There is one person though
I have to list separately. Of course I am talking about Debra Hatfield. 
The list of services she provided me with is so long that I will not even
attempt to mention all the things she has done for me. I am just stunned by
Deb's ability to solve my everyday problems and I will miss her.

There is yet another person missing who deserves a lot of credit and this person
is James W. York, Jr.!
I have been extremely lucky and privileged to have met Jimmy. His insight into
general relativity is almost impossible to match. Our private conversations
about general relativity, quantum gravity and other (sometimes more trivial) 
aspects of life encouraged me to keep working on my ideas.

Finally, I would like to apologize. I have never really worked with other people together
on a project so closely before. 
The problem with me is that I have my own ideas and ways of doing certain things.
I feel very strongly about them and about how physics ought to be done.
Ultimately, this makes me at odds with almost everybody, and I can only hope
Cornell will find better graduate students than me who are easier to work with in the future.

\begin{flushright}
Bjoern S. Schmekel
\end{flushright}

\begin{flushleft}
Ithaca, New York \\
July 2005
\end{flushleft}

\end{acknowledgements}

\contentspage
\figurelistpage

\preface
Plasma physics effects are ubiquitous. Most of space is made of plasma and many devices on earth
(including household appliances) generate plasmas, e.g. fluorescent bulbs, klystrons in microwave
ovens, electron beams in CRTs, particle accelerators or electron microscopes etc.
There are two recurring main questions which arise in plasma physics: How do plasmas evolve
in time and are there equilibria which are stable under the influence of small perturbations?
How much energy is lost due to radiation? Even though the mentioned applications do not seem
to have much in common, plasma instabilities and radiation are often caused by the same few
well-known (and some not so well-known) processes. This opens up possibilities for testing astrophysical
processes in the laboratory.

After covering a few basics, the beam-beam instability (a rather unpleasant instability encountered
in storage rings which severely limits the achievable luminosity) is reviewed. In some aspects
this instability resembles the two-stream instability which is currently considered to be responsible
for the radio emission of spinning neutron stars (radio pulsars). Chapter \ref{ch:CSR} deals with the Coherent Synchrotron
Radiation instability as an alternative to the two-stream instability in radio pulsars.  
According to the preceding paragraph it may not come as a surprise that this instability also plays
a crucial role in particle accelerators. In chapter \ref{ch:MHD} a simplified approach is presented which is based
on magnetohydrodynamics and the results of a computer simulation thereof can be found in chapter \ref{ch:PIC}.

It is hoped that one day a gifted experimenter will exploit these similarities and come up with
particle accelerator experiments which might greatly benefit the astrophysics community.

\chapter{Plasma Physics} \label{intro:plasma}

\normalspacing
\setcounter{page}{1}
\pagenumbering{arabic}
\pagestyle{cornell}

\section{Statistical Mechanics of Plasmas}

In classical mechanics the trajectory of a particle $\vec{x}(t)$ is completely determined
if the forces acting on the particle are known and if the position $\vec{x}_0$ and the
velocity $\vec{\dot x}_0$ (or alternatively the momentum $\vec{p}_0$) at an arbitrary
initial time are known. One needs two initial conditions because the equations of motion
are second order differential equations excluding some peculiar special cases. 
For a huge number of particles, e.g. gas molecules in a steel cylinder,
it is obviously not very practical to compute the trajectories of all those particles,
and it is not even necessary. Since for all practical purposes the gas molecules are indistinguishable
even for a ``classical'' observer one can ask instead how many particles 
$f(\vec{x},\vec{p},t) d^3 x d^3 p$ one could encounter at time $t$, position $\vec{x}$
and momentum $\vec{p}$ inside a phase space interval of size $d^3 x d^3 p$ assuming the normalization
\begin{eqnarray}
\int d^3 x \int d^3 p f(\vec{x},\vec{p},t) = N
\end{eqnarray}
where $N$ is the total number of particles. Of course any other normalization is equally good.
If all the forces acting on this collection of particles are known it should be possible
to find an operator acting on the distribution function $f$ which describes its time evolution.
Such an ``operator'' does exist and the resulting equation is known as the Vlasov equation. Note that a statistical
treatment based on a smooth distribution function \footnote{In simple models of shock formation the
distribution function may tend to a discontinuous limit.} eliminates certain features
known as discrete particle effects which can have a rather big impact on the distribution function.
Recovering these effects is usually rather difficult but nevertheless important. In plasma physics
for instance, continuous charge distributions like a rotating ring with completely uniform
charge density do not radiate, i.e. even synchrotron radiation (both coherent and incoherent)
is due to the discreteness of electric charge.

\section{Vlasov Equation} \label{sec:vlasov}

Consider the following ``microscopic distribution function''
\begin{eqnarray}
F(\vec{x},\vec{p},t) = \sum_{j=1}^N \delta [ \vec{x}-\vec{x}_j(t) ] \delta [ \vec{p}-\vec{p}_j(t) ]
\end{eqnarray}
which contains all the information about each individual particle. It satisfies the following equation of motion

\begin{eqnarray}
\sum_{j=1}^N \left \{ \frac{\partial}{\partial t} + \vec{v}_j \cdot \frac{\partial}{\partial \vec{x}} +
\frac{d}{dt} \vec{p}_j \cdot \frac{\partial}{\partial \vec{p}} \right \} \delta [ \vec{x}-\vec{x}_j(t) ] \delta [ \vec{p}-\vec{p}_j(t) ]=0
\end{eqnarray}
(Klimontovich equation). In a plasma the average force exerted on a particle by
all other particles (which acts like an external force) is bigger than the
force exerted by the nearest neighbors. With
\begin{eqnarray}
\nonumber
\left < F \right > = f
\\ \nonumber
\left < \vec{\dot x}_j \right > = \vec{v}
\\
\left < \vec{\dot p}_j \right > = \vec{\dot p} ~.
\label{averagingvlasov}
\end{eqnarray}
averaging the Klimontovich equation gives the Vlasov equation
\begin{eqnarray}
\left \{ \frac{\partial}{\partial t} + \vec{v} \cdot \frac{\partial}{\partial \vec{x}} +
\frac{d}{dt} \vec{p} \cdot \frac{\partial}{\partial \vec{p}} \right \} f = 0 ~.
\label{vlasov}
\end{eqnarray}
The last equation in (\ref{averagingvlasov}) is justified if binary correlations are assumed small.
Thus, all discrete particle effects like binary collisions have been removed. They can be recovered
by writing Eq.~(\ref{averagingvlasov}) as a sum of an average (external) part and an internal
part due to nearest neighbor interactions. In section \ref{sec:fokker} this will lead us to 
the Fokker-Planck equation.
For astrophysical plasmas the densities are usually so low that the plasma can be assumed to be
collisionless.

\section{Solving the Vlasov Equation}

For particles interacting electromagnetically one replaces $\frac{d}{dt} \vec{p}$ in Eq.~(\ref{vlasov})
with the Lorentz force
\begin{eqnarray}
\frac{d}{dt} \vec{p} = q \left ( \vec{E} + \vec{v} \times \vec{B} \right )
\end{eqnarray}
The fields are related to the sources by the Maxwell equations
\begin{eqnarray}
\nabla \cdot \vec{B} = 0
\\
\nabla \cdot \vec{E} = 4 \pi \rho
\\
\nabla \times \vec{E} = - \frac{\partial}{\partial t} \vec{B}
\\
\nabla \times \vec{B} = 4\pi \vec{j} + \frac{\partial}{\partial t} \vec{E}
\end{eqnarray}
where
\begin{eqnarray}
\rho = q \int d^3 p f(\vec{x},\vec{p},t)
\\
\vec{j} = q \int d^3 p \vec{v} f(\vec{x},\vec{p},t)
\label{vp}
\end{eqnarray}
Furthermore, velocities and momenta are related by
\begin{eqnarray}
\vec{v} = \frac{\vec{p}/m}{\sqrt{1+\vec{p}^2/m^2}}
\end{eqnarray}
where $m$ is the rest mass of the particles.
This system of equations is referred to as the relativistic Vlasov-Maxwell system of equations.
In the purely electrostatic case the Maxwell equations simplify to the Poisson equation and
the resulting system is known as the Vlasov-Poisson system.
The system of equations is nonlinear in $f$ and is therefore hard to solve analytically.
Eq.~(\ref{vp}) can usually be linearized either in the non-relativistic or in the ultrarelativistic case.
Finding an equilibrium distribution, i.e. a distribution with $\partial f / \partial t = 0$, simplifies
the system and this task is often doable under more or less realistic assumptions. 
It can be shown that any distribution function $f$ which only depends on the constants of motion
represents an equilibrium. Typically, the constants of motion are the Hamiltonian and the canonical
angular momentum. However, rewriting $f$ as a function of $\vec{x}$ and $\vec{p}$ still requires
solving a nonlinear equation. Once an equilibrium
is found one can ask whether this equilibrium is stable to small amplitude perturbations. This question
can be answered by linearizing the Vlasov equation about the equilibrium distribution. A variety of instabilities
has been analyzed this way.

\section{Solving the Linearized Vlasov Equation Using the Method of Characteristics}
As was pointed out earlier the nonlinear nature of the Vlasov equation makes it hard to find
analytical solutions for it, but looking for equilibrium solutions may be a successful endeavour
in many simplified situations. Once an equilibrium has been found it is possible to linearize
the Vlasov equation about the equilibrium. This provides valuable information about the
stability properties of the found equilibrium.  Writing the distribution function and the fields as
a sum of the equilibrium part and a perturbation
\begin{eqnarray}
\nonumber
f = f^0 + \delta f
\\ \nonumber
\vec{E} = \vec{E}^0 + \delta \vec{E}
\\
\vec{B} = \vec{B}^0 + \delta \vec{B}
\end{eqnarray}
and neglecting second order terms one obtains the linearized Vlasov equation
\begin{eqnarray}
\left \{ \frac{\partial}{\partial t} + \vec{v} \cdot \frac{\partial}{\partial \vec{x}} -e \left (\vec{E}^0 + \vec{v} \times \vec{B}^0 \right ) \cdot \frac{\partial}{\partial \vec{p}} \right \} \delta f =
- e \left ( \delta \vec{E} + \vec{v} \times \vec{B} \right ) \cdot \frac{\partial}{\partial \vec{p}} f^0
\end{eqnarray}
The linearized Vlasov equation can be rewritten by following a particle on an equilibrium orbit $(\vec{x'},\vec{p'})$
which passes through $(\vec{x},\vec{p})$ at time $t'= t$. The equilibrium orbits have to satisfy the equations of motion
\begin{eqnarray}
\frac{d}{dt'} \vec{x'}(t') = \vec{v'}(t')
\\
\frac{d}{dt'} \vec{p'}(t') = -e \left \{ \vec{E}^0(\vec{x'}(t')) + \vec{v'}(t') \times \vec{B}^0(\vec{x'}(t')) \right \}
\end{eqnarray}
Thus,
\begin{eqnarray}
\frac{d}{dt'} \delta f (\vec{x'}(t'),\vec{p'}(t'),t') = - e \left (
\delta \vec{E} (\vec{x'},t') + \vec{v'} \times \delta \vec{B}(\vec{x'},t') \right ) \cdot
\frac{\partial}{\partial \vec{p'}} f^0 (\vec{x'},\vec{p'})
\end{eqnarray}
Integrating the last equation an expression for $\delta f$ can now be obtained easily.

\section{Mathematical Properties of the Vlasov-Maxwell System}
Since it is hard to develop an intuition for how solutions
of the Vlasov-Maxwell system look like for a particular set of initial
data it is desirable to find theorems regarding general properties
of the system. Will the solution remain smooth at all times
for smooth initial data, i.e. will the momentum space carrier of the distribution
function remain compact in finite time? Will the solution remain symmetric for
symmetric initial data? 
For the three dimensional relativistic Vlasov-Maxwell equations 
the answer to the first question is still unknown, but lots of
incremental progress has been made. A very good
review article on this subject is \cite{Andreasson:2005qy}. 
Glassey and Schaeffer \cite{glassey1997} have shown that the answer is ``yes''
in two spatial and three momentum dimensions. The answer is important
for deciding whether shocks can form spontaneously or not.
The answer to the second question is ``yes'' for spherically symmetric initial data \cite{Andreasson:2005qy}.

\section{Magnetohydrodynamics}
The distribution function $f$ contains all the 
meaningful information one could possibly ask for
in a statistical treatment and it is determined by either
the Vlasov equation in the absence of discrete particle
effects or by the Boltzmann or Fokker-Planck equation
in the presence of discrete particle effects. These equations
are usually difficult to solve and solving them numerically
is usually not an option either - at least not in three
dimensions where one would have to deal with seven-dimensional
PDEs. Even though numerical solutions are being obtained by
Ellison and collaborators \cite{Venturini:2005wu} for lower dimensional problems
(one spatial dimension, two dimensions in phase space)
this is usually not a viable method.

The underlying problem is that the distribution function still contains 
a lot of information and one may wonder whether the problem simplifies
if one is content with less information. For many practical purposes
it may be enough to compute certain macroscopic quantities like

the number density
\begin{eqnarray}
n(\vec{x},t) = \int  f(\vec{x},\vec{p},t) d^3p ~,
\end{eqnarray}

the mean velocity
\begin{eqnarray}
\vec{u}(\vec{x},t) = n^{-1} \int \vec{v} f(\vec{x},\vec{p},t) d^3 p ~,
\end{eqnarray}

the mean momentum
\begin{eqnarray}
\vec{\bar p}(\vec{x},t) = n^{-1} \int \vec{p} f(\vec{x},\vec{p},t) d^3 p ~,
\end{eqnarray}

and the three dimensional stress tensor

\begin{eqnarray}
{\bf P} = m \int (\vec{v}-\vec{u}) \otimes (\vec{v}-\vec{u}) f(\vec{x},\vec{p},t) d^3 p ~,
\end{eqnarray}

i.e. moments of the distribution function. The definition of the stress tensor depends
on the previous definition of the mean velocity. 

More macroscopic quantities
can be obtained by multiplying the integrand by an arbitrary power
of momentum and / or velocity components. Starting from the Vlasov
equation one can find equations determining those moments easily.
Integrating Eq.~(\ref{vlasov}) over all momenta gives the continuity
equation. 

\begin{eqnarray}
\frac{\partial}{\partial t} n + \nabla \cdot (n \vec{u}) = 0
\end{eqnarray}

Multiplying Eq.~(\ref{vlasov}) by $\vec{v}$ and integrating
over all momenta results in the ``Euler equation''

\begin{eqnarray}
\left ( \frac{\partial}{\partial t} + \vec{u} \cdot \nabla \right ) \vec{\bar p} =
-n^{-1} \nabla \cdot {\bf P} + \vec{F}(\vec{x},t)
\end{eqnarray}

These are exactly the same equations one encounters in fluid dynamics
if the higher order moments, i.e. the stress tensor, is neglected.
Therefore, this approach is known as magnetohydrodynamics (MHD).
An arbitrary number of equations can be found this way. 
There is one serious problem, though. 
The moment expansion does not close in the absence
of collisions, i.e. each equation will couple to the next higher moment.
In fluid dynamics the equation of state can be used to close the system,
but in plasma physics such an equation is generally unavailable - at
least in the absence of collisions. 
However, under certain conditions it may be possible to guess
an equation of state or it may be possible to neglect the higher order
moments, e.g. for a cold plasma in the absence of a pressure gradient and heat flux.
Magnetohydrodynamics will usually give good results if the frequency which is
characteristic for the evolution of the distribution is much smaller
than the plasma frequency and the cyclotron frequency.

\section{Some Plasma and Fluid Instabilities}

\subsection{Negative Mass Instability}

A longitudinal bunching can occur in a beam executing circular motion if the effective mass of the particles is negative, i.e. if an increase in energy leads to a decrease in angular velocity ($d \dot \phi / dE < 0$ where $\dot \phi = |\vec{\dot \phi}|$).
The energy at which the sign of $d \dot \phi / dE$ changes is called
transition energy. In a weak focusing machine (e.g. charged particles moving perpendicular to an external magnetic field without gradient)
this condition is always satisfied.
Assume that due to an arbitrary initial perturbation the charge density
is higher at a certain point on the circle. The electrostatic potential
tries to repel particles away from the center of higher density. 
Particles in front of the region of higher density gain energy, but their angular velocity decreases. Similarly, particles behind the region of higher density lose energy and increase their angular velocity.  
Thus, neighboring particles are attracted to the region of higher
density. The negative mass instability can be compensated by a
sufficiently large energy spread. In the limit of zero energy spread the dispersion relation
is \cite{Entis1971}
\begin{eqnarray}
1 = \frac{N r_e}{2 \pi r_0} \frac{\eta}{\gamma^3} \frac{1}{(\Delta \tilde \omega)^2}
\end{eqnarray}
where
\begin{eqnarray}
\eta \equiv \frac{p}{\dot \phi} \frac{d \dot \phi}{dp} 
\end{eqnarray}
and
\begin{eqnarray}
\Delta \tilde \omega \equiv \frac{\omega - m \dot \phi}{m \dot \phi}
\end{eqnarray}
$N$ is the number of electrons, $r_0$ is the radius of the orbit, $r_e$ is the classical electron radius, 
and the azimuthal mode number is denoted by $m$.
Classical references are \cite{Nielson1959,Briggs1966,Entis1971}.
%

\subsection{Rayleigh-Taylor Instability}
The Rayleigh Taylor instability is a fluid instability which can develop if a less dense fluid with density
$\rho_1$ propagates in a denser fluid with density $\rho_2$. Clumps of gas observed in supernova remnants are often due to this instability.

\subsection{Kelvin-Helmholtz Instability}
The Kelvin-Helmholtz instability is a non-relativistic fluid instability which can form at the interface of two flows with different velocities $u_1$ and $u_2$. The ubiquitous water waves caused by wind blowing over the surface of a pond are a typical example. The Kelvin-Helmholtz instability may also be important to understand the observed patterns of astrophysical jets surrounded by the interstellar medium.

The complex frequency $\omega$ of a perturbation in a system unstable to both the Rayleigh-Taylor instability and the Kelvin-Helmholtz 
instability is given by \cite{Chandrasekhar1961}
\begin{eqnarray}
\frac{\omega}{k} = \frac{\rho_1 u_1 + \rho_2 u_2}{\rho_1 + \rho_2} \pm \sqrt{\frac{g}{k} \left ( \frac{\rho_1 - \rho_2}{\rho_1+\rho_2} \right ) - \frac{\rho_1 \rho_2 (u_1 - u_2)^2}{(\rho_1+\rho_2)^2} }
\end{eqnarray}
where $k$ is the wavenumber of the perturbation and $g$ is the gravitational acceleration. 

\subsection{Diocotron Instability}
The Diocotron instability is ubiquitous in the circular motion of a low density non-neutral plasma with $d \dot \phi / d r \neq 0$ and can be found in common microwave generating devices.
This electrostatic instability resembles the Kelvin-Helmholtz instability in the sense that it forms in the presence of shear. For equilibrium configurations with $\partial / \partial z = 0$ a sufficient condition for stability is that the number density $n(r)$ is a monotonically decreasing function, i.e. the maximum number density occurs at $r=0$. 

\subsection{Cyclotron Maser Instability}
A relativistic beam moving along a guiding magnetic field may be subject to the
cyclotron maser instability. Unlike the classical Diocotron or the negative mass
instability it is a transverse electromagnetic instability which is capable
of producing coherent electromagnetic waves. The instability is driven by 
an inverted population in the transverse (i.e. parallel to the magnetic field)
momentum distribution, i.e. the momentum distribution is sharp and its mean is
non-zero. The cyclotron maser instability is exploited in microwave generating devices.

\subsection{Two-stream Instability}
The two-stream instability is an electrostatic instability which occurs for a plasma consisting of two (or more streams) of (not necessarily the same species of) particles with different velocities. 
Its development requires a region in phase space with $\partial f / \partial p > 0$ and
the  momentum of the particles satisfying the latter inequality have to be large compared with
the momenta of the remaining particles.
This instability is of importance for the understanding of problems associated with the solar wind. In particle accelerators secondary emission of electrons from the beam pipe may provide a background of electrons which can trigger a two-stream instability.


\section{The Fokker-Planck equation} \label{sec:fokker}
The averaging employed in section \ref{sec:vlasov} was rather crude and removed all discrete
particle effects. In this section it is attempted to recover the effect of statistical processes like
radiation damping and quantum excitation occurring at random times $t_i$.
Instead of Eqs.~(\ref{averagingvlasov}) one now uses

\begin{eqnarray}
\dot x = g(x,p,t) + \sum_i \Delta x_i \delta (t-t_i)
\label{dotxg}
\end{eqnarray}
\begin{eqnarray}
\dot p = h(x,p,t) + \sum_i \Delta p_i \delta (t-t_i)
\label{dotph}
\end{eqnarray}

The probability for the occurrence of a perturbation in momentum space with $\Delta x_i$ and $\Delta p_i$
is given by the probability densities $P_x(\Delta x_i)$ and $P_p(\Delta p_i)$, respectively.
$P_x$ and $P_p$ are taken to be normalized to unity and symmetric in their arguments.
Without the former condition the distribution function would not remain normalized.
After a timestep $\Delta t$ the phase space element $\Delta x \Delta p$ changes by the factor
\begin{eqnarray}
1 + \left ( \frac{\partial g}{\partial x} + \frac{\partial h}{\partial p} \right ) \Delta t
\end{eqnarray}
as can be seen by expanding the evolution of $\Delta x \Delta p$ to first order in $\Delta t$ using
Eq.~(\ref{dotxg}) and (\ref{dotph}). Assuming the number of particles is conserved one obtains
\begin{eqnarray}
\nonumber
f(x + g \Delta t, p + h \Delta t, t + \Delta t) \Delta x \Delta p 
\left [ 1 + \left ( \frac{\partial g}{\partial x} + \frac{\partial h}{\partial p} \right ) \Delta t \right ] =
\\
\Delta x \Delta p \int_{-\infty}^{\infty} \int_{-\infty}^{\infty}  d(\Delta x) d(\Delta p) P_x(\Delta x) P_p (\Delta p)
f(x - \Delta x, p - \Delta p,t)
\end{eqnarray}
Expanding the distribution function inside the integral to second order in $\Delta x$ and $\Delta p$ allows one to
evaluate the integral. Making use of the properties of $P_x$ and $P_p$ mentioned above
\begin{eqnarray}
\frac{\partial f}{\partial t} + g \frac{\partial f}{\partial x} + h \frac{\partial f}{\partial p} -
\left ( \frac{\partial g}{\partial x} + \frac{\partial h}{\partial p} \right ) f
+ \frac{1}{2} \Xi_x \frac{\partial^2 f}{\partial x^2} + \frac{1}{2} \Xi_p \frac{\partial^2 f}{\partial p^2}
\label{fokkerplanck}
\end{eqnarray}
where the coefficients $\Xi_x$ and $\Xi_p$ are related to the second order moments of $P_x$ and $P_p$, respectively.
Eq.~(\ref{fokkerplanck}) is known as the Fokker-Planck equation.
It describes the evolution of a plasma under the additional influences of radiation damping due to
incoherent synchrotron radiation (in this case the parenthesis in Eq.~(\ref{fokkerplanck}) differs from unity)
and quantum excitation due to the statistical nature of the radiation process (the plasma emits ``discrete'' photons).
The former tends to increase the phase space density whereas the latter tends to decrease it.
Setting the parenthesis to unity and neglecting the excitation coefficients $\Xi_x$ and $\Xi_p$ the Vlasov
equation is recovered.

\section{A Simple Solution of the Fokker-Planck Equation in Beam Physics} \label{sec:fokkgauss}
Since radiation damping and quantum excitation counteract each other equilibria may exist, i.e.
distribution functions with $\partial f / \partial t = 0$. In action angle variables (cf. \S \ref{sec:emittance}) 
one can find the following $\phi$-independent equilibrium
\begin{eqnarray}
f = \frac{1}{2\pi \epsilon} e^{-\frac{J}{\epsilon}}
\label{GaussianBeamDistib}
\end{eqnarray}
Beams whose equilibrium distribution is given by Eq.~(\ref{GaussianBeamDistib}) are called Gaussian beams.
They can be encountered in electron-positron rings with significant synchrotron radiation.

\chapter{Physics of Particle Accelerators} \label{intro:accel}
\section{Applications and Limitations} \label{sec:applim}
Particle accelerators have become an invaluable tool for
high energy physics experiments. Due to the increasing complexity of
such machines particle accelerators themselves have become the
subject of detailed theoretical studies. Current machines can reach
center of mass energies of up to 2~TeV (Tevatron / Fermilab, as of 2001) and luminosities of
up to 6350 $pb^{-1}$ per year (CESR / Cornell, as of 2000). Such parameters give rise to
all kinds of instabilities. One can divide accelerators into
circular and linear machines. 
The best known linear accelerator (LINAC) is probably the Cathode Ray Tube which
accelerates electrons emitted from a filament by a large electrical
potential difference between the filament and a plate with a hole in it.
Due to the large potential differences needed for a high energy beam
such accelerators become technically unfeasible beyond 10MV.
In a Wideroe LINAC alternating current instead of direct current is used
to accelerate the particles. Charged particles travel through an array of 
conducting tubes with alternating polarity. Negatively charged particles 
in a gap between two tubes which leave a tube at negative potential are 
attracted by the next tube at positive potential. When the AC source
reverses the polarity the particles are inside a tube and are shielded
from the fields exerted by neighbouring tubes. To account for the 
increasing velocity of the particles the tubes have to increase in length.
Linear accelerators like the proposed Linear Collider tend to be rather 
long if high energies are desired. 
In a synchrotron the particles execute circular motion and can pass
the accelerating structure multiple times before reaching the desired
energy and being injected into a storage ring, for instance.
The increasing energy of particles passing through the accelerating
structure can be taken into account by adjusting the frequency of
the voltage applied to the structure. As the name suggests the purpose
of a storage ring is to store the accelerated particles. Usually there
is a rotating and a counter-rotating beam consisting of particles
and antiparticles, respectively \footnote{The never-completed SSC
was supposed to collide protons onto protons (instead of antiprotons)}.
At the interaction point (there may be multiple ones) the beams cross
each other and colliding particles may annihilate and produce new
particles which can be detected by a huge detector surrounding the
interacting point. However, the probability for such an event to happen
is small and huge amounts of energy would be wasted if the remaining
particles were just dumped. The idea of the storage ring is to keep
the particles circulating in the ring (possibly for many hours) until
they finally collide. 
There are two main problems with circular machines, though.  
Accelerated charges (in this case the acceleration stems from 
forcing the particles onto a circular orbit) emit synchrotron
radiation. The energy loss due to synchrotron radiation can make
the operation of such machines prohibitive at high energies.
At ultra-relativistic energies the radiated power is given by
\begin{eqnarray}
P_{\gamma} = \frac{c}{2\pi} C_{\gamma} \frac{E^4}{R}
\end{eqnarray}
where $C_{\gamma}$ is Sand's radiation constant for electrons
\begin{eqnarray}
C_{\gamma} = \frac{4\pi}{3} \frac{r_c}{(m_e c^2)^3} = 8.8575 \cdot 10^{-5}
{\rm m GeV^{-3}}~,
\end{eqnarray}
$r_c$ is the classical radius of the electron, $R$ is the radius of the ring
and $E$ is the energy of the electron. Thus, for high energies very large radii
are needed which makes accelerators expensive to build. Even if the energy loss
is not a concern the highest achievable energy is limited by the magnet technology.
Currently, the highest fields are provided by superconducting magnets, but
superconductivity breaks down at sufficiently high magnetic field strengths.
The current record for a continuous field is 45.1 Tesla measured at the National 
High Magnetic Field Laboratory at Florida State University \cite{NHMFL}.
The simplest conceivable circular accelerators consists only of dipole magnets which
bend the beam and two plates with a potential difference which accelerate the beam.
Such an accelerator is known as a weak-focusing machine. Its drawback is the beam size
which increases with increasing radius. Bigger and bigger machines were built until 
the apertures of the magnets became prohibitively big and expensive to produce.

\section{Strong Focusing}

The beam size was drastically reduced once strong-focusing machines were invented.
These machines contain quadrupole magnets with alternating gradients
(in addition to the dipole magnets). Therefore, the beam size and the force due
to the quadrupole magnets depends on the position $s$ in the ring which is a number
between zero and the circumference $2\pi R$ of the ring. The focusing force in the horizontal
and vertical plane, respectively, is  
\begin{eqnarray}
x'' = - K_x(s)x & \quad y'' = - K_y(s)y
\end{eqnarray}
with
\begin{eqnarray} \nonumber
K_x(s) & = & \frac{1}{R^2} + \frac{1}{B_{0y} R} \frac{\partial B_y}{\partial x} \\
K_y(s) & = & - \frac{1}{B_{0x} R} \frac{\partial B_x}{\partial y}
\end{eqnarray}
where $x$ and $y$ are the horizontal and vertical displacement from the design orbit, respectively.
Derivatives with respect to $s$ are denoted by a prime. 
\begin{figure}
\hspace{1in}
\includegraphics*[width=4in]{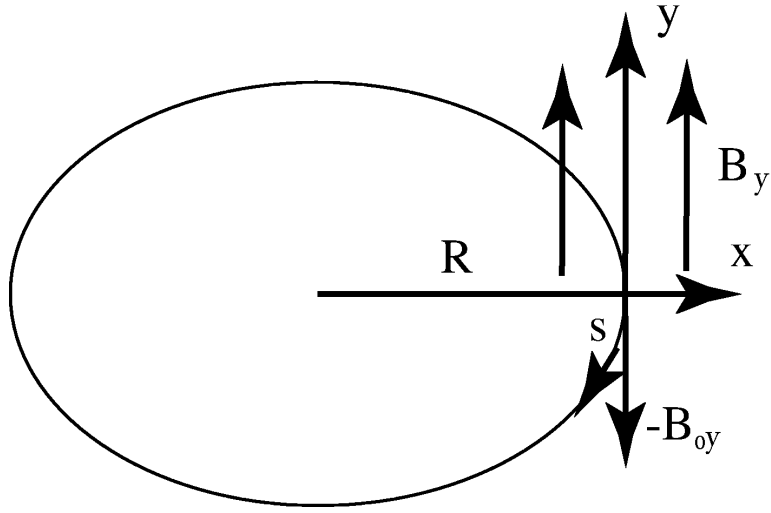}
\caption{Geometry of the strong focusing machine}
\label{fig:partacc}
\end{figure}
Let us focus on the equation for the horizontal motion. A solution which satisfies the initial conditions
$x(0)=x_0$, $x'(0)=x_0'$, $w(0)=w_0$, $w'(0)=w_0'$ and $\psi(0)=\psi_0$ is given by \cite{wienands2001,sands1970} 
\begin{eqnarray}
\nonumber
x = \left [ \left ( w_0' \sin \psi_0 + w_0^{-1} \cos \psi_0 \right ) x_0 - w_0 x_0' \sin \psi_0 \right ] w(s) \cos \psi(s) +
\\
\left [ - \left ( w_0' \cos \psi_0 - w_0^{-1} \sin \psi_0 \right ) x_0 + x_0' w_0 \cos \psi_0 \right ] w(s) \sin \psi(s)
\label{betatronsol}
\end{eqnarray}
where the width $w(s)$ of the beam is determined by the envelope equation 
\begin{eqnarray}
w(s)'' + K(s) w(s) + \frac{1}{w(s)^3} = 0 ~.
\label{envelope}
\end{eqnarray}
Furthermore, 
\begin{eqnarray}
\psi'(s) = w(s)^{-2} ~.
\end{eqnarray}
Individual particles oscillate 
(``betatron oscillations'') about the design trajectory $\nu$ times, but all particle 
orbits are contained in the ``envelope'' whose width is given by $w(s)$. 
The $w^{-3}$ term in Eq.~(\ref{envelope}) acts like a ``centrifugal barrier'' and gives
the beam envelope a non-zero width.
$\nu$ is called
the (machine) tune which is defined as
\begin{eqnarray}
\nu \equiv \frac{1}{2\pi} \oint \beta(s) K(s) ds
\end{eqnarray} 
where
\begin{eqnarray}
\beta(s) = w^2(s)
\end{eqnarray}
is called the betatron function and $\oint$ denotes the integration from zero to $2\pi R$.

\section{Weak Focusing}

In the case of weak focusing, i.e. $K(s)=K$, the above equations simplify dramatically. 
One obtains
\begin{eqnarray}
K = \beta^{-2} & \quad  \nu = 1 ~,
\end{eqnarray}
i.e. even in the absence of quadrupole magnets the beam executes one betatron oscillation per revolution.

\section{Emittance} \label{sec:emittance}

The forces exerted by dipoles, quadrupole and higher order magnets which may be needed
to correct for certain ``optical errors'' are conservative, i.e. the phase space density
occupied by the particles in a beam is constant. The phase space area in $(x,x')$ space
is $\pi \epsilon$ where $\epsilon$ is called the emittance. Rewriting Eq.~(\ref{betatronsol}) as
\begin{eqnarray}
x(s) = \sqrt{\epsilon \beta(s)} \cos \mu_0
\end{eqnarray}
the emittance is determined by
\begin{eqnarray}
\epsilon = \gamma_0 x_0^2 + 2 \alpha_0 x_0 x_0' + \beta_0 {x'_0}^2
\end{eqnarray}
where
\begin{eqnarray}
\alpha = - \frac{1}{2} \beta' \\
\gamma = \beta^{-1} ( 1 + \alpha^2 ) \\
\mu = \psi - \psi_0
\end{eqnarray}
Similarly, for $x'(s)$
\begin{eqnarray}
x'(s) = \sqrt{\frac{\epsilon}{\beta}} \left ( \sin \mu_0 - \alpha \cos \mu_0 \right )
\end{eqnarray}
Instead of using $x$ and $x'$ to describe the motion of a particle trajectory it is very often
advantageous to use the so-called ``action angle variables'' $\psi$ and $J=\epsilon/2$ 
because the latter is a constant of motion if the system is conservative.
Note that particles in an accelerator are subject to many non-conservative forces 
like synchrotron radiation damping and acceleration. 

\section{Beam-Beam Interaction}

All the limitations mentioned in section \ref{sec:applim} are well known, but there is a myriad of less obvious issues
that arise from the collective behavior of the particles which interact electromagnetically among themselves.
The most severe limit on the achievable luminosity is due to the beam-beam
interaction. If two oppositely charged beams which are slightly off-axis
collide head-on the rotating beam is deflected by the electromagnetic field
of the counter-rotating beam and vice versa. The beam-beam force is highly non-linear. Its presence causes
a tune shift $\xi$ at the interaction point. It is customary to estimate $\xi$ using the linear part of the
beam-beam force. $\xi$ is proportional to the number density of the beam and therefore $\xi$ is frequently used
to parametrize the strength of a beam. In $e^+e^-$ colliders the observed beam-beam limit is in the range
$0.02 \le \xi \le 0.1$ \cite{seeman,lep}. A further increase in $\xi$ increases the vertical emittance
and leads to particle loss. Attempts to cancel the beam-beam force have failed so far. In the DCI experiment
at SAL, Orsay, France, pairs of electron and positron beams were made to collide, i.e. both beams had zero net charge
\cite{leduff}. The result was disappointing. No significant improvement of the beam-beam limit was observed. This outcome
was explained by Derbenev in terms of a collective instability of the four-beam system \cite{derbenev}.
Therefore, it is reasonable to assume that a collective instability is responsible for the beam-beam instability
in the two-beam system as well. Indeed, collective oscillations are seen in computer simulations 
\cite{derbenev, Dikansky:1982jw, chao}. 
The linearized Vlasov equation has been used to study the stability of colliding beams \cite{derbenev, Dikansky:1982jw, chao}.
In \cite{chao} Chao and Ruth analyzed the stability properties of beams that are confined to motion in the vertical
direction (``flat beams''). They perturbed a ``water-bag'' equilibrium which has a uniform density within an ellipse in the phase-space
$(y,y')$. However, electron and positron beams tend to have a ``Gaussian'' distribution (cf. section \ref{sec:fokkgauss}). Very roughly 
speaking there are more particles in the inner region of the ellipse than in the outer region. In chapter \ref{ch:beambeam}
the stability properties of an electron beam colliding head-on with a positron beam are investigated.
They beams are assumed to be flat and have a ``Gaussian'' equilibrium distribution. Both angular and radial modes are considered.
Radial modes are modes which change the size of the ellipse. It is found that the radial modes have a profound 
influence on the stability of the system.

\section{Coherent Synchrotron Radiation}

The beam-beam interaction is ubiquitous in storage rings when beams collide, but it is by no means the only significant
instability. More recently an instability due to coherent synchrotron radiation (CSR) is being thoroughly investigated
which has been identified as a potential problem for the design of the proposed linear collider.
Since the particles in the beams have only one chance to collide one would like to decrease their emittances
as much as possible in order to achieve high luminosity. Such low emittance beams can only be produced in damping
rings where the emittance is reduced by emitting synchrotron radiation. The linear collider needs very short
bunches to operate, but beams in a damping rings are subject to other instabilities if their bunch length is too short.
The solution is to reduce the bunch length in a device known as a bunch compressor before the beam is injected
into the linear collider. Bunch compressors consist of an accelerating section and an arc section. In the arc
the beam emits synchrotron radiation whose wavelength may be close to the bunch length. In this scenario the electromagnetic
waves can modulate the beam in such a way that the bunches are equidistant. The radiation from individual bunches can
now interfere constructively and the incoherent radiation becomes coherent. For coherent radiation the total radiated power
scales as $N^2$ instead of $N$ where $N$ is the number of particles. The beam would lose all its energy almost instantly.
Therefore, it is important to know under which operating conditions one can avoid this effect. CSR has been observed
already in a couple of accelerator labs \cite{Loos2002,Kuske2003}.
In chapter \ref{ch:CSR} a simple model of a collisionless, relativistic, finite-strength, cylindrical layer of charged particles 
is presented which is capable of emitting coherent radiation. The particles interact with their retarded electromagnetic
self-fields in a way that allows them to clump together. Including the radial dynamics is difficult, and a small energy spread (which translates into a small non-zero thickness of the rotating layer) is one of the main requirements. It is shown that the betatron oscillations can lead to a significant decoherence which is responsible for the emission of a very characteristic spectrum. 
The stability properties are analyzed by solving the linearized Vlasov-Maxwell system of equations. The treatment resembles
work by Uhm, Davidson and Petillo \cite{Uhm1985} who examined the stability of a thin relativistic electron ring. However, their
interest is in the negative mass instability and their approximations are not suitable for electromagnetic effects like CSR.
A simpler model in which all particles were constrained to
move on a circle with fixed radius was presented in 1971 by Goldreich and Keeley \cite{GoldreichKeeley1971}. In this model the particles
initially move at constant speed, but they can gain or lose energy by interacting with the azimuthal component
of the electric field. However, no mechanism for fixing the radial degree of freedom is provided. It is unclear
whether (or under which conditions) the radial degree of freedom can be neglected. If, for instance, the circular
motion is due to an external magnetic field without gradient (``weak focusing'') an increase in energy translates into
an increase of the orbit radius whereas the velocity remains almost constant in the case of ultra-relativistic motion. 
This may not be very favorable for the development of a bunching instability.
A more realistic model was investigated by Heifets and Stupakov. In \cite{HeifetsStupakov2001,Stupakov2002} they analyze the stability of electrons executing circular motion. The radius of the individual particle orbit is determined by the energy of that particle, i.e. the radial motion and the relative longitudinal motion are coupled such that the problem has effectively only one degree of freedom. It is not entirely obvious under which conditions such an approach is valid. The model was extended by Byrd \cite{Byrd2002} to include the effect of a conducting beam
pipe which can serve as a cut-off of the allowed wavelengths. A conducting beam pipe can severely attenuate the CSR instability.
 
\chapter{Physics of Rotating Neutron Stars} \label{intro:neutron}
\section{Stellar Evolution}
A living star is supported against its own weight by the pressure 
it builds up as a result of heat generated in fusion reactions
inside the star. A young star generates its heat from the conversion of
hydrogen into helium by nuclear fusion. Once the supply of hydrogen
is exhausted in the core the star starts to shrink increasing
its temperature. This allows the star to burn the remaining hydrogen
in its shell. One has to distinguish two cases.
\subsection{$M < 8 M_{\sun}$ }
 In the red giant stage the shell expands leaving
behind the core which continues to shrink until a white dwarf
is formed. Fluid instabilities destroy the shell turning it
into a nebula.
This stage can be regarded as the end point of
the evolution of a light star. The white dwarf continues
to emit thermal radiation until it has completely cooled down.
\subsection{$M > 8 M_{\sun}$}
Heavier stars become hotter during contraction triggering fusion reactions
of heavier elements. Fusion stops once all material in the core
has been converted into iron \footnote{The element with the highest
binding energy is $^{62}$Ni and not $^{56}$Fe. Cf. an article
by Fewell \cite{fewell1995} on why iron is more abundant than nickel.}.
Like in the previous case burning continues in the outer shell.
Instead of a red giant a super red giant is formed with a radius bigger than 100 million
kilometers. The core is supported by the degeneracy pressure of non-relativistic electrons
and - as the star continues to contract - the electrons become relativistic and the
increase in pressure slows down. Furthermore, at relativistic electron energies
the protons can capture electrons which turns them into neutrons, thus reducing the 
degeneracy pressure of degenerate electrons. Photodissociation of iron leads to a 
polytropic index smaller than 4/3 rendering the core unstable to collapse \cite{Glendenning1997}.
The iron core implodes which generates a shock wave propagating outward.
The shock wave comes to a stop before it can leave the super red giant.
However, under certain conditions a bubble can form between the core and
the shock front. A small fraction of the binding energy of the star
is used to eject all the material of the star except the core in a supernova
explosion. Its mechanism is complicated, but it is believed to be caused
by convection and neutrinos transporting energy. The remnant is called a neutron
star because it is only supported by the degeneracy pressure of degenerate neutrons.
Sometimes the conditions under which a supernova explosion takes place are not
satisfied or an insufficient amount of matter is released. In this case
even the degenerate neutrons cannot prevent the star from collapsing even further
and a black hole is formed.
\section{Properties of Rotating Neutron Stars}
The radius of a typical neutron star is in the order of $R \sim 10~{\rm km}$ while its mass is in the order of $M_{\sun}$. 
In addition to a strong gravitational field on their surface they also posses a strong magnetic field which 
can be as strong as $10^8$ Tesla. The  field can be described by a magnetic dipole to a good 
approximation (cf. section \ref{brakingindex}). Like on earth the magnetic field may be created by the dynamo effect. Since charged particles are necessary to create a magnetic field a neutron star cannot consist entirely of neutrons. Indeed, it is believed that a neutron star contains a small fraction of electrons and protons in its core \cite{Shapiro1983}. 
Neutron stars born with a large amount of angular momentum are capable of emitting intense electromagnetic radiation in the radio frequency range. Such radio pulsars have rotation periods ranging from 1s down to 33ms for the Crab pulsar. In general the axis of rotation does not coincide with the alignment of the magnetic dipole moment. Therefore, the radiation sweeps out a cone about the axis of rotation.  Every time the observer's line of sight coincides with the magnetic axis a pulse of intense electromagnetic radiation is observed with a period equal to the period of the rotation of the star $T=2\pi/\Omega$ (lighthouse model). 

\begin{figure}
\includegraphics*[width=\columnwidth]{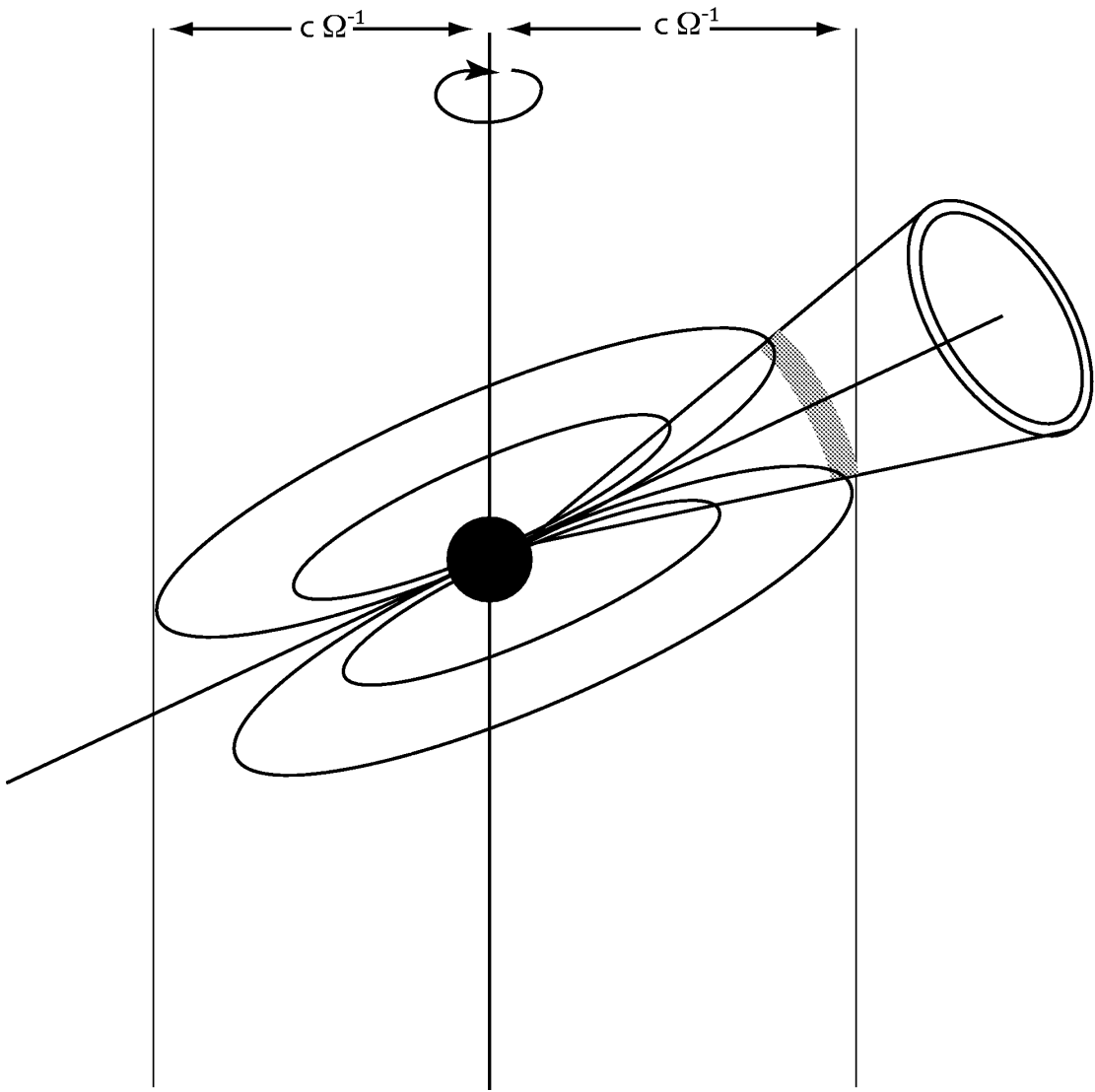}
\caption{Radio pulsars are spinning neutron stars with strong magnetic fields. At the magnetic poles their radiation follows the magnetic field lines. Since the field lines and the axis of rotation are misaligned the radiation sweeps out a cone.
The region in which the radiation is believed to be generated is shown in gray. A similar cone could be drawn on the other side
of the star.}
\label{fig:neutronstar}
\end{figure}

The discovery of millisecond pulsars ruled out white dwarfs as possible candidates for radio pulsars as can be seen from the following simple argument. For a stable star the centrifugal force exerted on particles at the surface of the star cannot exceed the force due to gravity. Thus,
\begin{eqnarray}
\Omega^2 R \le GM R^{-2}
\end{eqnarray}
White dwarfs are not sufficiently dense to satisfy the inequality above.

Due to its large mass and angular momentum the star posses a huge amount of kinetic energy which
powers the emission of the intense radiation. Therefore, as the pulsar continues to lose energy
its angular velocity has to decrease (``spin-down''). The details of the process converting kinetic energy to electromagnetic radiation are not completely understood yet. However, an argument based on conservation of energy suffices to relate some fundamental parameters of a pulsar. Assuming the energy loss is due to magnetic dipole radiation
\begin{eqnarray}
I \Omega \dot \Omega = \frac{dW}{dt} = P = \frac{1}{3} \Omega^4 R^6 B_0^2 ~,
\label{energyloss}
\end{eqnarray}
i.e. measuring the spin-down $\dot \Omega$ and equating the energy lost by a magnetic dipole to the loss of kinetic energy 
one can solve for the normal component $B_0$ of the magnetic field at the magnetic pole if the angular velocity $\Omega$, 
the radius $R$ and the moment of inertia $I$ are given. The spin-down can be measured very precisely. 
 
\section{Braking Index} \label{sec:brakingindex}
Eq.~(\ref{energyloss}) relates $\Omega$ to $\dot \Omega$
\begin{eqnarray}
\dot \Omega \propto \Omega^n
\label{brakingindex}
\end{eqnarray}
where the so-called braking index is denoted by $n$. According to Eq.~(\ref{energyloss}) the braking index is 3 if the
emission is due to dipole radiation. Deviations from $n=3$ would suggest that the simple model leading to Eq.~(\ref{energyloss})
is not completely accurate. Indeed, braking indices as low as $n=1.4$ have been measured.
The determination of the braking index is very simple if $\ddot \Omega$ is known. Differentiation of Eq.~(\ref{brakingindex})
and making use of the same equation again to eliminate the proportionality constant one obtains
\begin{eqnarray}
n = \frac{\ddot \Omega \Omega}{\dot \Omega^2}
\end{eqnarray}
\section{Some Fundamental Parameters of the Crab Pulsar}
We start by estimating how many charged particles could be present in the magnetosphere.
In a model by Goldreich and Julian \cite{Julian1969} the axis of rotation is assumed to coincide with the orientation of the magnetic dipole moment. Assuming the neutron star and its surrounding magnetosphere along the magnetic field lines are perfect conductors one obtains
\begin{eqnarray}
\vec{E} + (\vec{\Omega} \times \vec{r}) \times \vec{B} = 0 
\label{GJfields}
\end{eqnarray}  
Thus, the magnetosphere must have the Goldreich-Julian charge density
\begin{eqnarray}
n_{GJ} = (4\pi)^{-1} \nabla \cdot \vec{E} = \frac{\vec{\Omega} \cdot \vec{B}}{2\pi e} \sim 
10^{11} cm^{-3} (B/10^{12} G) (R/r)^3 [T / 1s]^{-1} 
\label{GJdensity}
\end{eqnarray}
at radius $r>R$ where $T=2\pi / |\vec{\Omega}|$.

\begin{figure}
\includegraphics*[width=\columnwidth]{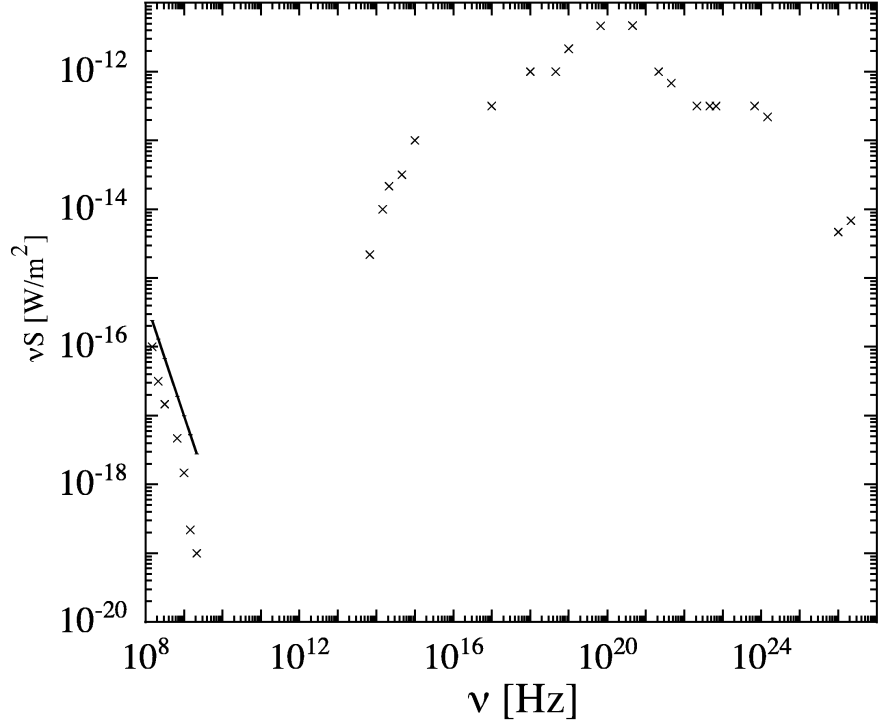}
\caption{The observed spectrum of the Crab pulsar extends from the radio regime to frequencies up to $10^{27} {\rm Hz}$ 
\cite{Thompson1996}.The straight line in the radio regime is proportional to $\nu^{-5/3}$}
\label{fig:spectrum}
\end{figure}

The energy loss can be determined from Eq.~(\ref{energyloss}) by measuring $\Omega$ and $\dot \Omega$.
\begin{eqnarray}
P = 4 \pi^2 I \dot T / T^3
\end{eqnarray}
It is in the order of $10^{39} {\rm erg~s}^{-1}$ for the Crab pulsar where \cite{lyne1998}
\begin{eqnarray}
\nonumber
M \sim 10^{31} {\rm kg} & R \sim 10^4 {\rm m} & I \sim 10^{33} {\rm kg~m}^2 \\ 
T = 33 {\rm ms} & \dot T = 4.22 \cdot 10^{-13} & B = 5.2 \cdot 10^{12} G
\end{eqnarray}

The highest detected frequency of the Crab pulsar is in the order of $10^{27} {\rm Hz}$, but the spectrum in
Fig.~\ref{fig:spectrum} starts to drop off significantly at $10^{24} {\rm Hz}$. 

\begin{figure}
\includegraphics*[width=\columnwidth]{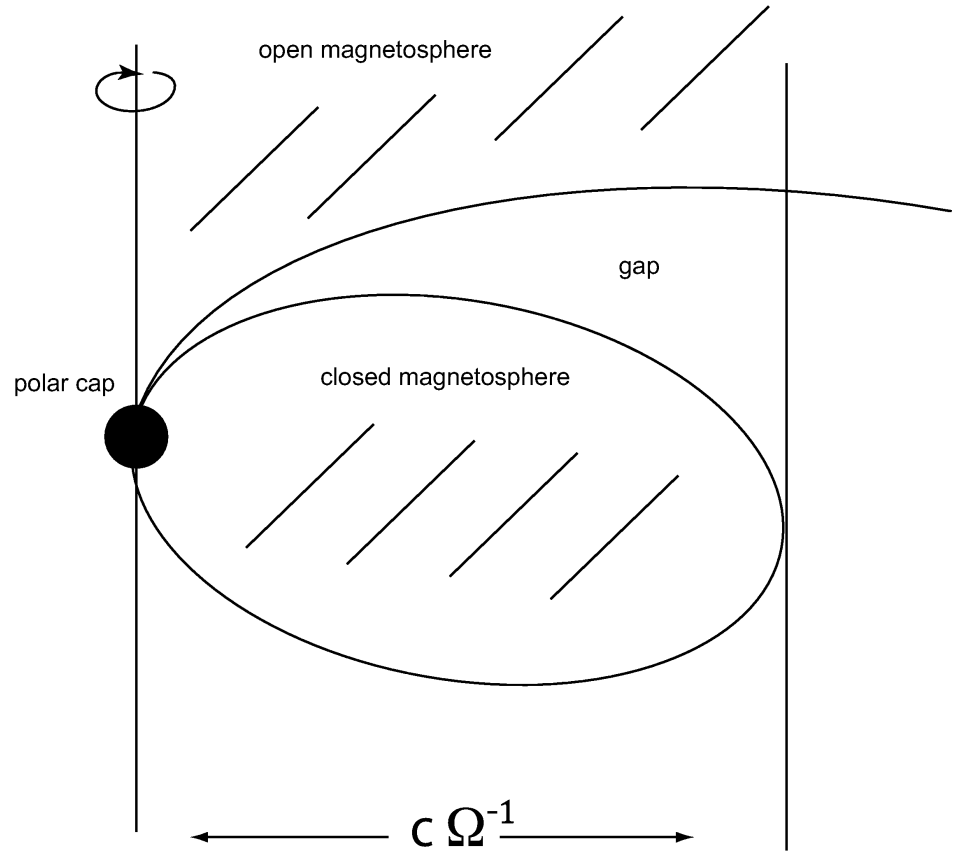}
\caption{Geometry of the regions surrounding a neutron star. The closed magnetosphere is followed by a gap in which strong
electric fields accelerate charged particles. The star is surrounded by a co-rotating magnetosphere which cannot extend
beyond the velocity-of-light cylinder, i.e. the radius at which the velocity of the particles would exceed the speed
of light at angular velocity $\Omega$ where $\Omega$ is the angular velocity of the star (and the co-rotating magnetosphere).}
\label{fig:magsphere}
\end{figure}

Integrating the Goldreich-Julian charge density from the surface of the surface to the velocity of light cylinder
(Fig.~\ref{fig:magsphere}) gives

\begin{eqnarray}
\nonumber
N & = & 10^{11} cm^{-3} (B/10^{12} G) [T / 1s]^{-1} R^3 \int_R^{c \Omega^{-1}} 4 \pi \cdot r^2 dr r^{-3} \\
  & = & 10^{11} cm^{-3} (B/10^{12} G) [T / 1s]^{-1} \cdot 4 \pi R^3 \ln \frac{c T}{2\pi R} \\
  & \sim & 10^{33} ~.
\end{eqnarray}
\section{Emission Mechanism}

Obviously, one would like to have a better understanding of how the rotational energy is converted into radiation and in particular how the radiation mechanism works. In this paragraph it is shown that incoherent synchrotron radiation cannot account for the observed brightness of the radio signal. The synchrotron radiation is partly reabsorbed by
the inverse Compton effect (cf. section \ref{sec:invCompton}). The ratio of the brightness temperature (cf. section \ref{CSR:brightnesstemp}) due to inverse Compton radiation to the brightness temperature due to synchrotron 
radiation is given by \cite{kellermann1969}
\begin{eqnarray}
L_{iC} / L_s \sim \frac{1}{2} (T_{max} / 10^{12} {\rm K})^5 (f_c / 1 {\rm MHz}) \left [ 1 + \frac{1}{2}
(T_{max} / 10^{12} {\rm K} )^5 (f_c / 1 {\rm MHz} ) \right ]
\end{eqnarray}
For $T_{max} < 10^{11} {\rm K}$ and an upper cutoff frequency $f_c \sim 10^5 {\rm MHz}$ in the radio regime 
this ratio is smaller than one, 
but for $T_{max} > 10^{12} {\rm K}$
\begin{eqnarray}
L_{iC} / L_s \sim (T_{max} / 10^{11} {\rm K})^{10}
\end{eqnarray}
Therefore, brightness temperatures exceeding $10^{12} {\rm K}$ are impossible to achieve with incoherent
synchrotron radiation ($P \propto N$) and some sort of coherent radiation mechanism ($P \propto N^2$) is required. 
The brightness temperature of the Crab pulsar is roughly $10^{31} {\rm K}$.

\section{Secondary Electron-Positron Plasma}
The strong magnetic field forces the electrons to move parallel to the field. Since magnetic fields cannot do any work a strong electric field with $\vec{E} \cdot \vec{B} \neq 0$ is necessary. Because Eq.~(\ref{GJfields}) implies 
$\vec{E} \cdot \vec{B} = 0$ the accelerating field must be due to a deviation from the Goldreich-Julian charge 
density \cite{ruderman1975}. Several effects accomplishing this have been suggested, e.g. general relativistic effects 
\cite{Muslimov1992} or the bending of the magnetic field lines \cite{arons1981}. 

Some photons emitted by the accelerated charges create secondary electron-positron pairs which screen the electric field except in compact regions called ``gaps''. The particles are accelerated in those gaps.
The plasma consisting of secondary particles has a distribution which differs significantly from the distribution
of primary particles. Instabilities of the primary plasma lead to the coherent emission of radio waves whereas 
instabilities of the secondary plasma lead to a non-thermal emission in the high frequency regime from
IR to $\gamma$-rays. This work focuses on instabilities found in the primary plasma (which may be induced by
the interaction with the secondary plasma).

\section{Free Electron Maser Emission}

Several mechanism were proposed to explain the emission of electromagnetic waves by the primary
plasma. This paragraph deals with the so-called free electron maser emission. It requires
a strongly modulated electric field parallel to the magnetic field. How such an electric
field could be generated in space is unknown. Rowe \cite{rowe1965} found that such a set-up is capable of self-amplification if the distribution is inversely populated, i.e. there is a region in phase space where 
the particle density increases as the energy increases. The electric field accelerates the
particles which therefore emit electromagnetic radiation. This radiation then modulates
the beam until bunches of particles radiate in phase (Actual free electron lasers
which are built in labs use a magnetic field from an undulator instead of an initially modulated
electric field. The undulator causes the particles to move on a helical path which then
emit synchrotron radiation.). Unfortunately, the growth rate as a function of energy decreases 
too rapidly to explain the high brightness temperatures that are observed.
  
\section{Two-stream Instability}
Particles of the secondary plasma might interact with those from the primary plasma
and (due to their very different distributions in phase space) trigger
a two-stream instability. All conditions for the development of a two-stream
instability are met. However, detailed calculations \cite{Kazbegi1987} show that the expected
growth rates are too low. Again, the high Lorentz factors and the low density
of the involved beams are the offending parameters. Another problem is posed
by the inability of the waves generated by the two-stream instability to
escape from the neutron star, i.e. they have to be converted into different
waves which can actually escape. This may involve some yet unknown non-linear
effects. Despite these shortcomings
the two-stream instability is considered to be the most promising candidate
for an explanation of the observed brightness temperatures by many authors \cite{usov2002}.
Two-stream instabilities due to electrons streaming against positrons in the secondary plasma
were also considered by several authors, e.g \cite{ChengRuderman1977}.
Again the reader is referred to the review article by Usov \cite{usov2002}.

\section{Curvature Radiation}

Curvature radiation was the first emission process
which was studied in the context of radio pulsars. The radiation
is due to the synchrotron radiation emitted by accelerated
charged particles where the acceleration originates from forcing
the particles to move along an arc. Since the observed radio waves
are polarized it is natural to attribute them to synchrotron
radiation. The coherence was explained by 
a maser-like mechanism by many authors. Most approaches based
on maser curvature emission ran into trouble because the conditions
for a self-amplifying maser instability were not satisfied.
On the other hand non maser-like mechanisms which started out
with a bunched distribution were heavily criticised because it was unclear
how the bunches could form and because of a lack of detailed models which
took the velocity spread of the distribution into account.

In chapter \ref{ch:CSR} one such model is presented. It is assumed that the radio 
emission is due to coherent curvature radiation which is produced
by small bunches of particles whose radiation interferes constructively.
For this approach the linear stability properties of a cylindrical, collisionless, relativistic layer made of charged particles whose axis of rotation is aligned with an external magnetic field are analyzed using the linearized Vlasov-Maxwell system.  The particles are allowed to interact with their own electromagnetic
self-fields. The bunches are seeded by arbitrarily small initial perturbations
which grow exponentially in time until the perturbations saturate. 
Knowledge of the saturation amplitude is a prerequisite for calculating
the intensity of the electromagnetic radiation and it can be estimated
considering the trapping of particles in the ``potential well of the wave''.
 

\section{Beaming}
If an isotropic emitter moves at relativistic speed an observer
at rest observes the radiation as if it was radiated into
a narrow cone pointing into the forward direction.
Its opening angle is approximately $\Gamma^{-1}$ where
$\Gamma$ is the Lorentz factor of the moving source.
This effectively increases the power measured by an observer
who can sample only a small solid angle.

\begin{eqnarray} 
\Delta \Omega^\prime = \Gamma^2 \Delta \Omega
\end{eqnarray} 
where a prime denotes quantities measured by the observer.

Furthermore, the emitted frequency undergoes a relativistic 
Doppler shift. For a source moving towards the observer
\cite{lightman1979}

\begin{eqnarray} 
\frac{\omega '}{\omega} = \sqrt{\frac{1+\beta}{1-\beta}} ~.
\end{eqnarray}

\section{Inverse Compton Radiation} \label{sec:invCompton}
It has been suggested that the radiation of the Crab pulsar
is caused by the inverse Compton effect. Low energy photons can be 
scattered by high energy electrons transferring energy from the
electron to the photon. No uniform magnetic field is needed
to initiate this process. However, the cross section for
sufficiently high energy transfers is too low to account for
the observed brightness temperature ruling out this radiation
mechanism.

\section{Self-absorption}
It is conceivable that the radiation emitted by a plasma
is partly reabsorbed. Indeed, this so-called ``synchrotron self-absorption''
is well known \cite{lightman1979} and can be derived for any source using
Einstein coefficients. Below the transition frequency
which corresponds to the mean particle energy the intensity of the observed
spectrum scales as $\nu^{5/2}$ regardless of the particular power
law obeyed by the source. Because in the radio regime the brightness temperature 
is many orders of magnitude bigger than the associated particle energy
this effect is irrelevant for the understanding of the radio spectrum
of a pulsar.

\chapter{Beam-Beam Interaction in Storage Rings$^\dagger$} \label{ch:beambeam}
\symbolfootnotetext[2]{This chapter appeared as a journal article \cite{Schmekel:2003cs}.
Reprinted in modified form with kind permission from the American Physical Society. 
\copyright ~ 2003 by the American Physical Society}

\section{Introduction}
Colliding particle bunches in a storage ring exert an electromagnetic 
force on each other. The beam-beam parameter $\xi$ is the tune shift exerted by 
one bunch on a particle near the center of the opposing bunch. It is a 
useful measure of the strength of the beam-beam interaction. 
A limiting value of $\xi_y$ is reached in an $e^+e^-$ 
collider when further increases in beam intensity lead to particle loss 
or to an increase in the vertical emittance of the beam.
In $e^+e^-$ colliders, where the action of radiation excitation and 
damping produces a flat beam, the observed vertical beam-beam parameter limit is in the 
approximate range $0.02 \le \xi_y \le 0.1$ \cite{seeman, lep}.
At present it is not known whether the
emittance increase is due to an incoherent, single-particle effect or to a
coherent, collective instability of the colliding beams.  The DCI 
storage rings at LAL, Orsay, France, used a pair of $e^+$ and $e^-$ beams to collide with 
another pair, in an attempt to cancel the beam-beam force~\cite{leduff}.  It was found, 
however, that the beam-beam limit in DCI was not significantly improved by the charge 
cancellation. Derbenev~\cite{derbenev} explained this result in terms of a 
collective instability of the four-beam system and in \cite{krishnagopal92} the performance
of DCI was analyzed numerically. This suggests that the beam-beam limit for 
two-beam $e^+e^-$ colliders may also be due to a collective instability. 
Simulations in \cite{podobedov, rogers, krishnagopal91} show collective oscillations
of the beam at the beam-beam limit.

In references \cite{derbenev, Dikansky:1982jw, chao} the stability of the 
colliding beams was examined by solving the Vlasov
equation for an equilibrium distribution with small perturbations. Chao and
Ruth \cite{chao} considered a beam-beam model in which motion was 
confined to the vertical plane, and in which the beam has a ``water-bag'' equilibrium 
distribution (uniform within an ellipse in phase space). Synchrotron radiation damping and 
excitation were not considered. When the Vlasov equation was solved for a linearized 
beam-beam force, coherent beam modes were found to be unstable near each resonance.
In \cite{Dikansky:1982jw} the stability of a Gaussian equilibrium distribution was
analyzed with the Vlasov equation for round beams where the beam-beam force
can be expanded in Bessel functions.
A flat beam model with a Gaussian distribution and synchrotron radiation was studied
in \cite{hirataPRL87, hirataPRD88} under the assumption that the distribution always remains Gaussian.
A similar approach was chosen in \cite{chaoSSC88} for a purely linear beam-beam force.
The findings of these models, e.g. flipflop solutions and period-n solutions are verified
numerically in \cite{krishnagopal96} where the behavior of flat and round beams is considered 
as well. 

In this paper we extend the model of Chao and Ruth to a Gaussian 
equilibrium distribution. In Section \ref{sc:eom} we set up the
equations of motion for the phase space distribution and its 
perturbations, and linearize the beam-beam force.  
In Section \ref{sc:radial} we solve the equations of motion for
radial and angular modes up to first order in the displacement from 
the design trajectory and discuss the implications of our results.

\section{Beam Evolution} \label{sc:eom}
We model the flat beam as a current sheet which is uniform in the horizontal direction, $x$,
and consider only motion in the vertical direction, $y$. Consider one-dimensional
phase space distributions $\psi_1(y,y',s)$ and $\psi_2(y,y',s)$ of the two beams which are normalized to unity.
Then the deflection from the second (first) beam on a particle in the first (second) beam is
\begin{eqnarray}
\Delta y'_{1,2}= - I_{\psi_{2,1}}(y,s) ,
\label{eq:beambeamkick}
\end{eqnarray}
where we define
\begin{eqnarray}
I_{\psi}(y,s) \equiv \frac{4 \pi N r_e}{\gamma}
\int_{-\infty}^{\infty} d\overline{y}\sgn(y-\overline{y})\int_{-\infty}^{\infty}d\overline
{y}^{\prime}\psi(\overline{y},\overline{y}^{\prime},s)
\label{eq:integral}
\end{eqnarray}
and $N$ is the number of particles per unit width in $x$ and $r_{e}$
is the classical radius of the electron. Both beams are assumed to
have the same number of particles per unit width. The equations
describing the motion of $\psi_{1,2}$ are given by the two Vlasov
equations
\begin{eqnarray}
\frac{\partial \psi_{1,2}}{\partial s} + y^{\prime} \frac{\partial \psi_{1,2}}{\partial y} -
K(s)y \frac{\partial \psi_{1,2}}{\partial y^{\prime}}
- \frac{\partial \psi_{1,2}}{\partial y^{\prime}} \delta_p(s) I_{\psi_{2,1}}(y,s) = 0
\label{eq:vlasov}
\end{eqnarray}
where the periodic delta function and the unperturbed focusing function are denoted by $\delta_p(s)$
and $K(s)$, respectively. We want to determine whether the beam is stable.
That is, we want to know if small perturbations of the phase space density grow. Thus, we choose
a perturbative ansatz
\begin{eqnarray}
\psi_{1,2} = \psi_0 + \Delta \psi_{1,2}
\label{eq:perturbative}
\end{eqnarray}
where $\psi_0$ is the equilibrium distribution, i.e. a solution of
Eq.~(\ref{eq:vlasov}) with $\psi_1(y,y',s)=\psi_2(y,y',s) =
\psi_0(y,y',s) = \psi_0(y,y',s+C)$, where the circumference of the
ring is denoted by $C$. Substituting Eq.~(\ref{eq:perturbative}) into
Eq.~(\ref{eq:vlasov}), subtracting Eq.~(\ref{eq:vlasov}) written for the
equilibrium distribution, and neglecting the term which contains a
product of two perturbations we find
\begin{eqnarray}
\frac{\partial \Delta \psi_{1,2}}{\partial s} + y^{\prime} \frac{\partial \Delta \psi_{1,2}}{\partial y} -
\frac{\partial \Delta \psi_{1,2}}{\partial y^{\prime}} F(y,s)
- \delta_p(s) \frac{\partial \psi_0}{\partial y^{\prime}} I_{\Delta \psi_{2,1}} = 0  ,
\label{eq:perturbed}
\end{eqnarray}
where
\begin{eqnarray}
F(y,s)=K(s)y + \delta_{p}(s) I_{\psi_0}(y)  .
\label{eq:F(y,s)}
\end{eqnarray}
If we approximate the beam-beam force as linear in $y$
\begin{eqnarray}
F(y,s) \approx F(s)y = K(s)y + \delta_p(s) I^1_{\psi_0}
\label{eq:F(s)y}
\end{eqnarray}
with
\begin{eqnarray}
I^1_{\psi_0} = I_{\psi_0}(0) + \left . \frac{\partial}{\partial y}
I_{\psi_0}(y) \right |_{y=0} \cdot y
\label{eq:I1} 
\end{eqnarray}
we can replace $K(s)$ by the perturbed focusing function $F(s)$ to compute
the perturbed Twiss parameters. In the next step we transform Eq.~(\ref{eq:perturbed})
to action-angle coordinates
\begin{eqnarray}
y=\sqrt{2 \beta J} \cos \phi
& \quad
y'=-\sqrt{\frac{2 J}{\beta}} \left ( \sin \phi + \alpha \cos \phi \right )  .
\label{eq:actionangle}
\end{eqnarray}
The betatron function is perturbed by the linearized beam-beam kick from
$\psi_0$.  We form the linear combinations for the $\sigma$- and the
$\pi$-mode
\begin{eqnarray}
f_{\pm} = \Delta \psi_1 \pm \Delta \psi_2 .
\label{eq:lincomb}
\end{eqnarray}
Then Eq.~(\ref{eq:perturbed}) can be decoupled and rewritten in
action-angle coordinates as
\begin{eqnarray}
\frac{\partial f_{\pm}}{\partial s} + \frac{1}{\beta} \frac{\partial
f_{\pm}}{\partial \phi} \mp \delta_p(s) \frac{\partial
\psi_0}{\partial y'} I^1_{f_{\pm}} = 0 .
\end{eqnarray}
The quantity $\frac{\partial \psi_0}{\partial y'} = - \sqrt{2 \beta J}
\left ( \sin \phi \frac{\partial}{\partial J} \psi_0 + \frac{\cos
\phi}{2J} \frac{\partial}{\partial \phi} \psi_0 \right )$ simplifies
since the linearization of the beam-beam force in Eq.~(\ref{eq:F(s)y})
leads to $\psi_0 = \psi_0(J)$ and we are left with
\begin{eqnarray}
\frac{\partial f_{\pm}}{\partial s} + \frac{1}{\beta} \frac{\partial
f_{\pm}}{\partial \phi} \pm \sqrt{2 \beta J} \sin \phi \delta_p(s)
\frac{\partial \psi_0}{\partial J} I^1_{f_{\pm}} = 0 .
\label{eq:fokkeractionangle}
\end{eqnarray}
In the following discussion we omit the label $\pm$.

\section{Solving the Equations of Motion} \label{sc:radial}
When the interaction term in Eq.~(\ref{eq:vlasov}) is not considered,
any differentiable distribution which depends solely on $J$ is an
equilibrium distribution.  In general, $\psi_0$ will be a function of
both $J$ and $\phi$. Fortunately, an arbitrary differentiable function
of $J$ is an equilibrium distribution, at least to linear order in $y$
after introducing the perturbed betatron function. We choose a
Gaussian equilibrium distribution
\begin{eqnarray}
\psi_{0}(J)=\frac{1}{2 \pi \epsilon} e^{-\frac{J}{\epsilon}}
\label{eq:equidist}
\end{eqnarray}
since in the presence of damping and quantum excitation the beam distribution naturally tends to
a Gaussian distribution.
The deflection of a particle due to the presence of a Gaussian beam can be obtained from Eq.~(\ref{eq:integral}), 
\begin{eqnarray}
I_{\psi_0}(y) =
\frac{4 \pi N r_e}{\gamma} 
\erf \left ( \frac{y}{\sqrt{2 \beta \epsilon}} \right ) .
\label{eq:deflection}
\end{eqnarray}
We expand the linearized version of Eq.~(\ref{eq:fokkeractionangle}) using the ansatz
\begin{eqnarray}
f(J,\phi,s) = \sum_{n'=0}^{\infty} \sum_{l'=-\infty}^{\infty} g_{n'
l'}(s) e^{-\frac{J}{\epsilon}} L_{n'} \left ( \frac{J}{\epsilon}
\right ) e^{i l' \phi} .
\label{eq:ansatz}
\end{eqnarray}
Since the perturbation must be periodic in $\phi$ we can express the
$\phi$ - dependence in terms of a Fourier series. The orthogonality
relation for the Laguerre polynomials comes with the convenient weight
factor $e^{-\frac{J}{\epsilon}}$ which simplifies working with
expressions that contain the Gaussian equilibrium
distribution. Furthermore, using the weight factor in the set of basis
functions, guarantees that the perturbation falls off as $J
\longrightarrow \infty$. 
We will refer to the modes represented by the first and second index in $g_{nl}$
as ''radial'' modes and ''angular'' modes, respectively, i.e. these words refer
to the two-dimensional phase space described by action-angle variables. 
With Eq.~(\ref{eq:ansatz}) the linearization in
Eq.~(\ref{eq:I1}) leads to
\begin{eqnarray}
I_f \propto \int_{-\infty}^{\infty}d \overline{y}' \left[ \int_{-\infty}^{y}
d\overline{y} f(\overline{y},\overline{y}',s) -
\int_{y}^{\infty}
d\overline{y} f(\overline{y},\overline{y}',s)\right]
\end{eqnarray}
\begin{eqnarray}
I^1_f \propto
\int_{-\infty}^{\infty} d \overline{y}'\left[ \int_{-\infty}^{0} d\overline{y} 
f(\overline{y},\overline{y}',s) - 
\int_{0}^{\infty} d\overline{y} f(\overline{y},\overline{y}',s)
+ 2y f(0,\overline{y}',s)\right]
\end{eqnarray}
Using $\int_{0}^{\infty} e^{-x} L_n(x) dx = \delta_{n0}$ the first
part of $I^1_f$ is given by
\begin{eqnarray}
\int_0^{\infty} dJ \int_{-\pi/2}^{\pi/2} d \phi \left [ f(J,\phi +
\pi,s) - f(J,\phi,s) \right ] = -4 \epsilon \sum_{l'=-\infty}^{\infty}
g_{0 (2l'+1)} \frac{(-1)^{l'}}{2l'+1} .
\end{eqnarray}
The second part is given by
\begin{eqnarray}
2 \int_{-\infty}^{\infty} d \overline{y}' f(0,\overline{y}',s) = 2
\int_{0}^{\infty} d J \frac{1}{\sqrt{2 \beta J}} [f(J,\pi/2,s) +
f(J,-\pi/2,s) ] = \nonumber\\ \sqrt{4 \pi} \sqrt{\frac{2
\epsilon}{\beta}} \sum_{n'=0}^{\infty} \sum_{l'=-\infty}^{\infty}
g_{n' 2l'} (-1)^{l'} \frac{(2n')!}{(2^{n'} n'!)^2} .
\end{eqnarray}
Here we have made use of
\begin{eqnarray}
\int_0^{\infty} \frac{1}{\sqrt{x}} e^{-x} L_n(x) dx =
\frac{(2n)!}{(2^n n!)^2} \equiv \sqrt{\pi} P_n.
\end{eqnarray}
Inserting $I^1_f$ into Eq.~(\ref{eq:fokkeractionangle}), projecting this
equation onto our chosen set of basis functions by means of the
orthogonality relation of the Laguerre polynomials
\begin{eqnarray}
\int_{0}^{\infty} e^{-x} L_n(x) L_m(x) dx= \delta_{nm}
\end{eqnarray}
and using 
\begin{eqnarray}
\int_0^{\infty} \sqrt{x} e^{-x} L_n(x) dx = - \frac{(2n)!
  \sqrt{\pi}}{2(2n-1)(2^n n!)^2} = -\frac{\sqrt{\pi}}{2(2n-1)} P_n
\end{eqnarray}
and
\begin{eqnarray}
\int_{0}^{\infty} x e^{-x} L_n(x) dx = \delta_{n0} - \delta_{n1}
\end{eqnarray}
we obtain
\begin{eqnarray}
\frac{\partial g_{nl}}{\partial s} + \frac{il}{\beta} g_{nl} = \mp
\delta_p(s) \xi \sum_{n'=0}^{\infty} \sum_{l'=-\infty}^{\infty}
M_{nl,n'l'} g_{n'l'}\ , \ \ \xi = \frac{N r_e}{\gamma} \sqrt{\frac{2
\beta^*}{\pi \epsilon}}\ ,
\label{eq:detgnl}
\end{eqnarray}
where
\begin{eqnarray}
\nonumber
\frac{1}{2n-1} P_n \! \left ( \delta_{l,1} - \delta_{l,-1} \right ) \!
\delta_{n',0} (-1)^{\frac{l'-1}{2}}\frac{1}{l'}a_{l'} \! + \! \left
(\delta_{n,0} - \delta_{n,1} \right ) \! \left (
\delta_{l,2}-\delta_{l,-2} \right ) \! P_{n'}(-1)^{\frac{l'}{2}} b_{l'}
\\
\equiv (2\pi i)^{-1} M_{nl,n'l'} \quad\quad\quad\quad
\label{eq:matrix}
\end{eqnarray}

The coefficients $a_l$ are $1$ for odd $l$ and $0$ for even $l$ and vice versa for the coefficients $b_l$. 
Each column and each row of the matrix $M$ refers to one particular combination of an $n$ and an $l$ value.

\section{Dynamic Tune}
We calculate the tune $\nu$ in terms of the unperturbed tune $\nu_0$ by 
means of Eq.~(\ref{eq:defdynamictune}).
\begin{eqnarray}
\nu - \nu_0 = \frac{1}{4 \pi} \oint \beta(s) (F(s)-K(s)) ds\ .
\label{eq:defdynamictune}
\end{eqnarray}
In order to obtain $F(s)-K(s)$ the deflection in
Eq.~(\ref{eq:deflection}) is linearized.  This gives
\begin{eqnarray}
\nu - \nu_0 = \frac{N r_e}{\gamma} \sqrt{\frac{2 \beta^*}{\pi
\epsilon}} \equiv \xi \ ,
\end{eqnarray}
where $\beta^*$ denotes the beta function at the interaction point.

\section{Coherent Beam-Beam Instability}
We solve the ODE (\ref{eq:detgnl}) and rewrite the solution in matrix form
such that the beam transport after one turn is described by a matrix $T$ which
acts on a column vector $G$ that contains all $g_{nl}$, i.e. $G(C) = T G(0)$. 
We parametrize the beam-current by the linear tune shift parameter $\xi$.
One obtains the following relation for the $g_{nl}$'s immediately
before and immediately after the interaction point by integrating
through the interaction point:
\begin{eqnarray}
G(0^+) - G(0^-) = \pm \xi M G(0^-)
\end{eqnarray}
There is no coupling among different Fourier components between collisions.
In this case Eq.~(\ref{eq:detgnl}) simplifies to
\begin{eqnarray}
\frac{\partial g_{nl}}{\partial s} + \frac{il}{\beta(s)} g_{nl} = 0 ,
\end{eqnarray}
which is solved by
\begin{eqnarray}
g_{nl}(C^-)=g_{nl}(0^+) e^{-il \int_0^C \frac{1}{\beta(s)} ds}=g_{nl}(0^+) e^{-2 \pi il \nu} .
\end{eqnarray}
The one-turn transfer matrix becomes
\begin{eqnarray}
T_{\pm} = R \left ( \mathbbm{1} \pm \xi M \right )
\end{eqnarray}
where $R$ is a diagonal matrix which has the elements $e^{-2 \pi il \nu}$ on its diagonal. 
The matrix $M$ has the following properties which follow immediately from Eq.~(\ref{eq:matrix}),
\begin{eqnarray}
M_{nl,n'l'} = 0 \textrm{ for $l+l'=$ odd}
\nonumber\\
M_{nl,n'-l'} = M_{nl,n'l'}
\nonumber\\
M_{n-l,n'l'} = -M_{nl,n'l'}
\nonumber\\
M^*_{nl,n'l'} = -M_{nl,n'l'}
\label{eq:properties}
\end{eqnarray}
In order to decide whether the system is stable or not we have to find
out what happens to an arbitrary initial perturbation after a large
number of turns, i.e. one needs to consider the limit $T^N$ where $N
\longrightarrow \infty$. Every matrix norm of the latter quantity
tends to infinity if the absolute value of one eigenvalue of $T$ is
bigger than $1$.  To analyze the stability for a given tune $\nu$ and
a beam-beam parameter $\xi$, we therefore compute the eigenvalue
$\lambda_{max}$ that has the largest modulus. In case of instability
we compute the corresponding eigenvector $G$ and find its component
$g_{nl}$ which has the largest modulus. This indicates that the
instability mainly drives the radial mode $n$ and angular mode $l$,
causing $f$ to be dominated by
$L_n(\frac{J}{\epsilon})e^{il\phi}$. Since the perturbation $f$ must
be real taking its complex conjugate must leave $f$ invariant which
gives the constraint $g_{nl}=g^*_{n-l}$. Indeed Eq.~(\ref{eq:detgnl})
is invariant under complex conjugation and replacing $l
\longrightarrow -l$. It follows that the coefficients of $T$ have the
property $T_{n-l,n'-l'}=T_{nl,n'l'}$, which also follows from
Eq.~(\ref{eq:properties}).  This requires that eigenvalues of $T$ are
either real or come in a pair with their complex conjugate: Let $S$ be
a matrix performing the transformation $l \longrightarrow -l$ then we
have $STSSG = \lambda SG$ and finally $T (SG^*) = \lambda^* (SG^*)$.
Therefore, the $l$-mode and the $-l$ mode are always excited
simultaneously with equal strength.

\section{Results and Discussion}
\begin{figure}
\includegraphics[angle=270,width=\columnwidth]{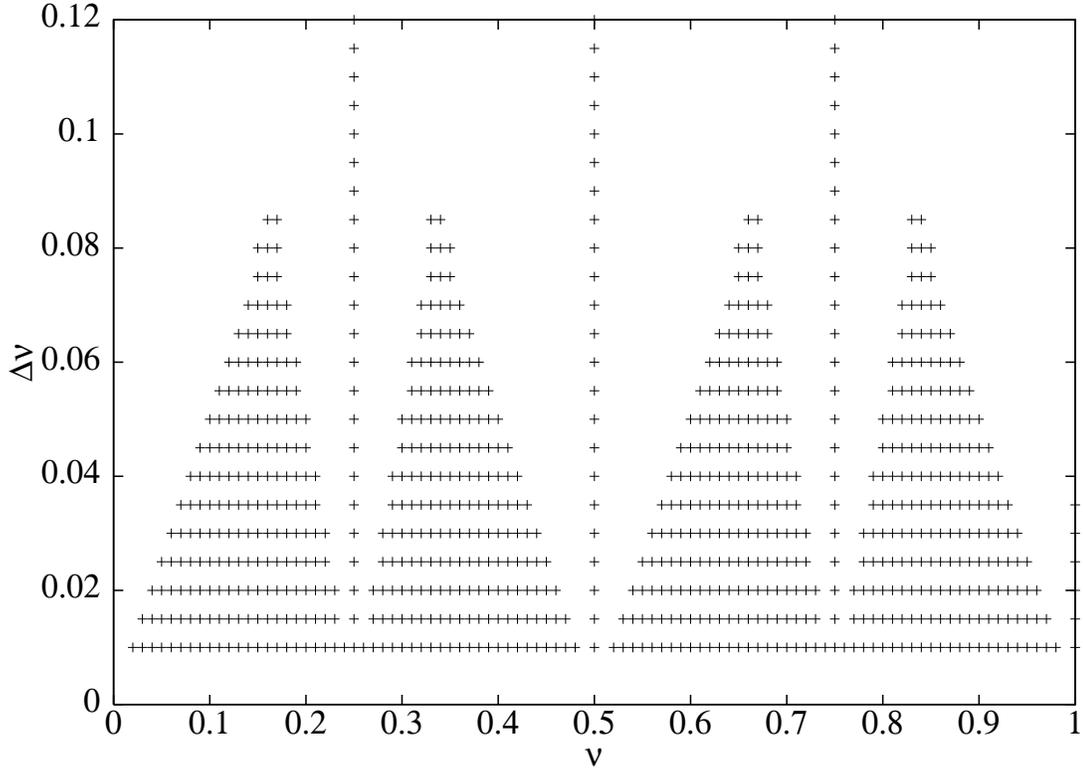}
\caption{Stability diagram for $n=0$, $l=-2 \ldots 2$} 
\label{fg:02radial} 
\end{figure}

\begin{figure} 
\includegraphics[angle=270,width=\columnwidth]{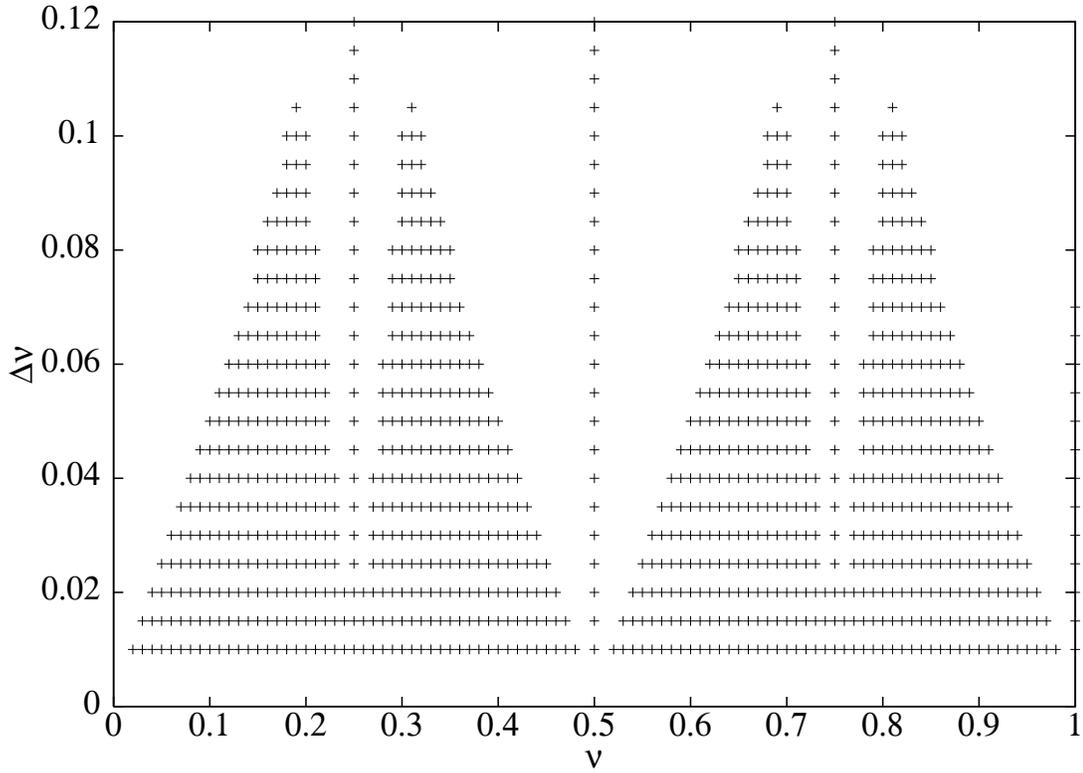}
\caption{Stability diagram for $n=0 \ldots 2$, $l=-2 \ldots 2$} 
\label{fg:22radial} 
\end{figure}

In Fig.~\ref{fg:02radial} and \ref{fg:22radial} we varied the tune
$\nu$ between $0$ and $1$ and the beam-beam parameter $\xi$ between
$0$ and $0.12$. A point has been plotted if the absolute value of all
eigenvalues of $T$ is smaller than or equal to $1$ for both the
$\sigma$- and the $\pi$-mode.  We truncated $T$ to the indicated
modes. In Fig.~\ref{fg:02radial} only the 5 modes $l=-2 \ldots 2$ for
$n=0$ were considered. In Fig.~\ref{fg:22radial} we included the same
angular modes for $n=0 \ldots 2$.  The first and second order resonances
can be recognized clearly. Resonances of orders higher than 2 cannot
be expected in our linearized model. It is interesting to note that
the inclusion of radial modes stabilizes the motion of the beam so
that a larger $\xi$ can be tolerated.

\begin{figure} 
\includegraphics*[width=\columnwidth]{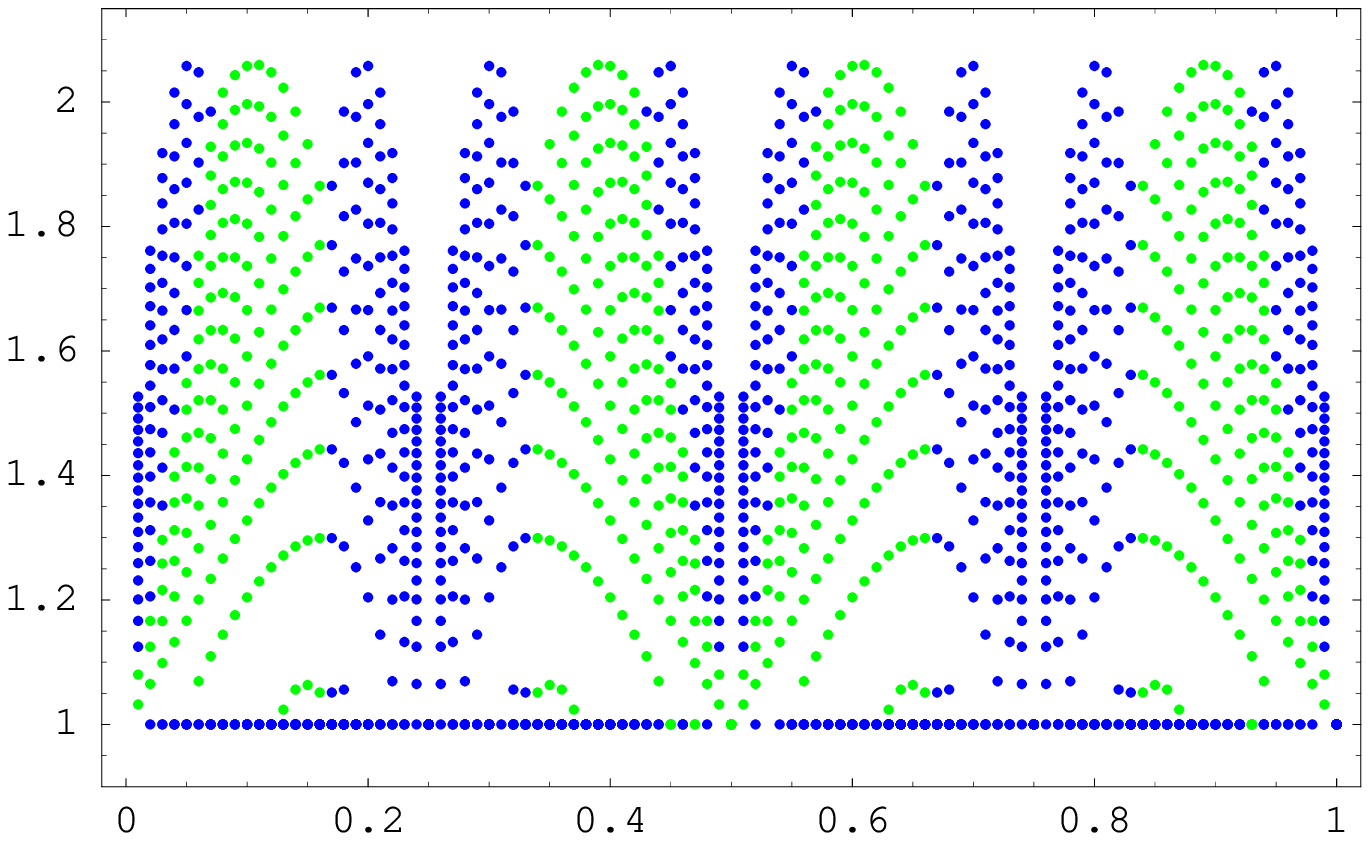}
\setlength{\unitlength}{1mm}
\flushleft\begin{picture}(0,0)(0,0)
\put(-15,61){$|\lambda_{{\rm max}}|$}
\put(76,4){$\nu$}
\end{picture}
\caption{Absolute value of the largest eigenvalue $\lambda_{max}$
vs.~tune.  Gray points indicate unstable $l=\pm1$ modes and black
points indicate unstable $l=\pm2$ modes.  The following modes
were included: $n=0$, $l=-2 \ldots 2$}
\label{fg:02radial000} 
\end{figure}

\begin{figure}
\includegraphics*[width=\columnwidth]{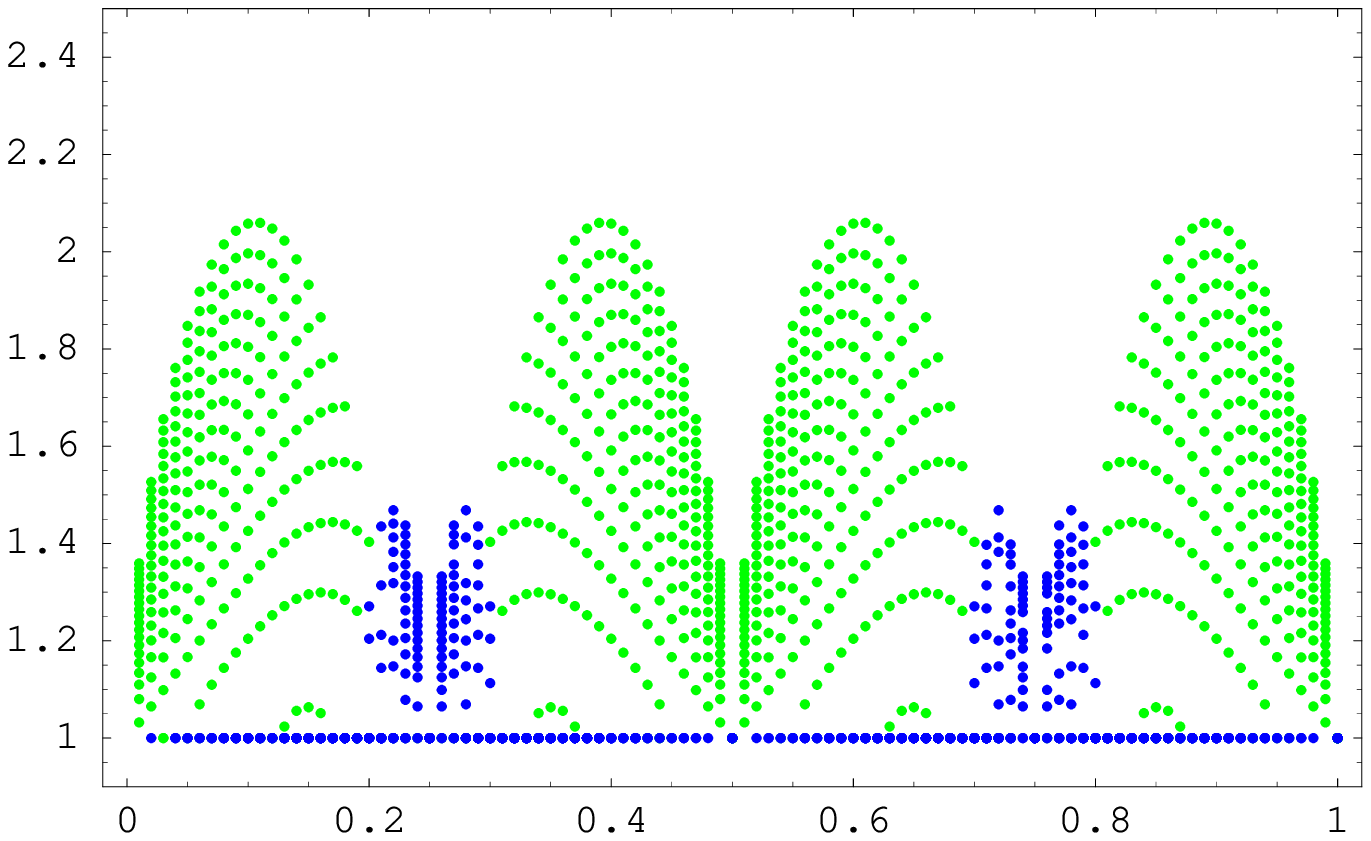}
\setlength{\unitlength}{1mm}
\flushleft\begin{picture}(0,0)(0,0)
\put(-15,61){$|\lambda_{{\rm max}}|$}
\put(76,4){$\nu$}
\end{picture}
\caption{Same as Fig.~\ref{fg:02radial000}, but for $n=0 \ldots 1$,
$l=-2 \ldots 2$}
\label{fg:22radial000} 
\end{figure}

In Fig.~\ref{fg:02radial000} and \ref{fg:22radial000} we again varied
$\nu$ and $\xi$ and plotted the largest eigenvalue $|\lambda_{max}|$
vs.~$\nu$ and determined which mode becomes unstable by selecting the
biggest component of the eigenvector which is associated with the
largest eigenvalue. The plot shows that in the absence of dynamics in
the radial direction $l=\pm 1$ and $l=\pm 2$ modes become unstable in
the vicinity of $\nu=0.5$, but in Fig.~\ref{fg:22radial000} only $l =
\pm 1$ modes are excited around $\nu=0.5$.  Furthermore, the unstable
$l= \pm 2$ modes which accumulate in the vicinity of $\nu=0.25$ and
$\nu=0.75$ are attenuated if the $n=1$ mode is included.  Therefore,
the radial motion leads to a damping of the $l= \pm 2$ modes.

\begin{figure} 
\includegraphics*[width=\columnwidth]{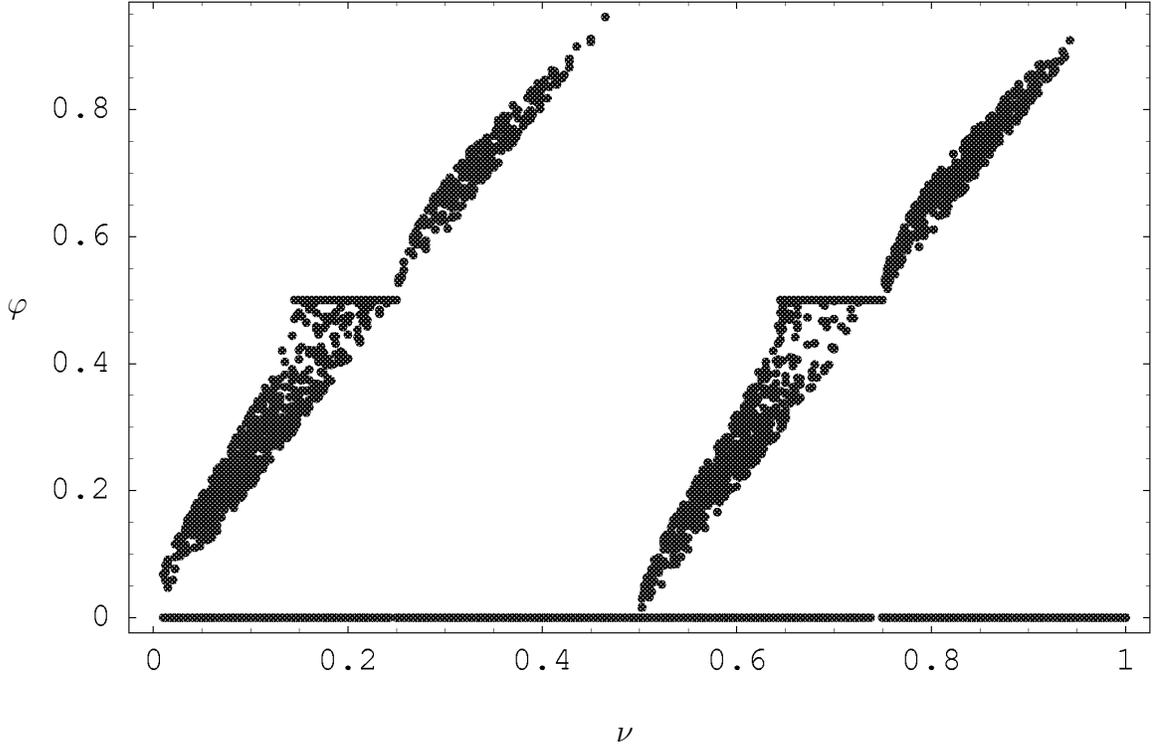}
\setlength{\unitlength}{1mm}
\flushleft\begin{picture}(0,0)(0,0)
\put(-5,61){$\varphi$}
\put(76,4){$\nu$}
\end{picture}
\caption{Phase vs.~perturbed tune for $n=0$, $l= \pm 2$ modes
($\pi$-mode only). }
\label{fg:02radial000+} 
\end{figure}

\begin{figure}
\includegraphics*[width=\columnwidth]{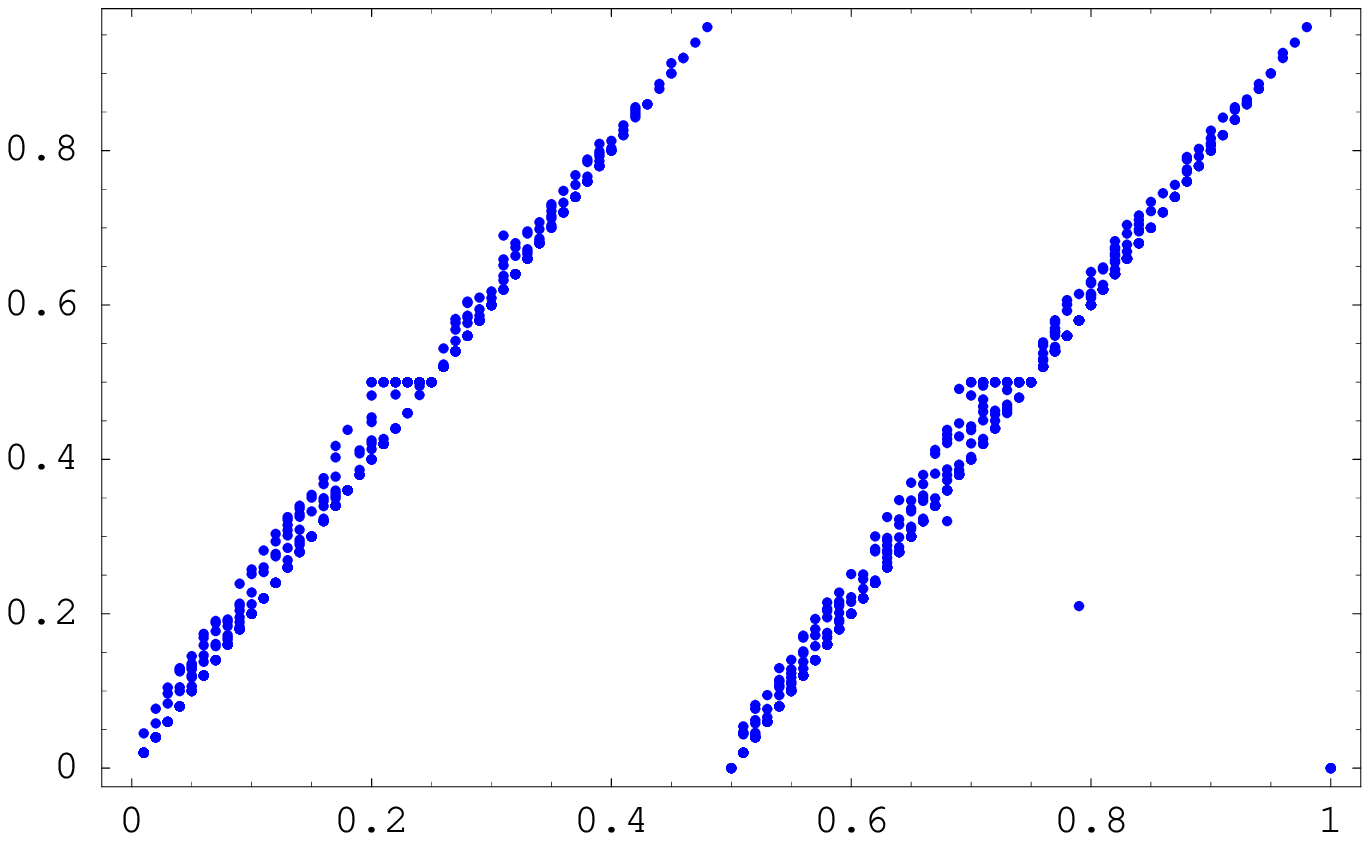}
\setlength{\unitlength}{1mm}
\flushleft\begin{picture}(0,0)(0,0)
\put(-5,61){$\varphi$}
\put(76,4){$\nu$}
\end{picture}
\caption{Phase vs.~perturbed tune for $n=0 \ldots 1$, $l= \pm 2$ modes ($\pi$-mode only). }
\label{fg:22radial000+} 
\end{figure}

In Fig.~\ref{fg:02radial000+} we computed the phase of the largest
eigenvalue of $l=\pm 2$ instabilities, corresponding to quadrupole
oscillations ($\pi$-mode only), versus the perturbed tune for
various $\Delta \nu$.  The slope of the two lower lines is $2$ which
indicates that the collective oscillation frequency of the quadrupole
mode is twice the single particle oscillation frequency for small
$\xi$.  The spread of the points for fixed $\nu$ shows how strongly
the beam-beam parameter $\xi$ influences the frequency of quadrupole
oscillations.  In Fig.~\ref{fg:22radial000+} this spread is
significantly lower which again shows that radial modes have a
stabilizing effect.

\begin{figure}
\includegraphics*[width=\columnwidth]{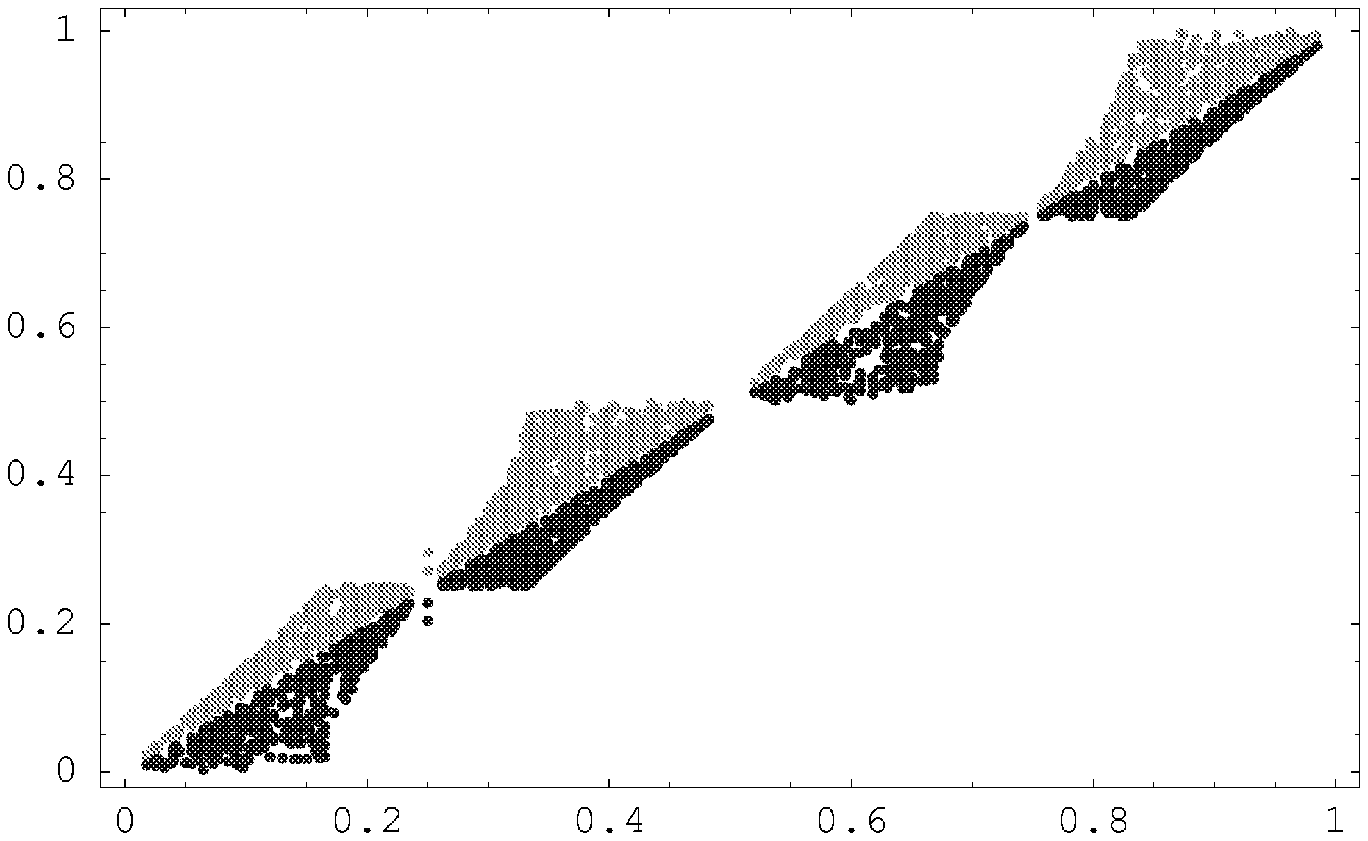}
\setlength{\unitlength}{1mm}
\flushleft\begin{picture}(0,0)(0,0)
\put(-10,61){$\frac{\nu_\pi-\nu_\sigma}{\xi}$}
\put(76,4){$\nu$}
\end{picture}
\caption{The dipole oscillation frequencies are plotted gray for
the $f_+$ distribution and black for the $f_-$ distribution
with $\xi=0$ to $0.2$ for $n=0$, $l= \pm 1$ modes. }
\label{fg:02radial000pm} 
\end{figure}

\begin{figure}
\includegraphics*[width=\columnwidth]{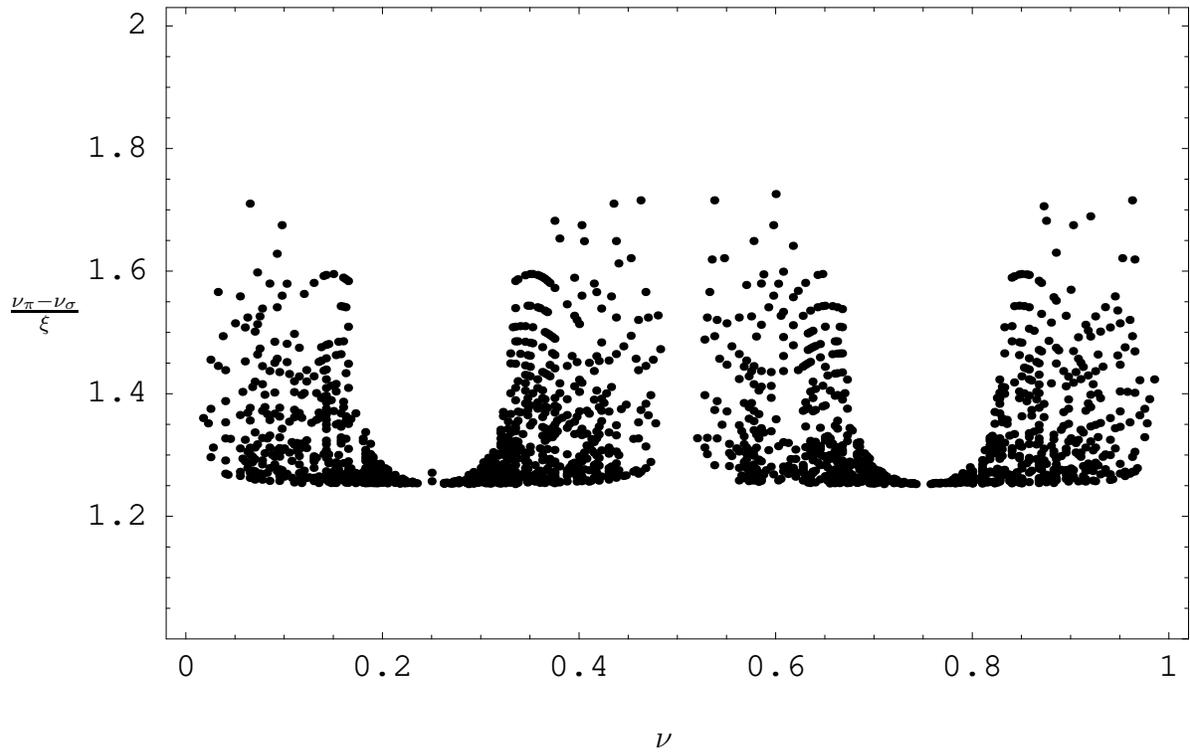}
\setlength{\unitlength}{1mm}
\flushleft\begin{picture}(0,0)(0,0)
\put(-10,61){$\frac{\nu_\pi-\nu_\sigma}{\xi}$}
\put(76,4){$\nu$}
\end{picture}
\caption{The Meller factor for stable motion in the region $\xi=0$ to
$0.2$ for $n=0$, $l= \pm 1$ modes. }
\label{fg:02meller000} 
\end{figure}

The dependence of this spread on $\nu$ can be understood analytically.
For simplicity we consider only the $n=0$ modes.  Close to a resonance
where $l \nu$ is integer, $g_{0l}$ and $g_{0-l}$ perturb the beams the
most. Thus, we content ourselves with the following 2x2 matrix
\cite{chao}
\begin{eqnarray}
T = \left (
\begin{array}{cc}
e^{-2\pi i l \Delta} & 0 \\
0                    & e^{2\pi i l \Delta}
\end{array} \right )
\left[\mathbbm{1}
\pm i \alpha \left ( \begin{array}{rr}1&1\\-1&-1\end{array} \right
) \right]\ ,
\end{eqnarray}
which satisfies all properties listed in Eq.~(\ref{eq:properties}) for
$i \alpha=\xi M_{0l,0l}$.  The imaginary parts of the eigenvalues of
the matrix $T$ vanish for eigenvalues whose absolute value is bigger
than $1$.  This leads to the plateaus at 0 and 0.5 in
Fig.~\ref{fg:02radial000+} and \ref{fg:22radial000+} at tunes $\nu$
where the $l=\pm 2$ mode becomes unstable in the
Fig.~\ref{fg:02radial000} and \ref{fg:22radial000}.
The difference between the dipole oscillation frequencies $\nu_\pi$
plotted in Fig.~\ref{fg:02radial000pm} gray and $\nu_\sigma$
plotted black of the $\pi$ and the $\sigma$ mode divided by the
beam-beam parameter $\xi$ is referred to as the Meller factor
\cite{meller81} or the Yokoya factor \cite{yokoya89}.  This factor is
plotted for all points of our computation for which both the $\pi$ and
the $\sigma$ mode indicate stable motion.  In
Fig.~\ref{fg:02meller000}, one can see that this factor is always
above 1.25 in our Gaussian flat beam model.

There are only a few points close to $\nu=0.25$ and $\nu=0.75$ since
the $l=2$ modes for these tunes are unstable for small $\xi$.

\section{Possible Extensions}

\subsection{Higher Order Resonances}
In order to study resonances of order higher than $2$
Eq.~(\ref{eq:integral}) must not be linearized, but rather the double
integral has to be expanded about $y=0$ to orders higher than $1$.
The expansion to 2nd order contains $y^2\int_{-\infty}^\infty d\bar y'
\frac{d}{dy} f(\bar y,\bar y',s)|_{\bar y=0}$.  Inserting the
expansion in Eq.~(\ref{eq:ansatz}) for $f$ and writing $\frac{d}{dy}$
in terms of $J$ and $\phi$ allows the evaluation of the integral.  The
resulting term $\frac{\partial\psi_0}{\partial
y'}y^2=-\frac{1}{2\pi\epsilon^2}(2\beta
J)^{\frac{3}{2}}e^{-\frac{J}{\epsilon}}\sin\phi\cos^2\phi $ in
Eq.~(\ref{eq:perturbed}) needs to be expanded in Laguerre polynomials
and gives rise to higher orders in radial modes.  The $n$-th order
term can be written in terms of powers of $\sqrt{J}$, $\cos n \phi$,
$\sin n \phi$ and lower frequency parts. Since the beam-beam force
acts only at a single point, its contribution is not averaged out in
the limit of a large number of turns if the tune matches the frequency
of one of the sine or cosine functions. This is the case if the tune
is a rational number, so higher order resonances would appear in
Fig.~\ref{fg:22radial}. Without truncating the series the model would
result in an infinite number of resonances since one can always find a
rational number between two irrational numbers.  However, this
procedure is complicated by the fact that Eq.~(\ref{eq:equidist}) is
not an equilibrium distribution anymore when nonlinear terms are
included.

When the length of the bunch and its longitudinal motion is included,
synchrobetatron resonances can occur \cite{krishnagopal89} when the
bunch length is in the order of the betatron function. Including these
resonances would require and extension of our treatment from two to four
dimensional phase space. This would be a worthwhile but tedious
continuation of our work.

\subsection{Damping by Synchrotron Radiation}
One can extend the presented model to account for damping by synchrotron radiation.
In order to obtain the equilibrium distribution in Eq.~(\ref{eq:equidist}) quantum excitation
must be included as well. This turns Eq.~(\ref{eq:vlasov}) into the Fokker-Planck
equation (\ref{eq:fokkerplanck}). In preliminary computations we found that the 
graphs we presented above remain unchanged for realistic values of the damping and excitation 
coefficients. To simplify the Fokker-Planck equation, we averaged over the phases
in the damping and excitation terms but not in the beam-beam interaction term.
This can be justified since the betatron phases in the terms for damping and
quantum excitation change during one turn while the phase in the interaction
term changes only once per turn. 
In Eq.~(\ref{eq:fokkerplanck}) $\lambda$ is the energy loss per turn due to 
synchrotron radiation divided by the energy of the particle, $\eta$ is the dispersion
and $D$ is the quantum excitation coefficient.
\begin{eqnarray}
\nonumber
\frac{\partial \psi_{1,2}}{\partial s} + y' \frac{\partial \psi_{1,2}}{\partial y} -
\left ( \frac{\lambda}{C} y' + K(s)y + 
\frac{4 \pi N r_e}{\gamma} \delta_p(s) I_{\psi_{2,1}}(y,s) \right )
\frac{\partial \psi_{1,2}}{\partial y'} 
\\
= \frac{\lambda}{C} \psi_{1,2} + D \left ( \eta \frac{\partial}{\partial y} + \eta'
\frac{\partial}{\partial y'} \right )^2 \psi_{1,2}
\label{eq:fokkerplanck}
\end{eqnarray}

\subsection{Different Tunes}
If the two beams have different tunes, Eq.~(\ref{eq:lincomb}) cannot
be used anymore to decouple the system. It is easier to work with the
uncoupled system and solve for the $g_{nl}$ of the two beams
separately. Introducing the column vector $G$ which contains the
$g_{nl}$ for both beams, one can proceed as before and describe the
beam transport for each turn by a matrix multiplication with a matrix
$T$. Introducing
\begin{eqnarray}
\tilde{R} = \left ( \begin{array}{cc} R(\nu_1) & 0 \\ 0 & R(\nu_2) \end{array} \right )
\end{eqnarray}
where $R(\nu)$ is a diagonal matrix which has the components $e^{-2 \pi il \nu}$ 
we can write the matrix $T$ as
\begin{eqnarray}
T = \tilde{R} \left [ \mathbbm{1} + \xi 
\left ( \begin{array}{cc} 0 & M \\ M & 0 \end{array} \right ) \right ]
\end{eqnarray}

\chapter{Coherent Synchrotron Radiation$^\dagger$} \label{ch:CSR}
\symbolfootnotetext[2]{This chapter appeared as a journal article \cite{Schmekel:2004jb}.
Reprinted in modified form with kind permission from the American Physical Society. 
\copyright ~ 2005 by the American Physical Society}

\section{Introduction}

The high brightness temperatures
of the radio emission of pulsars
($T_B \gg 10^{12} $K) implies
a coherent emission mechanism
\cite{Gold1968,Gold1969,GoldreichKeeley1971,
ManchesterTaylor1977,Melrose1991} and some part of
the radio emission of extragalactic jets
may be coherent \cite{GBKRL95}.
Recently, coherent synchrotron
radiation (CSR) has  been
observed in bunch compressors
\cite{Loos2002,Byrd2002,Kuske2003} which are a
crucial part of future particle accelerators.
       When a relativistic beam
of electrons interacts with its own
synchrotron radiation the beam may become modulated.
If the wavelength of the modulation is less than the
wavelength of the emitted
radiation, a linear instability may occur
which leads to exponential
growth of the modulation amplitude.
The coherent synchrotron
instability of relativistic electron rings
and beams has been investigated theoretically by
\cite{GoldreichKeeley1971,
HeifetsStupakov2001,Stupakov2002,Heifets2001,Byrd2003}.
Goldreich and Keeley analyzed the stability of a ring
of monoenergetic relativistic electrons
which were assumed to move on a circle of fixed
radius. Electrons of the ring gain or lose
energy owing to the tangential electromagnetic force and at
the same time generate the electromagnetic
field. \cite{Uhm1985} analyzed the stability of a relativistic
electron ring enclosed by a conducting beam pipe
in an external betatron magnetic field. A distribution function
with a spread in the canonical momentum was chosen for
their analysis. For simplicity the effect of the betatron oscillations
was not included in their treatment. They find a resistive wall
instability and a
negative mass instability. Furthermore, they find an instability
which can perturb the surface of the beam. \cite{Heifets2001}
analyzed the stability of a ring of relativistic electrons
in free space including a small energy spread which gives a range
of radii such that particles on the inner orbits can pass
particles on outer orbits. \cite{Byrd2003} has developed a similar
model which includes the effects of the conducting beam pipe.
Numerical simulations by \cite{Venturini2003} show the burst-like nature
of the coherent synchrotron radiation.

The present work analyzes the linear
stability of a cylindrical, collisionless, relativistic
electron (or positron) layer or E-layer \cite{Christofilos1958}.
Particle densities in pulsar magnetospheres are very
low, of order the Goldreich-Julian charge density
$n_{GJ}={\bf\Omega\cdot B}/2\pi ce\sim 
10^{11}\,{\rm cm^{-3}}(B/10^{12}\,{\rm G})(R/r)^3
[P({\rm sec})]^{-1}$ at radius $r>R$, where $R$ is the
stellar radius, $B=10^{12}B_{12}\,{\rm G}$ is the
surface field strength, and $P$ is the rotational period;
thus, the magnetospheric plasma is collisionless to an
excellent approximation \cite{Julian1969}.
          The particles in the layer
have a finite `temperature' and
thus a range of radii so that
the limitation of the Goldreich and
Keeley model is overcome.
Although we allow a spread in energies, we assume that it
is small, so the charge layer is also thin; efficient
radiation losses are probably sufficient to maintain rather
low energy spreads in a pulsar magnetosphere, although
the precise size of the spread is still not entirely
certain.
          Viewed from a moving frame the E-layer is
a rotating beam. The system is sufficiently simple
that it is relevant to electron
flows in pulsar magnetospheres (cf. \cite{Arons2004}).
          The analysis involves solving
the relativistic Vlasov equation using
the full set of Maxwell's equations and computing
the saturation amplitude due to trapping. The latter
allows us to calculate the energy loss due to
coherent radiation.

          In \S \ref{CSR:equilibria} we describe the considered Vlasov equilibria.
The first type of equilibrium (a) is formed by
electrons (or positrons) moving perpendicular to
a uniform magnetic field in the
$z-$direction so as to form a thin
cylindrical layer referred
to as an E-layer.
         The second type of equilibrium
(b) is formed by electrons moving almost parallel
to an external toroidal magnetic field and
also forming a cylindrical layer.
          \S \ref{CSR:linperturb} describes the method of solving the
linearized Vlasov equation which involves
integrating the perturbation force along the
unperturbed orbits of  the equilibrium.
           In \S \ref{CSR:firstapprox}, we derive the dispersion relation for
linear perturbations  for the case
of a radially thin E-layer and
zero wavenumber in the axial direction, $k_z=0$.
          We find that there is in general a short wavelength
instability.
            In \S \ref{CSR:nonlin} we analyze the nonlinear saturation
of the wave growth due to trapping of the
electrons in the potential wells of the
wave.
          This saturation allows the
calculation of the actual spectrum of
coherent synchrotron radiation.
         In \S \ref{CSR:kzneq0}, we derive the dispersion relation
for linear perturbations of a thin E-layer
including a finite axial wavenumber.
          The linear growth is found to occur
only for small values of the axial wavenumber.
          The nonlinear saturation due to trapping
is similar to that for the case where $k_z=0$.
In \S \ref{CSR:nonsatkzneq0} we consider the effect of the thickness
of the layer more thoroughly and include
the betatron oscillations.
           \S \ref{CSR:thick} discusses the apparent brightness
temperatures for the saturated coherent
synchrotron emission.
\S \ref{CSR:spectrum} discusses some implications on
particle accelerator physics.
\S \ref{CSR:brightnesstemp} gives conclusions of this work.

\section{Equilibrium Configuration} \label{CSR:equilibria}
\subsection{Configuration a}

\begin{figure}
\hspace{0.5in}
\includegraphics[width=4in]{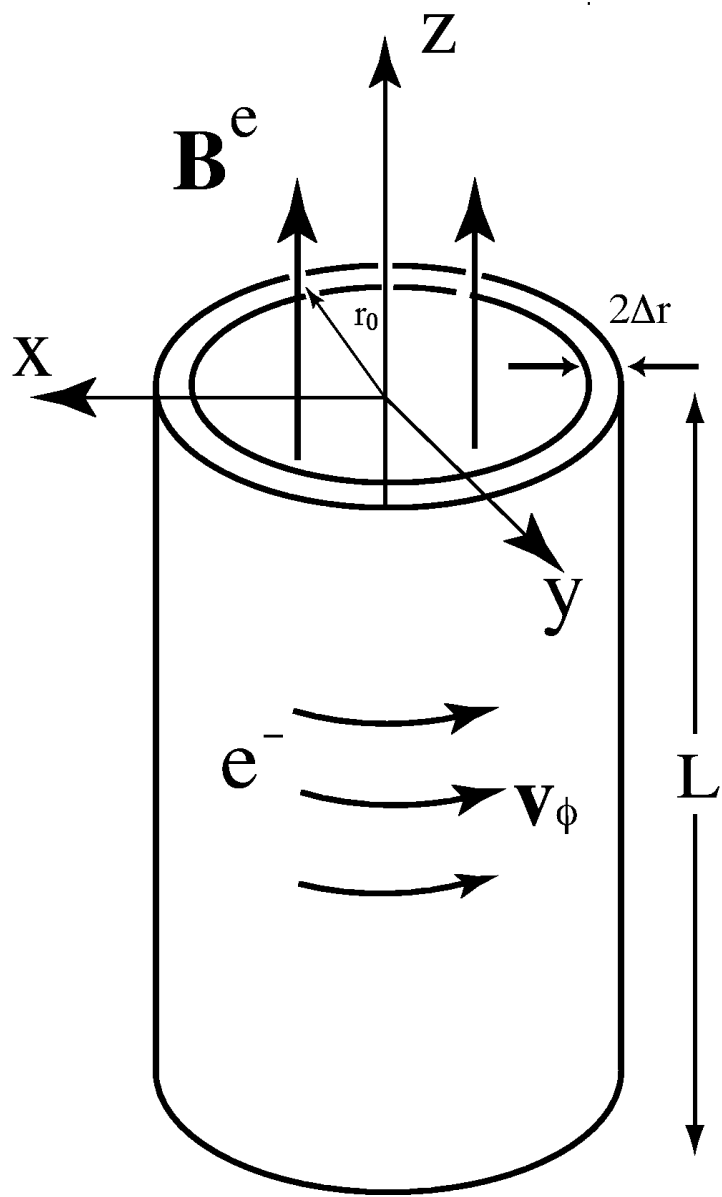}
\caption{
Geometry of relativistic E-layer for the case of a uniform
external axial magnetic field.
\label{configa}
}
\end{figure}

We first discuss the Vlasov equilibrium
for an axisymmetric, long, thin cylindrical
layer of relativistic electrons where
the electron motion is almost perpendicular
to the magnetic field.
This is shown in Fig.~\ref{configa}.  The case where
the electron motion is almost parallel to the
magnetic field is discussed below.
           The equilibrium has $\partial/\partial t=0,
~\partial/\partial \phi =0,~$ and $\partial/\partial z=0$.
           The configuration is
close to the non-neutral Astron E-layer of \cite{Christofilos1958}.
            The equilibrium distribution
function $f^0$ can be taken to be
an arbitrary non-negative function of the
constants of motion, the Hamiltonian,
\begin{eqnarray}
H \equiv \left (m_e^2 +p_r^2+ p_{\phi}^2
+ p_z^2 \right )^{1/2} - e\Phi^s(r)~,
\end{eqnarray}
and the canonical angular momentum,
\begin{eqnarray}
P_{\phi} \equiv r p_{\phi} - er A_{\phi}(r)~,
\end{eqnarray}
where $A_\phi =A_\phi^{e} + A_\phi^{s}$
is the total (external plus self) vector potential, $\Phi^s$
is the self electrostatic potential,
$m_e$ is the electron rest mass, $-e$ is
its charge, and the units are such that $c=1.$
          Here, the external magnetic field is assumed
to be uniform, ${\bf B}^e=B^e_z\hat{\bf z}$, with
$A_\phi^{e}=rB^e_z/2$, and $B^e_z>0$.
Thus we have $f^0 =f^0(H,P_\phi)$.
We consider the distribution function
\begin{eqnarray}
f^0 =K\delta(P_\phi-P_0)\exp\big[-H/T\big]~,
\label{equi}
\end{eqnarray}
where $K$, $P_0$, and $T$ are constants (see
for example \cite{Davidson1974}).
The temperature $T$ in energy units is assumed
sufficiently small that the fractional radial thickness of
the layer is small compared with unity.
Note that a Lorentz transformation in the $z-$direction
gives a rotating electron beam.

The equations for the self-fields are
\begin{eqnarray}
\frac{1}{r} \frac{d}{dr}
\left(r {d\Phi^s \over d r}
\right) = 4\pi e \int d^3 p~ f^0(H,P_{\phi})~,
\label{selfphi}
\end{eqnarray}
\begin{eqnarray}
\frac{d}{d r}\left( \frac{1}{r}
\frac{d( rA_\phi^{s})}{d r}\right)  =
4\pi e
\int d^3 p~ v_{\phi}~ f^0(H,P_{\phi})~,
\label{selfA}
\end{eqnarray}
where $v_\phi = (P_\phi/r +eA_\phi)/H$.

Owing to the small radial thickness of the layer,
we can expand radially near $r_0$
\begin{eqnarray}
\left[{P_\phi\over r}
+eA_\phi(r)\right]^2\!\!\!=\left[{P_\phi\over
r_0}\!\!+eA_\phi(r_0)\right]^2 \!\!+\delta r D_1 \!\!+\!{1 \over 2}
\delta r^2 D_2 ,
\label{expandp}
\end{eqnarray}
where $D_1$, $D_2$ are the derivatives evaluated at $r_0$,
and $\delta r \equiv r-r_0$ with $(\delta r/r_0)^2 \ll 1$.
We choose $r_0$ so as to eliminate the term linear in $\delta r$.
Thus,
\begin{eqnarray}
H=H_0 -e\Phi^s(r_0) +{1 \over 2 H_0}\big(p_r^2+p_z^2
+ H_0^2 ~\omega_{\beta r}^2~ \delta r^2\big)~,
\label{expandH}
\end{eqnarray}
where $\omega_{\beta r}$ is the radial betatron frequency,
and
\begin{eqnarray}
H_0
& \equiv&
m_e\left\{1+\left[{P_\phi\over r_0}+
eA_\phi(r_0)\right]^2\right\}^{1/2}~, \\ \nonumber
\gamma_0
& \equiv &
{H_0\over m_e}~, \\ \nonumber
v_{\phi0}
&\equiv&
{1\over H_0}\left[{P_\phi \over
r_0}+eA_\phi(r_0)\right]~.
\end{eqnarray}
          We assume $\gamma_0^2 \gg 1$ and $v_{\phi0} >0$ so
that $v_{\phi0} =  1-1/(2\gamma_0^2)$ to a good
approximation.
The ``median radius'' $r_0$ is determined by the condition
$$
{D_1\over 2 H_0}  -e ~{d\Phi^s\over dr}\bigg|_{r_0}=0~,
$$
or
\begin{equation}
{1\over H_0}\left({P_\phi\over r_0}+eA_\phi\right)
\left(-{P_\phi\over r_0^2}
+e~{dA_\phi\over dr}\right)\bigg|_{r_0}
=e~ {d\Phi^s\over dr}\bigg|_{r_0} .
\label{constraint}
\end{equation}
To a good approximation,
$$
r_0 ={m_e\gamma_0 v_{\phi0} \over (1-2\zeta)eB^e_z}\approx
{m_e\gamma_0  \over eB^e_z} (1+2\zeta ) \quad {\rm or}
$$
\begin{equation}
r_0^2=
{2P_\phi \over e B^e_z}[1+3\zeta+{\cal O}(\zeta^2)]~.
\label{r0}
\end{equation}
Here,
\begin{eqnarray}
\zeta\equiv-~{B_z^{s}(r_0)\over B^e_z}~,
\quad \quad{\rm with}\quad \quad
\zeta^2 \ll 1~,
\label{zeta1}
\end{eqnarray}
is the field-reversal parameter of Christofilos.
          For a radially thin E-layer of axial
length $L$ consisting of a total
number of electrons $N$, the surface density
of electrons is $\sigma =N/(2\pi r_0 L)$ and the surface
current density is $-ev_\phi \sigma$.
          Because $B_z^s(r_0)$ is one-half the full change of
the self-magnetic field across the layer, we have
$\zeta = r_e  N/(\gamma L)$, where $r_e=e^2/(mc^2)$ is
the classical electron radius.
           Notice that $N$, $\zeta$, and $\gamma L$ are
invariants under a Lorentz transform in the $z-$direction.

             The radial betatron frequency $\omega_{\beta r}$ is
given by
\begin{eqnarray}
H_0^2~\omega_{\beta r}^2={D_2\over 2} -{D_1^2 \over 4 H_0^2}
-H_0e{d^2\Phi^s \over dr^2}~.
\end{eqnarray}
          Using Eq.~(\ref{constraint}) gives
$$
\omega_{\beta r}^2 ={1-4\zeta \over 1-2\zeta}~
{v_{\phi0}^2\over r_0^2}~
+{e  v_{\phi0}\over \gamma_0 m_e}~
{d^2 A_\phi \over dr^2}\bigg|_{r_0}\!
-{e\over \gamma_0 m_e}~{d^2\Phi^s \over dr^2}\bigg|_{r_0}
$$
\begin{equation}
\approx
{1-2\zeta\over r_0^2}  -
{\sqrt{2/\pi}~\zeta \over   r_0\Delta r \gamma^2}~.
\label{omega2betar}
\end{equation}
The term $\propto 1/\Delta r$ is the
sum of the defocusing  self-electric force
and the smaller focusing self-magnetic force.
          For the layer to be radially confined we need
to have $\zeta <\sqrt{\pi/2}~\gamma^2(\Delta r/r_0)$.
         For $\zeta \ll \gamma^2(\Delta r/r_0)$ and
$\zeta^2 \ll 1$, we have $\omega_{\beta r} = 1/r_0$
to a good approximation.

The number density follows from Eq.~(\ref{equi}),
$$
n \approx n_0\exp\left(-{\delta r^2\over 2\Delta r^2}\right)
~~{\rm where}~~ \Delta r \equiv
\left({T \over H_0\omega_{\beta r}^2}\right)^{1/2}
$$
\begin{equation}
{\rm or} ~~ {\Delta r^2 \over r_0^2} \simeq
{v_{th}^2 \over  1-2\zeta -\sqrt{2/\pi}~
\zeta (r_0/\Delta r)/\gamma^{2} }
\end{equation}
where
\begin{eqnarray}
        v_{th}
\equiv \left({T \over \gamma_0 m_e}\right)^{1/2}  
\end{eqnarray}
and
\begin{eqnarray}
n_0=2\pi K H_0 T r_0^{-1} \exp \left ( - \frac{H_0 - e \Phi^s(r_0)}{T} \right ) ~.
\end{eqnarray}
           As mentioned we assume the layer to be radially
thin with $(\Delta r/r_0)^2 \ll 1$.
          Consequently, Eqs.~(\ref{selfphi}) and
(\ref{selfA}) become
\begin{eqnarray}
{ d^2 \Phi^s \over dr^2} & \approx &4\pi e n_0 ~
\exp\left(-{\delta r^2 \over 2\Delta r^2}\right)~,
\quad \quad  \nonumber \\
{ d^2 A_\phi^{s} \over dr^2} & \approx &4\pi en_0~ v_{\phi0}~
\exp\left(-{\delta r^2 \over 2\Delta r^2}\right)~.
\end{eqnarray}
          Thus we obtain
\begin{eqnarray}
          \zeta ={-B^{s}_z(r_0)\over B_z^e}={4\pi e n_0 v_{\phi0}\Delta r
\sqrt{\pi/2} \over B_z^e}
\end{eqnarray}
           The equilibrium is thus seen to be determined by
three  parameters,
\begin{eqnarray}
\zeta^2~,\quad v_{th}^2~,\quad {\rm and}\quad 1/\gamma_0^2~,
\end{eqnarray}
which are all small compared with unity.

\subsection{Equilibrium Orbits} \label{CSR:equiorbits}

           From the Hamiltonian of Eq.~(\ref{expandH}) we have
\begin{equation}
\!{d^2 \delta r \over dt^2}\!=
-   \omega_{\beta r}^2\delta r,~\rightarrow~
        \delta r(t^\prime)\!=
\delta r_i\sin[\omega_{\beta r}(t^\prime-t)+\varphi]~,
\label{rorbit}
\end{equation}
where $r-r_0=\delta  r_i \sin\varphi$.
For future use we express  the orbit so that
${\bf r}(t^\prime=t) ={\bf r}$, where $({\bf r},t)$
is the point of observation.
Also, we have
\begin{eqnarray}
{d\phi \over dt} ={ P_\phi +e r A_\phi(r) \over
m_e \gamma r^2} = \dot{\phi}(r_0) +
{d\dot{\phi} \over dr}\bigg|_{r_0} \delta r +..~,
\end{eqnarray}
so  that
\begin{eqnarray}
\phi(t^\prime)=\phi +(t^\prime -t)\dot{\phi}_0
+{1\over \omega_{\beta r}}
{\partial \dot{\phi}_0 \over \partial r}\bigg|_{r_0}\times
\nonumber \\
\bigg\{-\delta r_i\cos[\omega_{\beta r}(t^\prime-t)+
\varphi]+\delta r_i\cos(\varphi)\bigg\}~,
\label{phiorbit}
\end{eqnarray}
where $\partial \dot{\phi}/\partial r |_{r_0}
= -\dot{\phi}_0/r_0$.
            For $\zeta \ll \gamma^2(\Delta r/r_0)$ and
$\zeta^2 \ll 1$, we have
$\partial \dot{\phi}/\partial r|_{r_0}/\omega_{\beta r} = -1/r_0$
to a good approximation.
Because the E-layer is uniform in the $z-$direction,
\begin{eqnarray}
z(t^\prime)=z+(t^\prime -t)  v_z~.
\label{eqorbitz}
\end{eqnarray}
The orbits are necessary for the stability analysis.

\subsection{Configuration b}

          Here, we describe a Vlasov equilibrium
for an axisymmetric, long, thin cylindrical
layer of relativistic electrons where the
electron motion is almost parallel to the
magnetic field.
            The equilibrium distribution
function $f^0$ is again taken to be
given by Eq.~(\ref{equi}) in terms of
the Hamiltonian,
$H$, and the canonical angular momentum,
$P_{\phi} \equiv r p_{\phi} - er A_{\phi}(r),$
where $A_\phi =  A_\phi^{s}$.
          We make the same  assumptions as
above, $\gamma^2 \gg 1$,
$T/(m_e\gamma) \ll 1$, and $\Delta r^2/r_0^2 \ll 1$.
          In this case
there is no external $B_z$ field.
          Instead, we include an external toroidal magnetic
field $B_\phi^e$ with corresponding vector
potential $A_z^e$ and an external electric
field ${\bf E}^e$ with potential $\Phi^e$.
         The fields ${\bf B}^e$ and ${\bf E}^e$
correspond to the magnetic and electric fields of
a distant,  charged, current-carrying flow
along the axis.  Thus,  $|{ E}^e_r| <|{B}_\phi^e|$.
           The considered external field is of course
just one of a variety of fields which give
electron motion almost parallel with the
magnetic field.
          Note also that the distribution function
is restricted in the respect that it does
not include a dependence on the canonical
momentum in the $z-$direction $P_z=m_e\gamma v_z
-eA_z$.

          The distribution function (\ref{equi}) gives
$J_z=0$ so that there is no toroidal self
magnetic field.
          Thus the self-potentials in this case are
also given by Eqs.~(\ref{selfphi}) and (\ref{selfA}).
           Eqs.~(\ref{expandp}) - (\ref{constraint})
are also applicable
with the replacement of $\Phi^s$ by the
total potential $\Phi$.
          In place of Eq.~(\ref{r0}) we find
\begin{equation}
r_0= {m_e\gamma v_\phi^2 \over
(1-2\zeta)e E_r^e(r_0)}\approx
{m_e\gamma\over e E_r^e(r_0)}(1+2\zeta)~,
\label{r0confII}
\end{equation}
        where $
\zeta \equiv  {B_z^s(r_0) / E_r^s(r_0)}.$
          We again have
$\zeta = r_e  N/(\gamma L)$, where $r_e=e^2/(mc^2)$ is
the classical electron radius and $L$ is the
axial length of the layer.
          Because $d^2 \Phi^e/dr^2 =-(1/r)d\Phi^e/dr$,
the radial betatron frequency is again
given by Eq.~(\ref{omega2betar}) (with $\Phi$ now
the total potential) so that the equilibrium orbits
given in \S \ref{CSR:equiorbits} also apply in this case.
          The electron motion is almost parallel to
the magnetic field in that $(B_z^s/B_\phi^e)^2 =
\zeta^2 (E_r^e/B_\phi^e)^2 < \zeta^2 \ll 1$.
          Notice that Eq.~(\ref{r0confII}) for $r_0$ is formal in
the respect that $E_r^e \propto 1/r$.
          Therefore, $r_0$ is in fact arbitrary
in this case.
          Because the wavelengths of the unstable
modes are found to be small compared with
$r_0$,
        it may be interpreted as local radius
of curvature of the magnetic field.

\begin{figure}
\hspace{0.5in}
\includegraphics[width=4in]{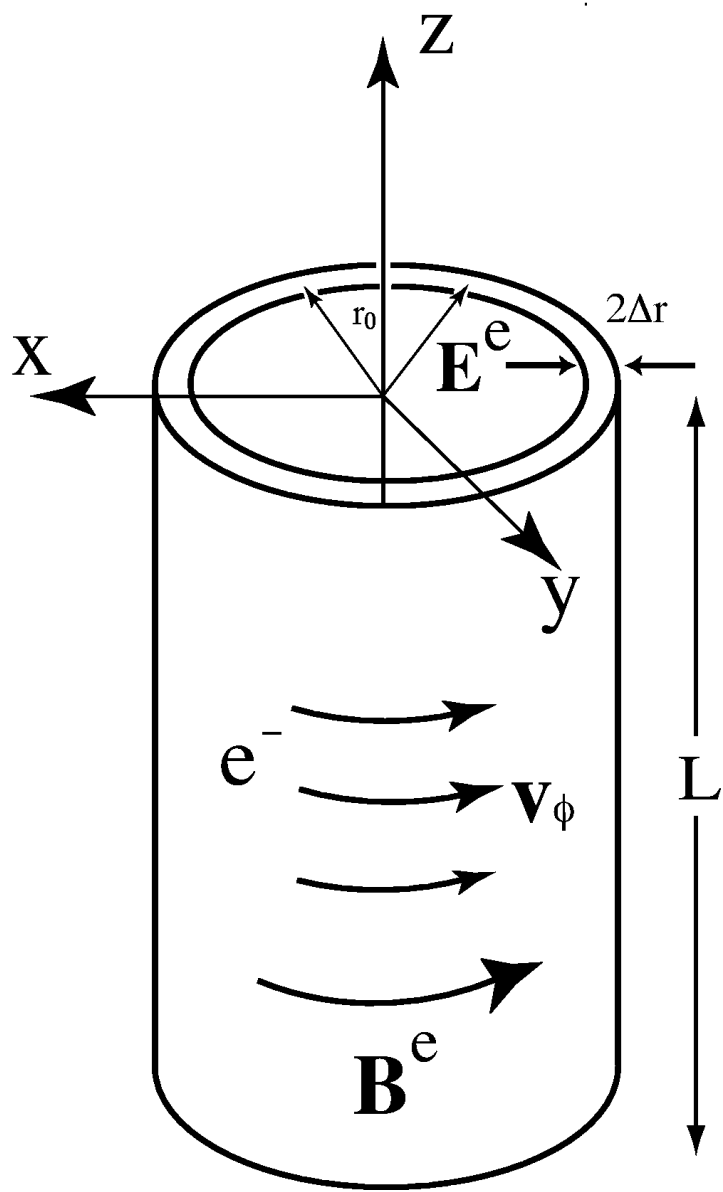}
\caption{
Geometry of relativistic E-layer for the case of 
an external toroidal magnetic field with an
external radial electric field.
\label{configb}
}
\end{figure}

\section{Linear Perturbation} \label{CSR:linperturb}

We now consider a general perturbation of the Vlasov
equation with $f({\bf r, p},t)=f^0({\bf r,p}) +
\delta f({\bf r,p},t)$.  To first order in the
perturbation amplitude $\delta f$ obeys
\begin{eqnarray}
\left({\partial \over \partial t}
+ {\bf v \cdot}{\partial \over \partial {\bf r}} +
{d{\bf p} \over dt}~{\bf \cdot}~
{\partial \over \partial {\bf p}}\right)\delta f
\equiv{D\delta f \over Dt}=
e(\delta {\bf E} + {\bf v \times} \delta {\bf B})~
{\bf \cdot}~{\partial f^0 \over \partial {\bf p}}~,
\label{exprdf}
\end{eqnarray}
where $\delta{\bf E}$ and $\delta {\bf B}$ are
the perturbations in the electric and magnetic fields.
All scalar perturbation quantities are considered
to have the dependencies
\begin{eqnarray}
F(r)\exp(im\phi +ik_z z -i\omega t)~,
\label{ansatz}
\end{eqnarray}
where the angular frequency $\omega$ is taken to
have at least a small positive imaginary part which corresponds
to a growing perturbation.
          This allows for a correct initial value treatment of
the problem \cite{Landau1946}.
           For a perturbation taken to vanish as $t\to -\infty$,
\begin{equation}
\delta f({\bf r, p},t) =
e\int_{-\infty}^t dt^\prime
\bigg\{\delta {\bf E}[{\bf r}(t^\prime),t^\prime] +
{\bf v}(t^\prime){\times}
\delta {\bf B}[{\bf r}(t^\prime),t^\prime]\bigg\}~{\bf
\cdot}~{\partial f^0 \over \partial {\bf p}}~,
\label{dfLandau}
\end{equation}
where the integration
follows the orbit $[{\bf r}(t^\prime),{\bf p}(t^\prime)]$
which passes through the phase-space point $[{\bf r},{\bf
p}]$ at time $t$.
            For the considered axisymmetric equilibria,
\begin{eqnarray}
\frac{\partial H}{\partial \vec{p}} = \frac{\vec{p}}{H}
\\
\frac{\partial}{\partial \vec{p}} = \frac{\vec{p}}{H} \frac{\partial}{\partial H
} = \frac{\vec{p}}{H} \frac{\partial P_{\phi}}{\partial H} \frac{\partial}{\partial 
P_{\phi}}
\\
\frac{\partial f}{\partial \vec{p}} = \frac{\vec{p}}{H} \frac{\partial P_{\phi}}{
\partial H} \left ( \left . \frac{\partial f}{\partial P_{\phi}} \right |_H +
\frac{\partial H}{\partial P_{\phi}} \left . \frac{\partial f}{\partial H} \right
 |_{P_{\phi}} \right )
\end{eqnarray}
since $f \equiv f(H,P_{\phi})$ and $H \equiv H(P_{\phi},\ldots)$
\begin{eqnarray}
\frac{\partial}{\partial \vec{p}} = \vec{v} \left . \frac{\partial}{\partial H} 
\right |_{P_{\phi}} + r \hat e_{\phi} \left . \frac{\partial}{\partial P_{\phi}} \right |_H
\\
{\partial f^0 \over \partial {\bf p}} =
{{\bf p}\over H} \left . {\partial f^0 \over \partial H} \right |_{P_{\phi}}
+ r\hat{\rvecphi~} \left . {\partial f^0 \over \partial P_\phi} \right |_{H}~,
\label{fHP}
\end{eqnarray}
where the partial derivatives are to be evaluated at constant $P_{\phi}$
and $H$, respectively.
Thus, the right-hand side of Eq.~(\ref{exprdf}) becomes
\begin{eqnarray}
e\left(
-{d\delta \Phi \over dt}
+i\omega(\dot{\phi}~\delta \Psi-\delta \Phi)
        + i \omega {\bf v}_\perp\cdot\delta {\bf A} \right)
{\partial f^0  \over \partial H} +
        \nonumber\\
e\left( -{d \delta \Psi \over dt}
+im(\dot{\phi}~\delta \Psi -\delta \Phi)
+im{\bf v}_\perp \cdot \delta{\bf A} \right)
{\partial f^0 \over \partial P_\phi}~,
\end{eqnarray}
where $\delta {\bf E}= -{\bf \nabla}\delta \Phi
-\partial {\bf \delta A}/\partial t$ and $\delta {\bf B}
={\nabla \times}{\bf \delta A}$, $\delta \Psi \equiv
r\delta A_\phi$ is the perturbation in the
flux  function, ${\bf v}_\perp =(v_r, v_z)$, and $d/dt =
\partial/\partial t + {\bf v}\cdot {\bf \nabla}$.
         We assume the Lorentz gauge ${\bf \nabla \cdot}\delta
{\bf A} + \partial \delta \Phi/\partial t=0$.

         Evaluating Eq.~(\ref{dfLandau}) gives
\begin{eqnarray}
\delta f = & e & {\partial f^0 \over \partial H}
\left[-\delta \Phi
+i\omega\int_{-\infty}^t dt^\prime
\left(\dot\phi^\prime \delta \Psi^\prime
-\delta \Phi^\prime +{\bf v}^\prime_\perp
\cdot \delta {\bf A}^\prime\right)\right]
\\
+ & e & {\partial f^0 \over \partial P_\phi}
\left[-\delta \Psi
+im\int_{-\infty}^t dt^\prime~
\left(\dot{\phi}^\prime \delta \Psi^\prime -\delta
\Phi^\prime+{\bf v}^\prime_\perp \cdot \delta
{\bf A}^\prime\right)\right]~,
\label{dfLandaupots}
\end{eqnarray}
where the prime indicates evaluation at
$[{\bf r}(t^\prime),t^\prime]$.
         The integration is along the unperturbed particle orbit so
that $\partial f^0/\partial H$ and $\partial f^0/\partial
P_\phi$ are constants and can be taken outside the
integrals.
Note also that $d/dt$ acting on a function of $({\bf r},t)$
is the same as $D/Dt$.

\section{First Approximation} \label{CSR:firstapprox}

As a  starting approximation we neglect (i) the radial
oscillations in the orbits [$(\Delta r/r_0)^2 \ll 1$], (ii) the
self-field corrections to orbits proportional to $\zeta$,  (iii)
the terms in $\delta f$ proportional to $v_\perp^2$ ($v_{th}^2
\approx (\Delta r/r_0)^2 \ll 1$),
(iv) we take $k_z=0$ and (v) we assume the layer is very thin.
         Owing to approximation (iii), we can neglect
the terms $\propto {\bf v}_\perp \cdot \delta {\bf A}$ in
Eq.~(\ref{dfLandaupots}) in the evaluation of $\delta \rho$
and $\delta J_\phi$.
          This is because these
terms give contributions to $\delta f$ which
are odd functions of $v_r$ and $v_z$. Therefore, their average
contribution can be neglected. 

         Evaluation of Eq.~(\ref{dfLandaupots}) gives
\begin{equation}
\delta f =- e{\partial f^0 \over \partial H}\bigg|_{P_\phi}
\!\!\!
{\dot{\phi}(\omega \delta \Psi-m\delta \Phi) \over
\omega - m\dot{\phi}}
-e {\partial f^0 \over \partial P_\phi}\bigg|_{H}
{\omega \delta \Psi - m \delta \Phi
\over \omega - m \dot{\phi}}~,
\label{dfpot}
\end{equation}
where $\dot{\phi}=\dot{\phi}(r_0)$.
          The approximations lead to a closed system
with potentials $(\delta \Phi,~ \delta \Psi)$
and sources $(\delta \rho,~ \delta J_\phi)$.

We have
$$
\delta \rho =
-e\int d^3p~ \delta f = -{e \over r_0}
\int dp_r dp_zdP_\phi ~ \delta f~,~~~
$$
\begin{equation}
\delta J_\phi =
-e\int d^3p~v_\phi \delta f = -{e \over r_0}
\int dp_r dp_z dP_\phi ~ v_\phi \delta f~.
\label{psints}
\end{equation}
          For the considered distribution function,
Eq.~(\ref{equi}), $\partial f^0/\partial H = -f^0/T$.
          The $\partial f^0/\partial P_\phi$ term
in Eq.~(\ref{dfpot}) can be integrated by parts.
Furthermore, note that $\partial H / \partial P_\phi =
\dot \phi$ and
$\partial \dot{\phi}/\partial P_\phi =
- (\dot{\phi})^2/H$, which corresponds to an
effective ``negative mass'' for the particle's
azimuthal motion \cite{Kolomenskii1959,Nielson1959,Lawson1988}.
          From the partial integration
the small term proportional to
$\partial v_\phi/\partial P_\phi = v_\phi/(r_0H^3)$
is neglected.
Also note that $H$ is not a constant when performing
the integration over momenta. Evaluating this term by an integration by parts
with a general function $g(P_{\phi})$ in the integrand gives
\begin{eqnarray}
\nonumber
& \ & \int d P_{\phi} \left .
\frac{\partial f^0}{\partial P_{\phi}} \right |_H g(P_{\phi}) =
\\ \nonumber
& - & K \int d P_{\phi} \delta(P_{\phi}-P_0)
\frac{\partial}{\partial P_{\phi}} \left [g(P_{\phi}) e^{-H/T} \right ] =
\\ \nonumber
& - & K \int d P_{\phi} \delta(P_{\phi}-P_0)
\frac{\partial}{\partial P_{\phi}} \left [g(P_{\phi}) \right ] e^{-H/T}
\\
& + & \frac{K}{T} \int d P_{\phi} \delta(P_{\phi}-P_0) g(P_{\phi}) e^{-H/T}
\frac{\partial H}{\partial P_{\phi}}~.
\end{eqnarray}
That is, the integration produces
an additional term which cancels the $1/T$-term.
Thus,
\begin{equation}
\int d P_{\phi} \delta f =
-e \int d P_{\phi} {f^0 \over H}~
{m \dot{\phi}^2(\omega \delta \Psi - m \delta \Phi)
\over (\Delta\omega)^2 }~,
\label{dfpots}
\end{equation}
where $\Delta \omega \equiv \omega -m \dot{\phi}$.
Integrating over the remaining momenta gives
\begin{equation}
(\delta \rho, ~\delta J_\phi)=
(1, v_\phi)~e^2~ n(r)
{m\dot{\phi}^2 \over H} \cdot
{(\omega \delta \Psi -m\delta \Phi ) \over
(\Delta \omega)^2}~.
\label{drhodpot}
\end{equation}

           For a radially thin E-layer we may take 
\begin{eqnarray}
n(r)=n_0 \exp(-\delta r^2/2\Delta r^2) \rightarrow 
n_0\sqrt{2\pi}\Delta r ~\delta(\delta r).
\end{eqnarray}
   We  comment on this approximation below in more detail
when we include the radial wavenumber $k_r$ of the perturbation.
Then Eqs.~(\ref{dpot1}) and (\ref{dpot2}) can be written as
\begin{equation}
[\delta \Phi(r_0),~\delta \Psi(r_0)]=
        \big[1,~r_0 v_\phi (1 + \Delta \tilde \omega) \big]~2\pi^2r_0~ Z
        \int dr ~\delta \rho(r)~,
\label{dpotdrho}
\end{equation}
where $Z \equiv iJ_m(\omega r_0) H_m^{(1)}(\omega r_0)$, 
$\tilde{\omega}\equiv \omega/(m\dot{\phi})$ and
$\Delta \tilde{\omega}\equiv \Delta \omega/(m\dot{\phi})$.
Integrating Eq.~(\ref{drhodpot}) over the radial extent of the E-layer
and canceling out the field amplitudes
gives the dispersion relation
\begin{equation}
1=2\pi^2 r_0~
[n_0 e^2 \sqrt{2\pi} \Delta r]~Z
{m\dot{\phi}^2 \over H} \cdot
{\omega r_0v_\phi (1+\Delta \tilde \omega) -m \over
(\Delta \omega)^2}~.
\label{dispdim}
\end{equation}
In terms of dimensionless variables this becomes
\begin{equation}
1= \pi~ \zeta~ Z~
\left(2 \Delta \tilde{\omega} - {1 \over \gamma^2}\right)
{1 \over (\Delta \tilde{\omega} )^2}~,
\label{disp}
\end{equation}
where $Z=iJ_m(m\tilde{\omega}v_\phi)
H_m^{(1)}(m\tilde{\omega}v_\phi)$,
$H^{(1)}_m=J_m +i Y_m$,
and the field-reversal parameter
$\zeta=4\pi e n_0v_\phi\Delta r\sqrt{\pi/2}/B^e_z$ as given
by Eq.~(\ref{zeta1}).

For $m\gg 1$ approximation (\ref{BessAiry}) can be used to give
\begin{eqnarray}
\nonumber
J_m(m\tilde{\omega}v_\phi)\approx (2/m)^{1/3}{\rm Ai}(w)
\\
Y_m(m\tilde{\omega}v_\phi)\approx -(2/m)^{1/3}{\rm Bi}(w), 
\end{eqnarray}
where
\begin{eqnarray}
w =(m/2)^{2/3}(\gamma^{-2} -2 \Delta \tilde{\omega}).
\end{eqnarray}
Thus we have 
\begin{eqnarray}
Z = i J_m H_m^{(1)} \approx (2/m)^{2/3} [\Ai(w) \Bi(w) + i {\Ai}^2(w)].
\end{eqnarray}
Occasionally, $Z_m(w)$ is denoted by $Z$.   
For $|w|^2 \gg 1$ using (\ref{expAiry})
\begin{eqnarray}
Z \approx (2/m)^{2/3}/(2\pi |w|^{1/2}).
\end{eqnarray}
and for $|w|^2 \lesssim 0.5$ using (\ref{linAiry})
\begin{eqnarray}
Z \approx (2/m)^{2/3} [\sqrt{3}(c_1^2-c_2^2 w^2)+i(c_1-c_2 w)^2].
\end{eqnarray}
For $|w|^2 \ll 1$, 
\begin{eqnarray}
Z \approx (0.347+0.200~i)/m^{2/3}.
\end{eqnarray}

\subsection{Range of Validity}

We are interested in the regime where the wavelength of the
emitted radiation is comparable to the ``bunch length'', i.e.
$\omega \approx m$ or equivalently $\Delta \tilde \omega \ll 1$.
However, Eq.~(\ref{disp}) is only valid if 
$\Delta \tilde \omega \ll \gamma^{-2}$. Since we neglected
$\delta J_r$ and $\delta J_z$ we obtain from the continuity equation
$\delta J_{\phi} = \frac{\omega r_0}{m} \delta \rho$. Due to this
approximation the factor on the right hand side can become bigger 
than the speed of light if $\Delta \tilde \omega > \gamma^{-2}$
which leads to unphysical results. In the latter case 
$\delta J_{\phi} = v_{\phi} \delta \rho$ is a better approximation.
Fortunately, $\Delta \tilde \omega \ll \gamma^{-2}$ is the most
interesting case and in the remainder of this paper we will
always work in this limit.
Furthermore, for the continuum approximation to be valid the mean particle
distance has to be much smaller than the wavelength.

\subsection{Growth Rates}

\def\zm{Z_m}
\def\Zm{{\cal\Zeta}_m}
\def\omtil{\tilde\omega}
\def\domtil{\Delta\omtil}
\def\domtilvert{\vert\domtil\vert}
\def\vth{v_{th}}
\def\sigvert{\vert\sigma\vert}
\def\wvert{\vert w\vert}
\def\Aisq{{\rm Ai^2}(w)}
\def\gvth{\gamma^2\vth^2}
\def\gvthth{\gamma^3\vth^3}

It will prove useful to define two characteristic
values of $m$: $m_1\equiv\zeta^{3/2}\gamma^3$ and
$m_2=2\gamma^3$, and therefore $m_1=\zeta^{3/2}m_2/2$.
We can obtain approximate solutions to Eq.~(\ref{disp}) in two
different cases.
There may be solutions with small values
of $\gamma^2\domtil$, so that $w\simeq(m/m_2)^{2/3}$.
In this case, Eq.~(\ref{disp}) becomes a simple quadratic
equation, which can be solved for $\domtil$. We can simplify
the solution somewhat by changing variables to $\sigma\equiv\gamma^2
\domtil$ in which case Eq.~(\ref{disp}) can be written in the form
\begin{eqnarray}
\nonumber
1={\pi\zeta\zm\gamma^2\over\sigma^2}
(\sigma-1)\approx
-{\pi\zeta\zm\gamma^2\over\sigma^2},
\end{eqnarray}
where we have neglected $\sigma$ compared to one in the
approximate version of this equation.
We find that
\begin{eqnarray}
\sigma\simeq\sqrt{-\pi\zeta\zm\gamma^2}~.
\label{sigsmall}
\end{eqnarray}
For case I let us assume that $m\ll m_2 $, in which case
Eq.~(\ref{sigsmall}) implies
\begin{eqnarray}
\nonumber
\sigma\simeq\pm1.121(m_1/m)^{1/3}~e^{i(7\pi/12)}
\\
=1.121(m_1/m)^{1/3}(-0.2588+0.9659i)~.
\end{eqnarray}
so $\sigvert\ll 1$ for $m\gg m_1$. The growth rate of the
unstable mode is
\begin{equation}
\omega_i \simeq {1.083\zeta^{1/2}m^{2/3}\dot\phi\over\gamma}
\end{equation}
in this regime. For case II we assume that $m\gg m_2$, in
which case Eq.~(\ref{sigsmall}) implies
\begin{eqnarray}
\nonumber
\sigma=\pm{i\zeta^{1/2}\gamma^{3/2}\over m^{1/2}}=
\pm i\zeta^{1/2}(m_2/2m)^{1/2}
\\
\omega_i \simeq{\zeta^{1/2}m^{1/2}\dot\phi\over\gamma^{1/2}}
~;
\end{eqnarray}
note that the growth rates in cases I and II match almost
exactly at $m=m_2$, where $\sigvert\approx\zeta^{1/2}$.

\begin{figure}
\includegraphics[width=\columnwidth]{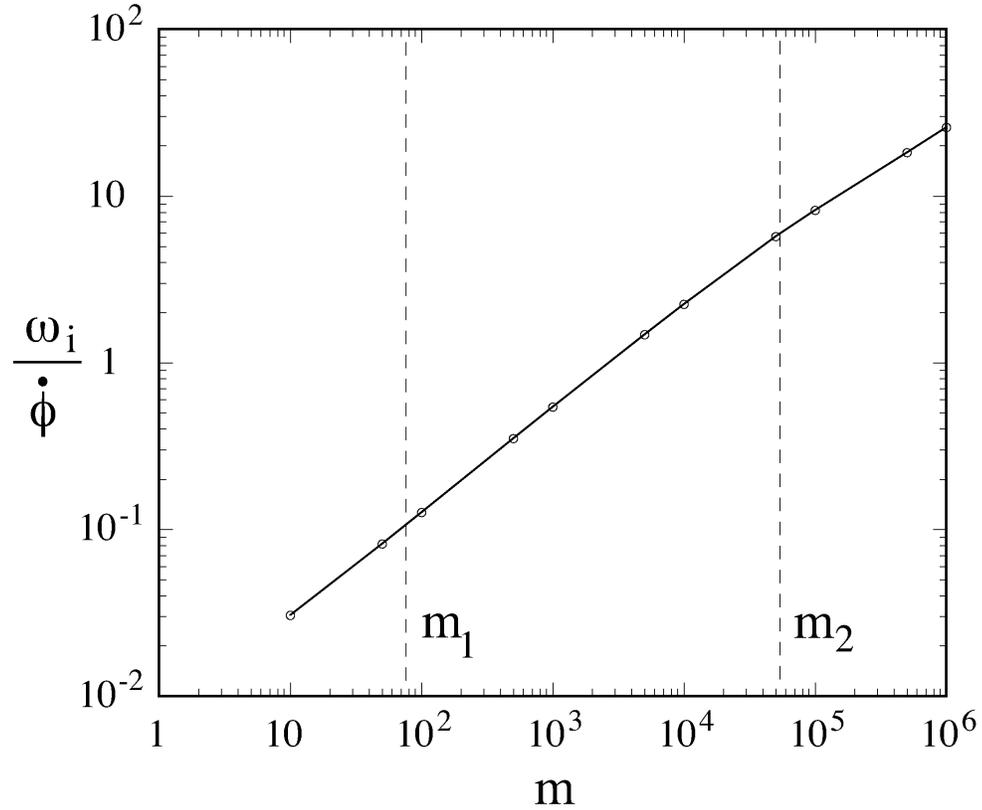}
\caption{The graph shows
the frequency dependence of the growth
rate
for a sample case where $\gamma=30$ and
$\zeta =0.02$ obtained from our approximations
for Eq.~(\ref{disp}). For these parameters,
$m_1\approx 10^2$ and $m_2 \approx 2.7 \times 10^4$.
}
\label{growthnobeta}
\end{figure}

Note that $m_2 \dot{\phi}$ is the approximate frequency
of the peak of the single particle synchrotron
radiation spectrum.
For more accurate results we employ a numerical method for
solving Eq.~(\ref{disp}) outlined in \cite{Botten1983}.
This method also allows us to count the
number of roots which are enclosed
by a contour. The basic idea is that for a null-homotopic
cycle $\Gamma$ which does not cross any poles or roots 
and a meromorphic function $f$ which is not constant \cite{FischerLieb}
\begin{eqnarray}
N(0) = \frac{1}{2\pi i} \int_{\Gamma} \frac{f'(\zeta)}{f(\zeta)} d \zeta
\end{eqnarray}
where $N(0)$ is the number of roots minus the number of poles enclosed by $\Gamma$
(an n-th order root or pole counts as n roots or n poles, respectively).
So far we have no numerical evidence of the existence of more than one
solution with a positive real part.
The numerical results agree very well with our approximations
even if $m < m_1$ and are shown in Fig.~\ref{growthnobeta}.

\subsection{Comparison with Goldreich and Keeley}

       Goldreich and Keeley
\cite{GoldreichKeeley1971} find a
radiation instability in
a thin ring of relativistic,
monoenergetic, zero temperature
electrons constrained to
move in a circle of fixed radius. Under
the condition $1 \ll m^{1/3} \ll
\gamma$ their growth rate is
$\omega_i \approx 1.16\dot{\phi}~
m^{2/3} [r_eN/(\gamma^3r_0)]^{1/2}$
which is close to our growth
rate with $L$ replaced by $r_0$.

\section{Nonlinear Saturation} \label{CSR:nonlin}

           Clearly the rapid exponential growth of the
linear perturbation can continue only for a
finite time.
          We analyze this by studying
the trapping of electrons in the moving
potential wells of the perturbation.
For $(\Delta r/r_0)^2 \ll 1$, the electron orbits
can be treated as circular.
           The equation of motion is
\begin{eqnarray}
{d P_\phi \over dt} = r \delta F_\phi~,\quad
\delta F_\phi = -e[\delta E_\phi +
({\bf v \times }\delta {\bf B})_\phi]~,
\label{eom}
\end{eqnarray}
where $P_\phi$ is the canonical angular momentum, where
\begin{eqnarray}
\delta F_\phi=
-e \delta E_{\phi 0} \exp(\omega_i t)
\cos(m\phi -\omega_r t) ~,\!\!
\end{eqnarray}
where $\delta E_{\phi 0}$ is the initial value of
the potential,
        $\omega_r \equiv {\rm Re}(\omega)$, and
$\omega_i \equiv {\rm Im}(\omega)$.

For a relativistic particle in a circular orbit,
\begin{eqnarray}
\delta P_\phi = m_{e*} r_0^2 \delta \dot{\phi},~~
{\rm where}~ m_{e*}={- m_e \gamma^3 \over \gamma^2-1}
\approx - m_e \gamma,
\label{negmass}
\end{eqnarray}
where $m_{e*}$ is the ``effective mass,'' which
is negative, for the azimuthal motion of the electron
(\cite{Kolomenskii1959,Nielson1959} or \cite{Lawson1988}, p.68).
Combining Eqs.~(\ref{eom}) and (\ref{negmass}) gives
\begin{equation}
{d^2 \varphi \over dt^2}= -\omega_T^2(t)
\sin \varphi~,
\end{equation}
where $
\varphi \equiv m\phi -\omega_t t + \frac{3}{2} \pi$,
$ \omega_T\equiv \omega_{T0}\exp(\omega_i t/2),$ and $
\omega_{T0} \equiv [ e m \delta E_{\phi 0} / ( m_e \gamma
r_0) ]^{1/2}$,
where $\omega_T$ is termed the ``trapping frequency.''
At the ``bottom'' of the potential well of the
wave, $\sin\varphi \approx \varphi$.
An electron  oscillates about the bottom of the
well with an
angular frequency $\sim \omega_T$.
          This is of course a nonlinear effect of the
finite wave amplitude.
          A WKBJ solution of Eq.~(\ref{negmass})
gives
\begin{equation}
\varphi \propto\omega_{T0}^{-1/2} \exp(-\omega_i t/4)
\sin\big\{(2\omega_{T0}/\omega_i)[\exp(\omega_i t/2)
-1]\big\}~.
\end{equation}
         The exponential growth
of the linear perturbation will cease at the time
$t_{sat}$ when the particle is turned around in the
potential well.
         This condition corresponds
to $\omega_T(t_{sat}) \approx \omega_i$.
         Thus, the saturation amplitude is
\begin{eqnarray}
\big|\delta E_{sat}\big|^2 =
\left ( \frac{m_e\gamma }{ e r_0 m } \right )^2
\left({\omega_i(m) \over \dot{\phi}}\right)^4~,
\label{satamp}
\end{eqnarray}
where $|\delta E_{sat}|\equiv
|\delta E(t_{sat})|=|\delta E_0|\exp(\omega_i
t_{sat})$.

\section{First Approximation with $k_z \neq 0$} \label{CSR:kzneq0}

Here, we consider $k_z \neq 0$ but keep the
other approximations. Our ansatz for $\delta f$
is general enough to handle this case since it
retains the biggest contribution to the Lorentz force in
the $z$-direction which is of the order $v_{\phi} B_r$.
In place of Eq.~(\ref{dfpots}) we obtain
\begin{equation}
\int d P_{\phi} \delta f =
-e \int d P_{\phi} {f^0 \over H}~
{m \dot{\phi}^2(\omega \delta \Psi - m \delta \Phi)
\over (\omega-m\dot{\phi} - k_z v_z)^2 }~,
\end{equation}
where we assume without loss of generality $k_z>0$
and $k_z \ll m/r_0, \omega$.
         In place of Eq.~(\ref{disp}) we find
\begin{equation}
\varepsilon(\omega,k_z)=1\! +
k_zA(\omega,k_z)\int_{-\infty}^\infty \!\!dv_z
~{\exp(-v_z^2/2v_{th}^2)
\over \sqrt{2\pi}~ v_{th}}\big[...\big]
=0,
\end{equation}
where
$$
\big[...\big] \equiv
-{m \dot{\phi} \over (\omega-m\dot{\phi}-k_zv_z)^2}~.
$$
Here, $\varepsilon$ acts as an effective dielectric
constant for the E-layer,  and
$$
A(\omega,k_z) \equiv \pi ~\zeta ~Z(\omega,k_z)~
\left(u-{k_\phi \over k_z \gamma^2}\right),
~ ~ u\equiv
{\omega - m\dot{\phi} \over k_z}~,
$$
\begin{equation}
Z \equiv iJ_m[r_0(\omega^2-k_z^2)^{1/2}]~
H_m^{(1)}[r_0(\omega^2-k_z^2)^{1/2}]~,
\end{equation}
and $k_\phi = m/r_0$ is the azimuthal wavenumber.
The expression for $Z$ is from \S \ref{CSR:firstapprox}.
          An integration by parts gives
\begin{eqnarray}
\varepsilon(u)=1 +
A(\omega)\int dv_z~{\exp(-v_z^2/2v_{th}^2)
\over \sqrt{2\pi}~ v_{th}^3}
{m\dot{\phi} v_z/k_z \over v_z- u} ~,
\end{eqnarray}
where the $k_z$ dependence of $\varepsilon$
and $A$ is henceforth implicit.
          We can also write  this equation as
\begin{eqnarray}
\varepsilon(u)=1 + B(u) \left [ 1 +
\frac{u}{v_{th}}
F\left({u\over v_{th}}\right) \right ]~,
\label{epsu}
\end{eqnarray}
where
\begin{eqnarray}
\nonumber\\
B(u) &\equiv& {\pi \over v_{th}^2}~\zeta~Z~{k_\phi \over k_z}
\left(u- {k_\phi \over k_z \gamma^2}\right)~,
\end{eqnarray}
and
$$
         F(z) \equiv
{1 \over \sqrt{2\pi}}\int_{-\infty}^\infty
dx ~{\exp(-{x^2 / 2}) \over x-z}~,
$$
for ${\rm Im}(z)>0$, and
$$
         F(z) \equiv
{1 \over \sqrt{2\pi}}\int_{-\infty}^\infty
dx ~{\exp(-{x^2 / 2}) \over x-z}+i\sqrt{2\pi}
\exp\left(-{z^2 \over 2}\right)~,
$$
${\rm for~Im}(z)<0$.
        The second expression for $F(z)$ is the
analytic continuation of the first expression
to ${\rm Im}(z)<0$ which corresponds to wave
damping
(see, e.g., \cite{Montgomery1964}, ch. 5).
Note that terms of order $\Delta \tilde \omega$
have been omitted.

          For $m\gg1$, the factor $Z=iJ_m(J_m+iY_m)$ can
be expressed in terms of Airy functions in a way
similar to that done in \S \ref{CSR:nonlin}.
           One finds
$J_m[r_0(\omega^2-k_z^2)^{1/2}] \approx
(2/m)^{1/3}{\rm Ai}(w)$,
$Y_m[r_0(\omega^2-k_z^2)^{1/2}] \approx
-(2/m)^{1/3}{\rm Bi}(w)$,
\begin{equation}
        Z_r\approx \left({2\over m}\right)^{2/3}{\rm Ai}(w){\rm
Bi}(w)~,~ Z_i\approx \left({2\over
m}\right)^{2/3}{\rm Ai}^2(w)~,
\label{ZriAiBi}
\end{equation}
        where
$$
        \tan\psi \equiv {k_z \over k_\phi}~,~
w \equiv \left({m \over 2}\right)^{2/3}
\left({1\over \gamma^2} +\tan^2\psi
-2u\tan\psi\right)~.
$$
It is clear that $\varepsilon$ has in general a rather
complicated dependence on $u=u_r+iu_i$ and $\tan \psi$.
          Note that the expression for $w$ goes
over to our earlier $w$ for $\psi=0$
noting that $u\tan \psi \rightarrow \Delta
\tilde{\omega}$.

\begin{figure}
\hspace{0.5in}
\includegraphics[width=4.5in]{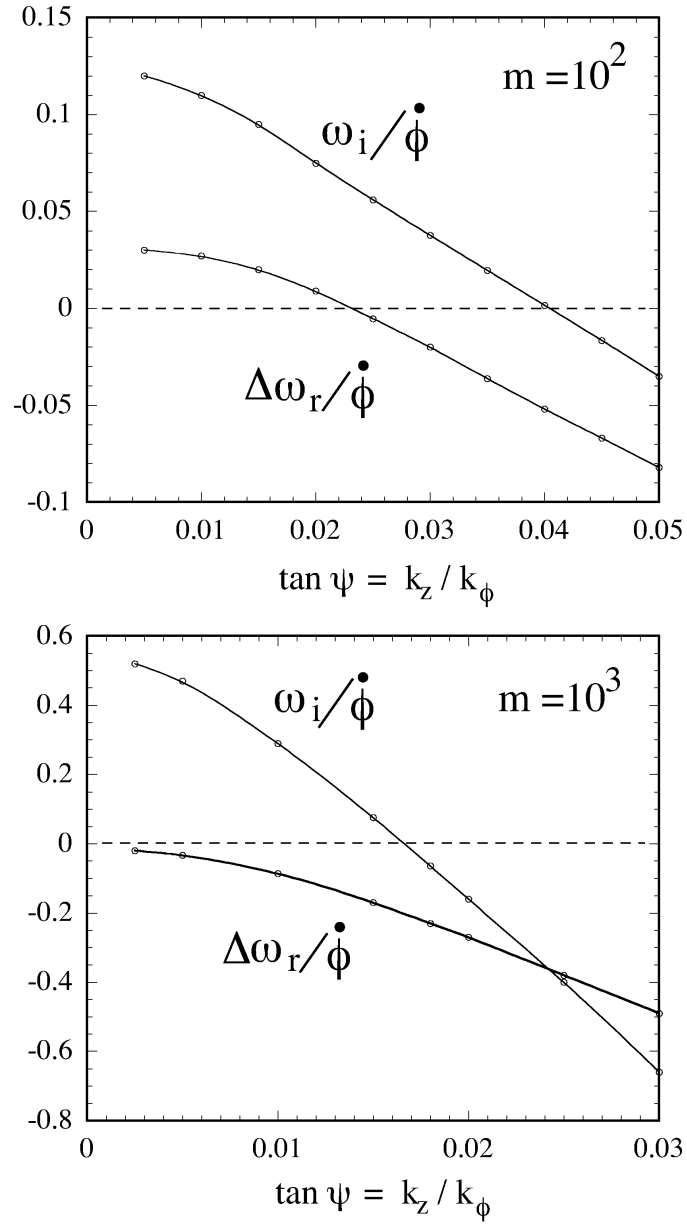}
\caption{The figure shows
the growth / damping rate $ \omega_{i}$
and real part of the
frequency $\Delta \omega_{r}=\omega-m\dot{\phi}$
in units of $\dot{\phi}$ as
a function of $\tan\psi=k_z/k_\phi$ for $m=100$
and $m=1000$ for an E-layers
with $\gamma=30$, $\zeta=0.02$ and $v_{th}=30/\gamma^2$.
In the region of damping $\omega_i<0$, the
second expression for $F(z)$ in Eq.~(\ref{epsu}) is used.}
\label{growthkz}
\end{figure}

          A limit where Eq.~(\ref{epsu}) can be solved
analytically is for $|u|^2= |\Delta \tilde{\omega}|^2/
\tan^2\psi \gg v_{th}^2$, that is, for
sufficiently small $\tan \psi$.
          In this limit Eq.~(\ref{epsu}) can be expanded
as an asymptotic
series $F(z)=-1/z - 1/z^3 - 3/z^5 -..$.
          Keeping
just the first three terms of the expansion gives
\begin{equation}
\varepsilon =1 + \pi \zeta Z
\left(1+{3 v_{th}^2 \tan^2\psi \over (\Delta \tilde{\omega})^2}\right)
{{\gamma^{-2}}\over(\Delta \tilde{\omega} )^2} = 0.
\label{dispkzapprx}
\end{equation}
         For $\tan\psi \rightarrow 0$ and $\Delta \tilde \omega \ll \gamma^{-2}$, this is
the same as Eq.~(\ref{disp}) as it should be.
In general Eq.~(\ref{dispkzapprx}) will have more than one unstable mode.
In the remainder of this paragraph we will only study the largest unstable solution
for which we recover the growth rates found in \S \ref{CSR:firstapprox} in the limit $\tan\psi \rightarrow 0$.
Fig.~\ref{growthkz} shows some sample solutions.
        For the case
shown the $u$ dependence of $Z$ is negligible.

         General solutions of Eq.~(\ref{epsu}) can be obtained
using the Newton-Raphson method (\cite{Teukolsky1989},
ch. 9) where an
initial guess of $(u_r,u_i)$ gives $(\epsilon_r,\epsilon_i)$.
This guess is incremented by an amount
\begin{eqnarray}
          \left[\begin{array}{c} {\delta u_r}
\\ \noalign{\medskip}{\delta u_i} \end{array}\right]
~= ~\left [\begin {array}{cc}
{\partial \epsilon_r/\partial u_r}&{\partial
\epsilon_r/\partial u_i}
\\\noalign{\medskip}{\partial \epsilon_i/\partial u_r}&
{\partial \epsilon_i/\partial u_i}
\end {array}\right ]^{-1}
\left[\begin{array}{c} {-\epsilon_r}
\\ \noalign{\medskip}{-\epsilon_i} \end{array}\right]~,
\end{eqnarray}
and the process is repeated until $\varepsilon_r=0$
and $\varepsilon_i=0$.
         Fortunately, the convergence
is very rapid and gives
$|\varepsilon| < 10^{-10}$ after a few iterations.

\begin{figure}
\includegraphics[width=\columnwidth]{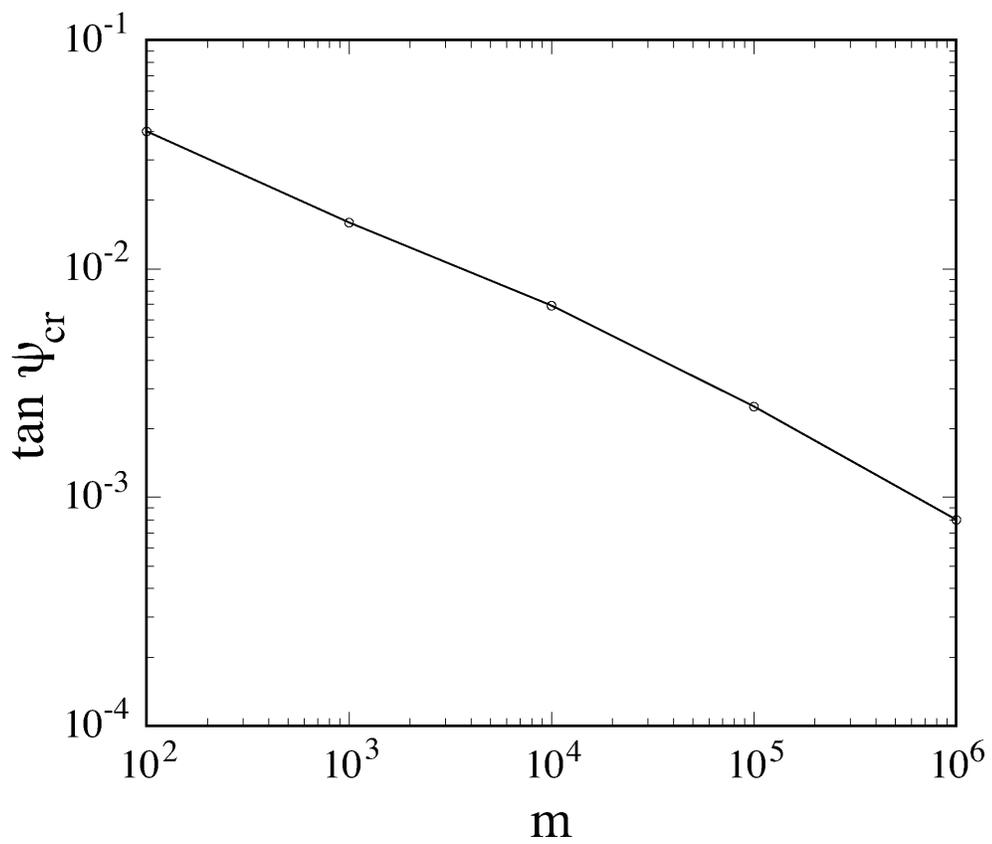}
\caption{Critical angle for
$\gamma=30$, $\zeta=0.02$ and $v_{th}=30/\gamma^2$.}
\label{critang}
\end{figure}

         Fig.~\ref{growthkz} shows the dependence of the
complex wave frequency on the tangent
of the propagation angle,
$\tan \psi = k_z/k_\phi$, for a sample cases.
          The maximum growth rate is for $\psi =0$
or $k_z=0$.
          With increasing $\psi$ the growth rate
decreases, and for $\psi$ larger
than a critical angle $\psi_{cr}$ there is
damping.  For the damping the second expression
for $F$ in Eq.~(\ref{epsu}) must be used.
          Roughly, we find that the critical angle
corresponds to having the wave phase velocity
in the $z-$direction of the order of the thermal
spread in this direction,  that is,
$u_r = \Delta \omega_r/k_z \sim v_{th}$.
This gives
\begin{eqnarray}
\tan\psi_{cr} \sim {\sqrt{\zeta}
\over v_{th} \! ~ \! \gamma ~ m^{1/3}}=
\left({ r_e N \over v_{th}^2
\gamma^3 L}\right)^{1/2} \!
{1\over m^{1/3}} \le \frac{1}{\gamma^2 v_{th}}~,
\label{tanpsi}
\end{eqnarray}
for $m_1<m<m_2$. Note that the dimensionless parameter
which determines the cut-off at $\tan \psi_{cr}$ is $\gamma^2 v_{th}$.
Our numerical calculations of
$\psi_{cr}$ give a slightly
faster dependence, $\tan \psi_{cr}
\propto 1/m^{0.40}$
for this range of $m$. Fig.~\ref{critang} shows
the $m$-dependence of the critical angle.
It is reasonable to assume that in a particle accelerator
the weak focusing in the z-direction sets a low limit on $k_z$.

\section{Nonlinear Saturation for $k_z \neq 0$} \label{CSR:nonsatkzneq0}

We generalize the results of \S \ref{CSR:kzneq0} by
including the axial as well as the azimuthal
motion of the electrons in the wave.
          The axial equation of motion is
\begin{eqnarray}
m_e\gamma {d^2 z \over dt^2}
&=&-e\left[\delta E_z +({\bf v}
\times \delta {\bf B})_z\right]
\nonumber\\
& \approx & -e \delta E_{z 0}
\exp (\omega_i t) \cos(m\phi+k_z z -\omega t)~.
\label{axialeom1}
\end{eqnarray}
         The approximation involves neglecting the force
$\propto v_r \delta B_\phi$ which is valid for
a radially thin layer ($\Delta r^2/r_0^2 \ll 1$).
           Following the development of \S \ref{CSR:kzneq0},
the azimuthal equation of motion
is
\begin{eqnarray}
m_e\gamma r_0{d^2 \phi \over dt^2}
=-e \delta E_{z 0} \cos(m\phi+k_z z -\omega t)~.
\label{axialeom2}
\end{eqnarray}
Combining Eqs.~(\ref{axialeom1}) and (\ref{axialeom2}) gives
\begin{eqnarray}
{d^2\varphi \over dt^2}=-~{e~ m \delta E_{\phi 0} \over
m_e \gamma r_0}\left(1+\tan^2\psi\right)
\sin\varphi~,
\end{eqnarray}
where $\varphi\equiv m\phi+k_zz-\omega t +\frac{3}{2} \pi$
and $\tan\psi = k_z/k_\phi$.
          Because $\psi^2 \ll 1$ for wave growth
(Eq.~(\ref{tanpsi})), the saturation
wave amplitude $\delta E_{sat}$
is again given by Eq.~(\ref{satamp}).

\section{Thick Layers Including Radial Betatron Oscillations} \label{CSR:thick}
\subsection{The Limit $k_r \Delta r \gg 1$}

In this section we include the
small but finite radial thickness of the E-layer.
         We keep the other approximations mentioned
at the beginning of \S \ref{CSR:firstapprox}.
        In particular we consider
$k_z=0$.
         In order to include the layer's radial thickness, we
consider the wave equations within the E-layer,
\begin{eqnarray}
(\nabla^2 + \omega^2)\delta \Phi
&=& -4\pi \delta \rho~,
\nonumber \\
(\tilde{\nabla}^2 +\omega^2)\delta \Psi
&=& -4\pi r \delta J_\phi~,
\label{laplacians}
\end{eqnarray}
where
\begin{equation}
\tilde{\nabla}^2 \equiv {\partial^2 \over \partial r^2}
-{1 \over r}{\partial \over \partial r} - \frac{m^2}{r^2}
+{\partial^2 \over \partial z^2}~,
\end{equation}
is the adjoint Laplacian operator.

        Within the E-layer, we assume
that the potentials can be written
in a WKBJ expansion as
\begin{equation}
(\delta \Phi,~\delta \Psi) =
(K_\Phi,~K_\Psi)\exp\big[im\phi
+ik_r (r-r_0) -i\omega t\big]~,
\label{WKBJ}
\end{equation}
where $k_r$ is the radial wavenumber with
$(k_r \Delta r)^2 \gg 1$
$(K_\Phi,K_\Psi)$ are constants.
        This is equivalent to assuming
that the charge density
is constant between $r_0-\Delta r$
and $r_0+\Delta r$ and zero elsewhere.
Evaluation of the time integrals in
Eq.~(\ref{dfLandaupots}) for $r=r_0$  gives

\begin{eqnarray}
e^{-iz \sin t} = \sum_{n=-\infty}^{\infty} e^{-int} J_n(z)
\end{eqnarray}

\begin{eqnarray}
\nonumber
\int_{-\infty}^{t} dt' \delta \Phi e^{-i \omega t' + im \phi' + i k_r r'} =
\quad\quad\quad\quad
\\ \nonumber
\int_{-\infty}^{t} dt' \delta \Phi \exp \left \{
- i \omega t' + im \left [ \phi + (t'-t) \dot \phi_0 - 
\right . \right . \\ \nonumber \left . \left .
\frac{\delta r_i}{r_0}
\left ( - \cos [ \omega_{\beta r} (t'-t) ] +1 \right ) \right ] + 
i k_r \delta r_i \sin ( \omega_{\beta r} (t' - t) ) + i k_r r_0 \right \}
\\
= \int_{-\infty}^{t} dt' \delta \Phi e^{-i \omega t' + im \dot \phi_0 + i k_r r_0 - im \frac{\delta r_i}{r_0} } \sum_{n=-\infty}^{\infty} e^{-in \omega_{\beta r} t'} J_n(-k_r \delta r_i )
\end{eqnarray}

\begin{eqnarray}
\int_{-\infty}^t dt^\prime \delta \Phi^\prime =
\delta \Phi(r_0,t)
\sum_{n=-\infty}^\infty {J_n(k \delta r_i)i^n
\exp(-ik_\phi \delta r_i -in\psi)
\over i(m \dot{\phi} + n \omega_{\beta r} -\omega)}~,
\label{Psiint}
\end{eqnarray}
where $n$ is an integer, $k\equiv (k_r^2+k_\phi^2)^{1/2}$,
with $k_\phi = m/r_0$, and $\tan \psi \equiv k_r/k_\phi$.
       There is an analogous expression for the integral
of $\delta \Psi$.
          We have used Eq.~(\ref{rorbit})
for the radial motion with $\varphi =0$
assuming $\zeta^2 \ll 1$ and
$\zeta \ll \gamma^2(\Delta r/r_0)$ so
that $\omega_{\beta r} =1/r_0$, and 
Eq.~(\ref{phiorbit}) for the $\phi$-motion with
       $ \partial \dot{\phi}_0/\partial r|_{r_0}/\omega_{\beta r}
= - 1/r_0$.
Using Eqs.~(\ref{dfLandaupots}) and (\ref{Psiint}), the momentum space
integrals (\ref{psints}) can be done to give
\begin{eqnarray}
\nonumber
\left ( eK e^{-H/T} \right )^{-1} \int  d P_{\phi} \delta f = -
\frac{m \dot \phi^2 }{H} \frac{\omega K_{\Psi}-m K_{\Phi}}{(m \dot \phi - \omega)^2} + \quad\quad\quad\quad
\\ \nonumber
\frac{1}{T} (K_\Psi\dot \phi -K_\Phi)\left \{
J_0(k \delta r_i) - 1 + (m \dot \phi - \omega) \sum_{n=-\infty}^{~~\infty~~\prime}
\frac{i^n e^{-in \psi -i k_{\phi} \delta r_i} J_n(k \delta r_i)}
{m \dot \phi + n \omega_{\beta r} - \omega}  \right \} -
\\ 
\frac{m \dot \phi^2 }{H} m( K_{\Psi} \dot \phi - K_{\Phi} ) \left \{
\frac{J_0(k \delta r_i) - 1}{(m \dot \phi - \omega)^2} +
\sum_{n=-\infty}^{~~\infty~~\prime} \frac{i^n e^{-in \psi -i k_{\phi} \delta r_i}
J_n(k \delta r_i)}{(m \dot \phi + n \omega_{\beta r} - \omega )^2} \right\}
~,
\end{eqnarray}
and finally if $\Delta \tilde \omega \ll \gamma^{-2}$
\begin{eqnarray}
\nonumber
\delta \rho \approx {e^2 n_0  m \dot{\phi}^2K_\Phi \over H}
\bigg( { r_0^2  \dot{\phi} \omega
-m[1-(1-F_0)/\gamma^2]\over
(\omega - m\dot{\phi})^2}  \! -  \! {m\over \gamma^2} \! \!
\sum_{n=-\infty}^{~~\infty~~\prime}
{F_n  \over (m\dot{\phi}+n \omega_{\beta r} -\omega)^2} \bigg)~.
\\
\label{drhotthick}
\end{eqnarray}
The prime on the sums indicate
that the $n=0$ term is omitted.
Here,
\begin{equation}
F_n\equiv {i^n \exp(-in\psi)\over \sqrt{2\pi} \chi}
\int_{-\infty}^\infty d\xi~
J_n(\xi) \exp\left(- {\xi^2 \over 2 \chi^2}
-i{k_\phi  \xi \over k} \right)~,
       \end{equation}
with
\begin{equation}
\chi \equiv k\Delta r ~.
\end{equation}
The $1/T$ terms in Eq.~(\ref{drhotthick}) do not cancel exactly.
      They may  be neglected if
\begin{equation}
|\Delta \tilde \omega|^2 \ll {F_0}v_{th}^2
\end{equation}
for the $n=0$ term or if
\begin{equation}
{|\Delta \tilde \omega|}|n/m - \Delta \tilde \omega| \ll {v_{th}^2}
\end{equation}
for the $n \neq 0$ terms.
      
For  weak E-layers
we have for
$\chi \rightarrow 0$,  $F_0 \rightarrow 1$
and $F_{n\neq 0} \rightarrow 0$.
     In this limit we
recover the results of \S \ref{CSR:firstapprox}.
For $\chi \gg 1$ and
$1 \ll  k_r r_0 \ll  k_{\phi} r_0$, the Gaussian
factor in the integrand of $F_n$
   can be neglected so that one obtains
$$
F_n \approx
i^n e^{-in \psi} \frac{1}{\sqrt{2 \pi} \chi}
\frac{2k}{|k_r|} \cos
\left ( \frac{n \pi}{2} \right )~,~~
{\rm even}~  n~,
$$
\begin{equation}
F_n \approx- i^n e^{-in \psi}
\frac{1}{\sqrt{2 \pi} \chi} \frac{2ik}{|k_r|} \sin
\left ( \frac{n \pi}{2} \right )~,~~
{\rm odd} \ n~.
\end{equation}

\begin{figure}
\includegraphics[width=\columnwidth]{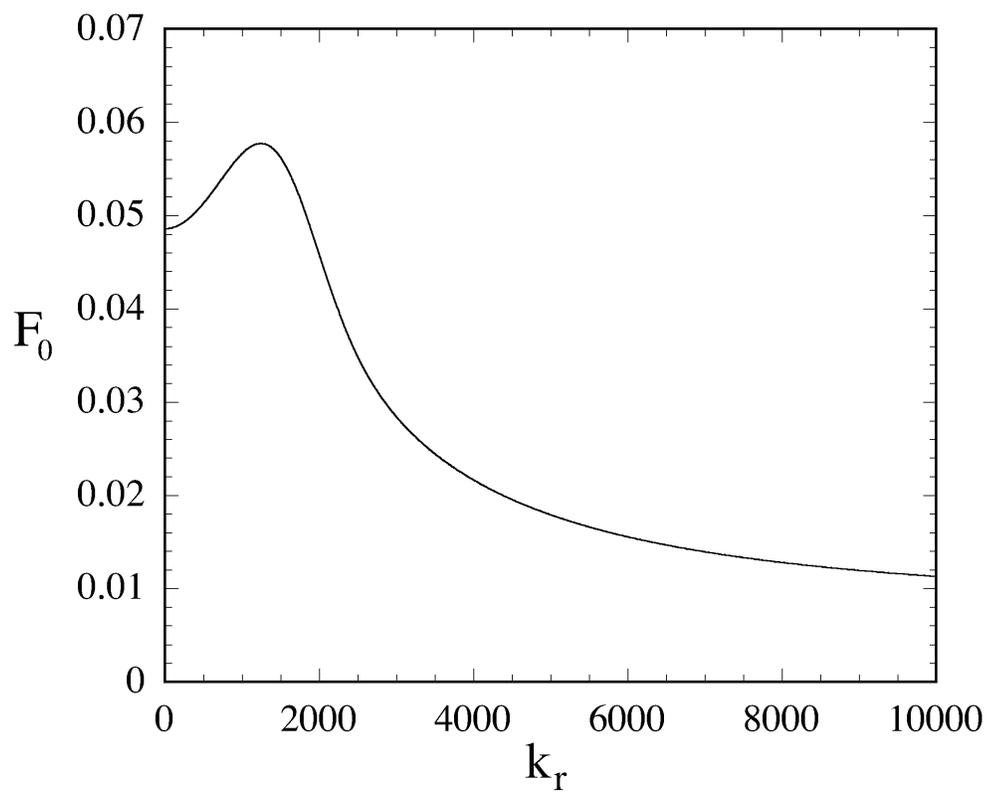}
\caption{$F_0$ for $v_{th}=0.01$ and $ k_{\phi}r_0=10^4$}
\label{figF0}
\end{figure}

     An alternative approximation
for $F_n$ can be obtained
by using the
   integral representation of
the Bessel function.
The remaining integral can then
be computed numerically more easily.
In this way we find
    \begin{equation}
F_n = \frac{i^n e^{-in \psi}}{2\pi} \int _{-\pi }^\pi
d\theta \exp\left[-in \theta -(\chi^2/2)  ( k_{\phi}/k - \sin
\theta  )^2\right]~.
\label{Fnnum}
    \end{equation}
For $\chi \gg 1$, $1 \ll (k_{\phi}/k_r)^2$ and $|n| < \sqrt{\chi}$
we can approximate $\sin \theta$ in the exponent by a parabola
at its maximum. We obtain
     \begin{equation}
F_n \approx \frac{i^n e^{-in \psi}}{2^{3/4}
\Gamma\left(\frac{3}{4}
\right ) \sqrt{\chi}}~.
\end{equation}
     In general $ F_n/(i^n e^{-in \psi})$
decreases as $\chi$ and $n$ increase.
     This acts to prevent the
unlimited increase of
the growth rate as $m \longrightarrow \infty$,
and it ensures that the sums over $n$ converge.
Fig.~\ref{figF0} shows a plot of $F_0$ obtained
by numerical evaluation of
   Eq.~(\ref{Fnnum}).

Within the E-layer, Eq.~(\ref{laplacians}) gives
\begin{eqnarray}
\nonumber
k_r^2=\omega^2-{m^2 \over r_0^2}
+{4\pi e^2 n_0  m \dot{\phi}^2 \over H}
\quad\quad\quad\quad\quad\quad\quad
\\
\times \bigg({r_0^2\dot{\phi}\omega -m[1-(1-F_0)/\gamma^2] \over
(\omega - m\dot{\phi})^2}
-~ {m\over \gamma^2}~{\sum_n}^\prime {F_n \over
(m\dot{\phi}+n\omega_{\beta r} -\omega)^2} \bigg)~.
\label{calckr}
\end{eqnarray}
In terms of dimensionless variables this equation becomes
\begin{eqnarray}
\nonumber
{\bar k_r}^2 = 2m^2 \Delta
\tilde \omega -\frac{m^2}{\gamma^2} + 
\frac{\zeta \sqrt{2}}{v_{\phi}
v_{th} \sqrt{\pi}} \quad\quad\quad\quad\quad\quad\quad\quad
\\
\times \left (\frac{(1+ \Delta \tilde \omega)
(1 - \gamma^{-2})
- [1+(F_0 - 1)/ \gamma^2]}{(\Delta \tilde
\omega)^2}
- \frac{1}{\gamma^2} {\sum_n} '
\frac{F_n}{( v_{\phi}^{-1} n/m -
\Delta \tilde \omega)^2} \right )~,\quad~~
\label{calckrbar}
\end{eqnarray}
where $\bar k_r \equiv r_0 k_r$,
$ \bar k_{\phi} \equiv r_0 k_{\phi}$,
$\bar k \equiv r_0 k$, and $\chi = \bar k v_{th}$.

         Notice that
Eq.~(\ref{WKBJ}) can also be written
as
\begin{equation}
\delta \Phi = C_2\sin\big[k_r(r-r_0)\big]
+C_3\cos\big[k_r(r-r_0)\big]~,
\end{equation}
for $r_0-\Delta r \leq r \leq r_0+\Delta r$.
         For $r\leq r_0-\Delta r$, we have
\begin{equation}
\delta \Phi = C_1 J_m(\omega r)~,
\end{equation}
since the potential must be well behaved
as $r \rightarrow 0$.
         For $ r\geq r_0+\Delta r$, we must have
\begin{equation}
\delta \Phi = C_4 \big[J_m(\omega r) + i Y_m(\omega r)\big]~.
\end{equation}
This combination of Bessel functions gives
$\delta \Phi(r\rightarrow \infty) \rightarrow 0$
for the assumed conditions where ${\rm Im}(\omega)>0$.
Note that these potentials are just the solutions
of Eq.~(\ref{laplacians}) in our approximation for
$\delta \rho$. The eigenvalue problem can now be solved
by matching the boundary conditions. However,
we have not solved the full eigenvalue problem.
        Instead we consider unstable solutions
with the restriction that
$k_r \Delta r \gg 1$.
        Under this condition
we can interpret Eq.~(\ref{calckrbar}) as a local
dispersion relation.  Unstable modes found
from Eq.~(\ref{calckrbar}) will need a slight
correction in order to satisfy the boundary conditions.

We expect that Eq.~(\ref{calckrbar}) has solutions near
each betatron resonance at $\Delta \tilde \omega = {\pm n}/{m}$.
This is a familiar concept in the treatment of resonances in storage
rings (cf. \cite{chao} or \cite{Schmekel:2003cs}).
We extract each solution by summing over a single value
of $n$ and $-n$ only and obtain from Eq.~(\ref{calckrbar})
for the case $n \neq 0$ and $\Delta \tilde \omega \ll \gamma^{-2}$
\begin{eqnarray}
\nonumber
\Xi \equiv - \frac{\gamma^2 v_{th} \sqrt{\pi} 
\left ( \bar k_r^2 +m^2 \gamma^{-2}\right )}{\zeta \sqrt{2}} = 
\frac{F_{-n}}{\left ( \frac{n}{m} + \Delta \tilde \omega \right )^2} +
\frac{F_{n}}{\left ( \frac{n}{m} - \Delta \tilde \omega \right )^2} ~.
\end{eqnarray}
Thus,
\begin{eqnarray}
\Delta \tilde \omega \approx \frac{F_{-n}-F_n}{\Xi} \pm \frac{n}{m}
\end{eqnarray}
for sufficiently big $\Xi$, i.e. we expect the imaginary part of
$\Delta \tilde \omega$ to be negligible for the $n \neq 0$ modes.
Despite a lot of effort we were not able to prove this statement
under more relaxed conditions.

We can easily find an analytic
solution of Eq.~(\ref{calckrbar})
for the case where the $n=0$ term is dominant.
If $|\Delta \tilde \omega| \ll 1 / \gamma^2$ and
$|\Delta \tilde \omega| \ll F_0 / \gamma^2$, we obtain
\begin{eqnarray}
\Delta \tilde \omega = \pm \frac{2^{1/4} \sqrt{-\zeta F_0}
}{\pi^{1/4} \sqrt{v_{th} (m^2+\gamma^2 \bar k_r^2)}}
\label{soln0}
\end{eqnarray}
The dependence of the growth rate on $k_r$ becomes
significant when
$\gamma^2 \bar k_r^2 / m^2$ is comparable to unity.
For $m\sim m_2\sim \gamma^3$, we see that this happens
when $(k_r\Delta r)^2/\gamma^4v_{th}^2\sim 1$,
which involves the combination $\gamma^2 v_{th}$ again.

\begin{figure}
\includegraphics[width=\columnwidth]{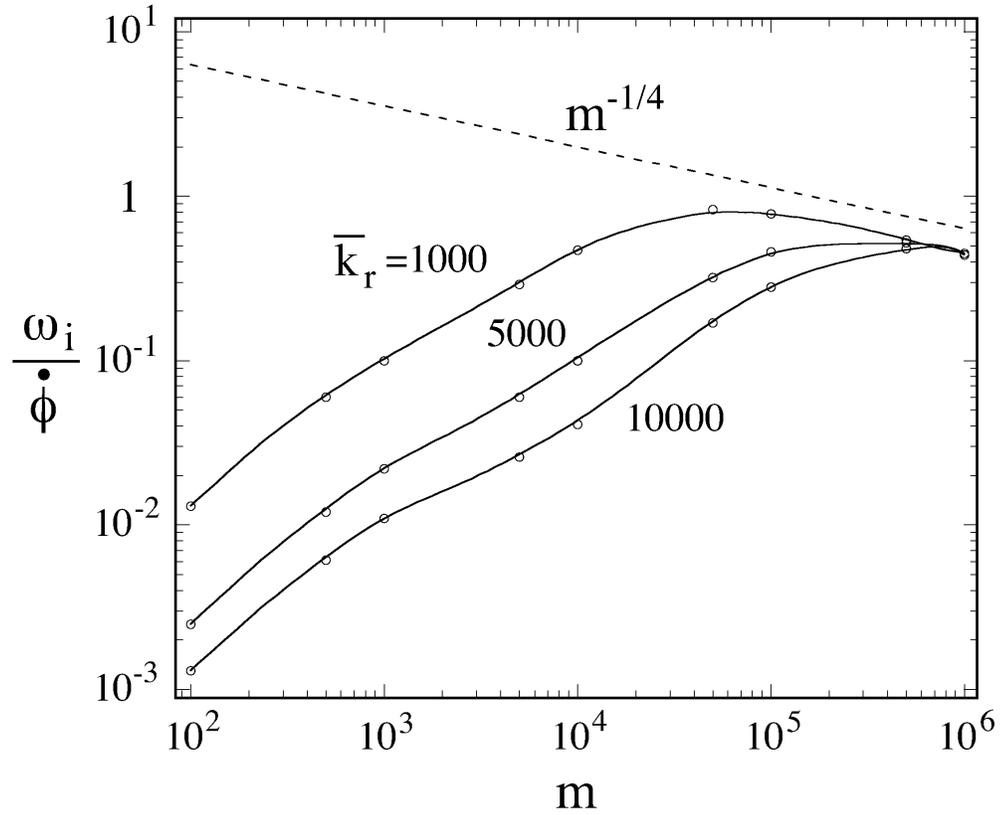}
\caption{Growth rates in the limit $\bar k_r v_{th} \gg 1$
for our reference case $\gamma=30$, $\zeta=0.02$ and
$v_{th}=1/\gamma^2$ and various values of $\bar k_r$. The line
proportional to $m^{-1/4}$ is shown for comparison.}
\label{krbvthgg1}
\end{figure}

The growth rate of Eq.~(\ref{soln0}) is proportional to $\sqrt{\zeta}$.
     This  implies from \S \ref{CSR:nonlin} that
the emitted power scales as the square of
the number of particles in the E-layer
which corresponds to coherent radiation.
       Sample results are
shown in Fig.~\ref{krbvthgg1}.
      We conclude that
the main effect of the betatron oscillations
is an indirect one.
    The radial motion itself is
unimportant for the interaction.
    However, the influence of the radial motion
on the time dependence of
the azimuthal angle
$\phi$ of a particle is important since a
shift in $\phi$ can take the particle  out
of coherence with the wave.
      This effect is accounted for
by $F_0$.

\subsection{Qualitative Analysis of the Effect of the Betatron Motion}

Let us suppose that $v_{th} \gg 1/\gamma^2$, and that
$|\Delta \tilde \omega|$ is not necessarily small compared with $v_{th}$
(We can still assume $|\Delta \tilde \omega| \ll 1$ without requiring the
more restrictive condition $\gamma^2 | \Delta \tilde \omega| \ll 1$.).
The key effect of the betatron oscillations is to ``wash out'' the 
phase coherence
of the response within the layer; for a cold layer, all orbiting particles
move in ``lock step'', which is particularly favorable for a bunching 
instability.
Let us suppose that $|\Delta \tilde \omega|$ has a real part that is
substantially larger than $1/\gamma^2$. The response in the layer scales as
an Airy function with argument $w(1+\xi)$ where $|\xi| < v_{th}$.
The phase accumulated across the layer thickness $\sim v_{th} r_0$ is
$\eta \sim m v_{th}^{3/2}$ if $\Delta \tilde \omega_r \ll v_{th}$
and $\eta \sim m v_{th}^{3/2} (\Delta \tilde \omega_r / v_{th})^{1/2}$
if  $\Delta \tilde \omega_r \gg v_{th}$.
Large $\eta$ ought to imply substantial decoherence of the response 
in the layer.
We see that this is likely irrespective of the value of
$\Delta \tilde \omega_r / v_{th}$ provided that $m \gg v_{th}^{-3/2}$, i.e. for
$m/\gamma^3 \gg (\gamma^2 v_{th})^{-3/2}$. At large values of 
$\gamma^2 v_{th}$, phase
smearing should suffice to suppress - if not eliminate - the bunching 
instability at
frequencies near the synchrotron peak. Moreover, if $\gamma^2 v_{th} 
\gtrsim \zeta^{-1}$,
the instability should be suppressed over the entire range $m \gtrsim 
\zeta^{3/2} \gamma^3$
for which we found unstable modes in \S \ref{CSR:firstapprox}. Large $\Delta \tilde 
\omega_r / v_{th}$
would merely accentuate the smearing. At a given value of $m$, we see that
$\Delta \tilde \omega_r \gtrsim (m^2 v_{th}^2)^{-1}$, i.e.
$\gamma^2 \Delta \tilde \omega_r \gtrsim (m/\gamma^3)^{-1}(\gamma^2 
v_{th})^{-2}$
suffices for large phase decoherence in the layer.

\subsection{The Limit $k_r \Delta r \ll 1$}
\begin{figure}
\includegraphics[width=\columnwidth]{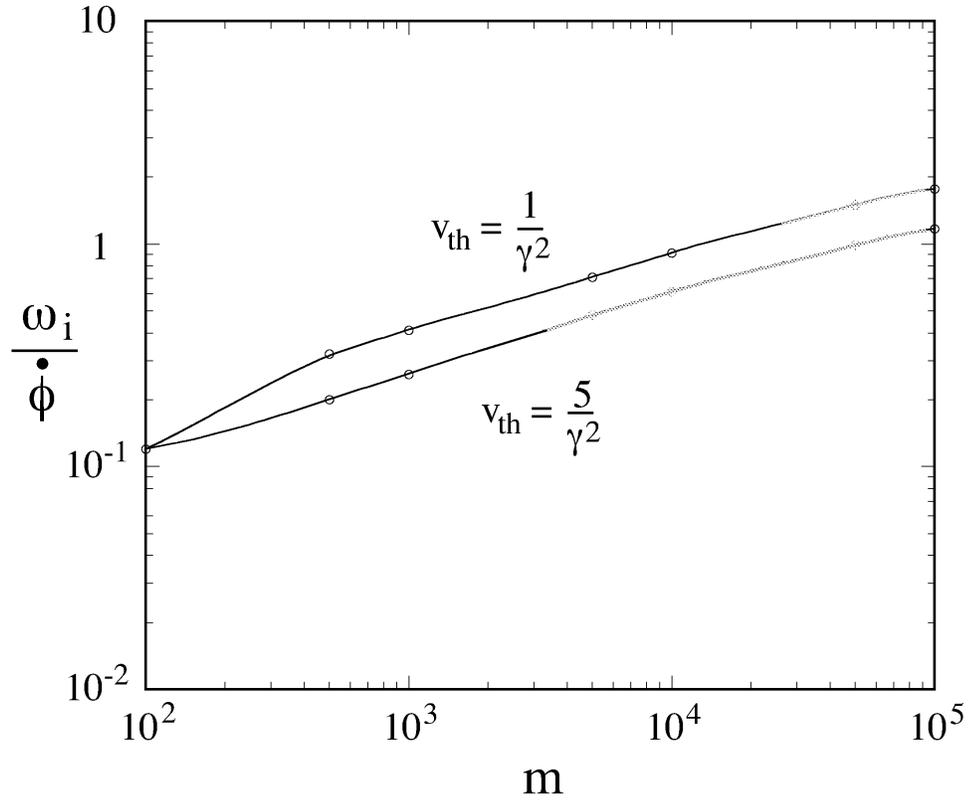}
\caption{Solutions of the dispersion
relation in the presence
of betatron oscillations in
the limit $\bar k_r v_{th} \ll 1$,
$\gamma = 30$, $\zeta = 0.02$.
Points which do not satisfy the inequalities
$m \gg 1$, $m^{2/3} v_{th} < 1$ and
$|\Delta \tilde \omega|^2 < F_0 v_{th}^2$
are plotted in gray.  }
\label{thinbetatron}
\end{figure}

\begin{figure}
\includegraphics[width=\columnwidth]{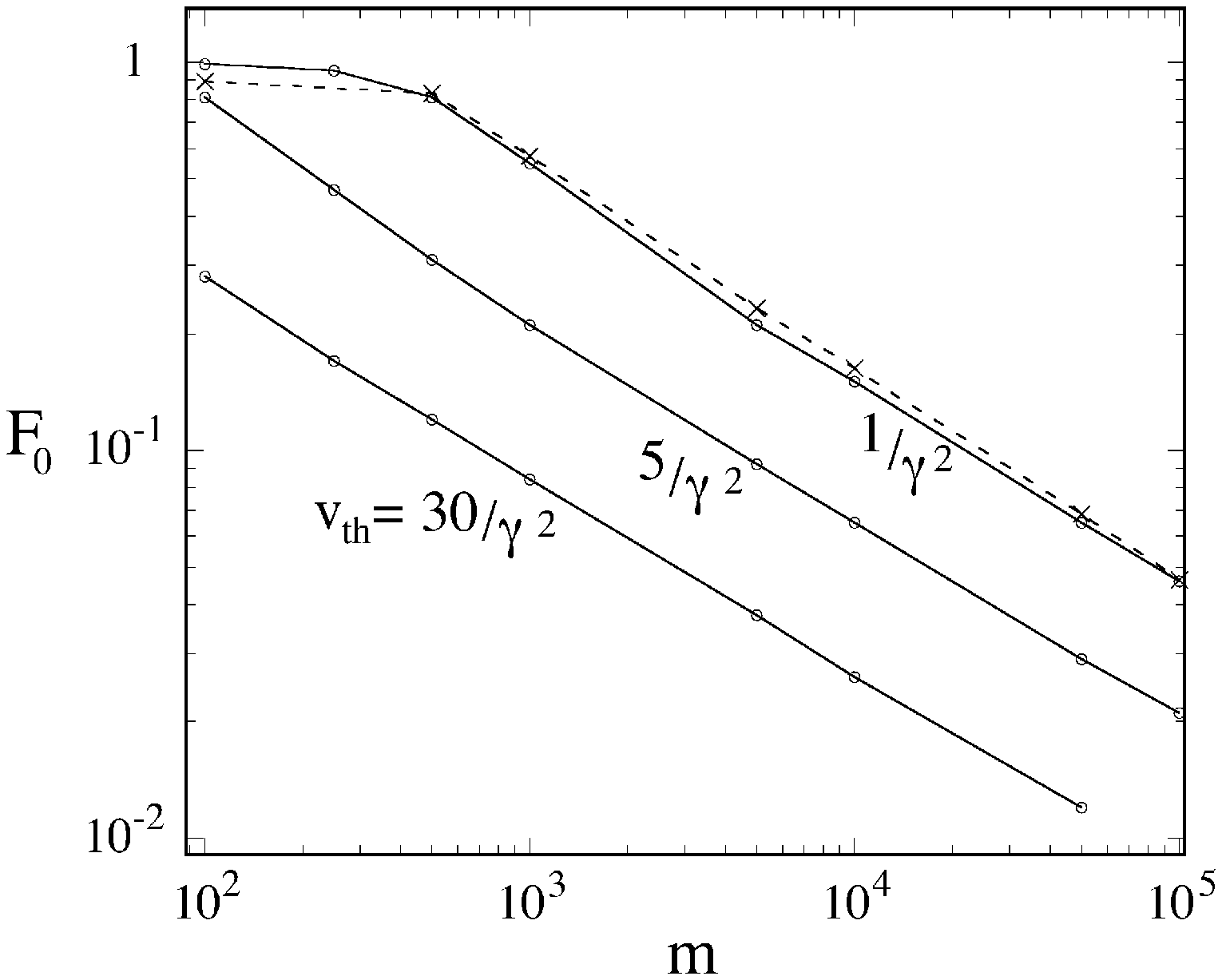}
\caption{$F_0$ as a function of $m$ for 
various values of $v_{th}$ and the
squared ratio of the growth rates
from Fig.~\ref{growthnobeta}
and Fig.~\ref{thinbetatron} (dashed line)}
\label{ratioF0}
\end{figure}

In order to determine the lowest
allowed value for $k_r$ and the highest possible growth rate
the full eigenvalue problem has to be solved.
We estimate the result by evaluating Eq.~(\ref{dpot1})
in the thin approximation again.
Looking at Eq.~(\ref{dpot1}) and replacing the Bessel functions
by their Airy function approximations for the case
$m \gg m_1$ and $m \ll m_2$ we see that the thin approximation
is justified if $\bar k_r v_{th} \ll 1$ and  $m^{2/3} v_{th} \ll 1$.
It starts to fail completely if $m^{2/3} v_{th} \gtrsim 1$, i.e. once
we start integrating over the oscillating and/or the exponentially 
damped/increasing
part of the Airy function, which implies we would like to have
$m^{2/3} | \Delta \tilde \omega | \ll \sqrt{F_0}$ with $|\Delta 
\tilde \omega|^2 \ll F_0 v_{th}^2$
from the previous paragraph. However, for real values of $k_r$ we expect
that the thin approximation will still give us an upper bound of the 
growth rate
because it is easier to maintain coherence if all the radiation is emitted
from the same orbit. With Eq.~(\ref{drhotthick}) we obtain in the limit $\Delta \tilde \omega \ll \gamma^{-2}$
\begin{eqnarray}
1= - \pi~ \zeta~ Z~ \frac{F_0 \gamma^{-2}}
{(\Delta \tilde{\omega} )^2}~,
\label{dispF0}
\end{eqnarray}
The growth rates can be found as before. For $m \gg m_1$ we
obtain
\begin{equation}
\omega_i \simeq {1.083\zeta^{1/2}m^{2/3}\dot\phi\over\gamma} \sqrt{F_0}
\label{analyticm1}
\end{equation}
and
\begin{eqnarray}
\nonumber
\omega_i \simeq{\zeta^{1/2}m^{1/2}\dot\phi\over\gamma^{1/2}} \sqrt{F_0}
\end{eqnarray}
for $m \gg m_2$, i.e. there is an additional factor of $\sqrt{F_0}$.
The results for our reference case are plotted in Fig.~\ref{thinbetatron} which
were computed numerically. In Fig.~\ref{ratioF0} the function $F_0$ is plotted
which we compare with the squared ratio of our new growth rates to the ones
evaluated previously without betatron oscillations.
%

We could also study the effect
of the non-zero thickness alone without
betatron oscillations setting
$F_0=1$ and $F_{n \neq 0}=0$ and solving the
full eigenvalue problem.
Due to the complicated nature of the
dispersion relation we have
not done this yet.
Note that the thin approximation
will suppress certain modes,
e.g. the negative mass instability cannot be
expected to be present with the
   fields having been evaluated at one radius
only, cf. \cite{Briggs1966}.

\section{Spectrum of Coherent Radiation} \label{CSR:spectrum}

\begin{figure}
\includegraphics[width=\columnwidth]{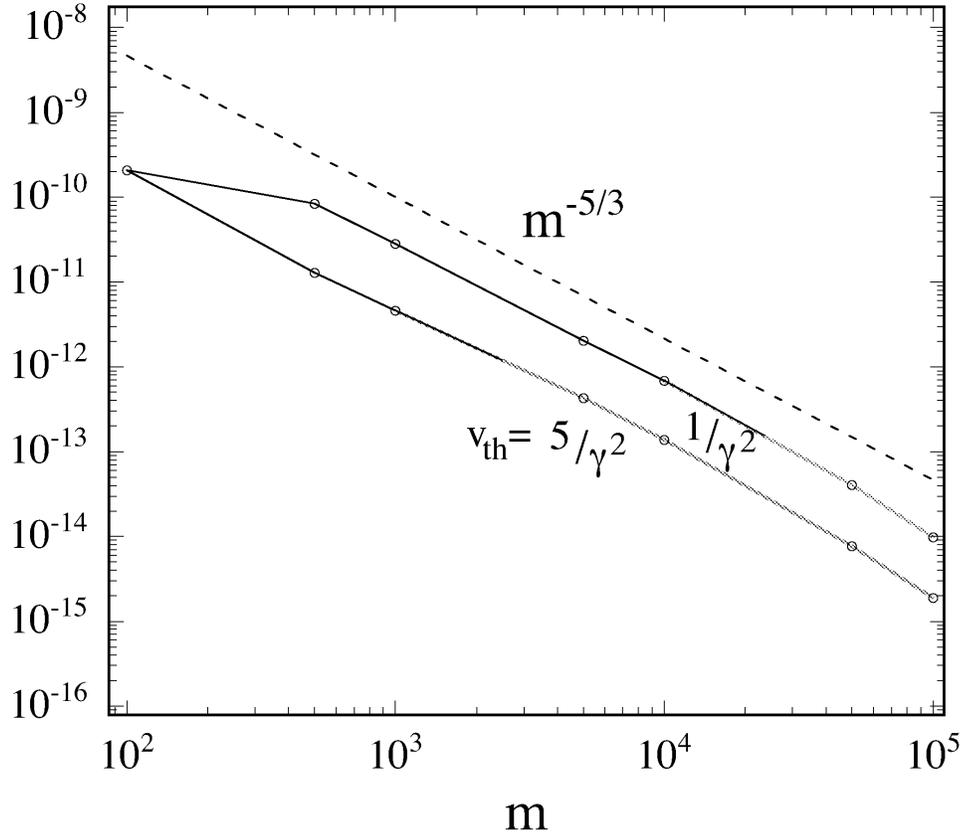}
\caption{Radiated power $(m^{-3} (\omega_i / \dot \phi)^4)$
for $\gamma = 30$ and $\zeta = 0.02$
in arbitrary units. The straight line is proportional
to $m^{-5/3}$ and is shown for comparison.
Points which do not satisfy the inequalities
$m \gg 1$, $m^{2/3} v_{th} < 1$ and
$|\Delta \tilde \omega|^2 < F_0 v_{th}^2$
are plotted in gray. }
\label{radpower}
\end{figure}

Having computed the growth rate
and the saturation amplitude,
the radiated power can now be calculated.
Starting from Eq.~(\ref{dIdOmega1}) we now have
\begin{eqnarray}
P_m = {\pi \over 2}  L \omega r_0^4~
\big| \delta J_{\phi 0}\big|^2
\left | \int
\xi d \xi e^{i \bar k_r \xi}
J_m^{\prime}(\omega r_0 \xi) \right |^2~,
\end{eqnarray}
where $\xi \equiv r/r_0$ and the integration is
over the thickness of the layer.
        The Bessel function can be expressed
approximately in term
of an Airy function as done before.
       We take the linear approximation to the
Airy function
as discussed previously, and this gives
\begin{eqnarray}
P_m =  \frac{\pi L \omega}{2} r_0^4 c_2^2
\left ( \frac{2}{m} \right )^{4/3}
\big| \delta J_{\phi 0} \big|^2
\left | \int_{1-v_{th}}^{1+v_{th}}
\xi d \xi e^{i \bar k_r \xi} \right |^2~,
\label{powintegral}
\end{eqnarray}
where $c_2\approx 0.259$.
This is valid for sufficiently big values of
$\gamma$ and low $m$.
The largest values occur for $\bar k_r v_{th} \ll 1$,
where this quantity
is simply $4 v_{th}^2$.
This is enough motivation
for us to work in this limit.
Thus,
\begin{eqnarray}
P_m \le 2\pi L \omega r_0^4 c_2^2 v_{th}^2
     \left | \delta J_{\phi 0}
\right |^2
\left ( \frac{2}{m} \right )^{4/3}~.
\end{eqnarray}
Because we calculated our
growth rates in the thin approximation
for $k \approx k_{\phi}$ it is consistent to use
$\delta \phi = 4 \pi^2 v_{th}
v_{\phi}^{-1} Z r_0 \delta J_{\phi 0}$.
Furthermore, we set $\omega \rightarrow
m \dot \phi$. This is consistent
even for large growth rates since
the exponential growth has stopped.
With our expression for the
saturation amplitude we obtain
\begin{equation}
P_m \le \frac{Lc_2^2v_{\phi}^2 m_e^2}{8 \pi^3 r_0 e^2}
\frac{\gamma^6}{|Z|^2}\left ( \frac{2}{m} \right )^{4/3}
    {1 \over m^3}
\left ( \frac{\omega_i(m)}{\dot \phi} \right )^4~.
\end{equation}
Since the number of particles
$N$ is proportional to $\zeta$ and the growth
rates are proportional
to $\sqrt{\zeta}$ for $m > m_1$
the radiated power scales like $N^2$.
This suggests that the
emitted radiation is coherent.
In Fig.~\ref{radpower} we plotted
the radiated power in arbitrary
units having evaluated $F_0$ numerically. For large
$m$ the curve scales as $m^{-5/3}$.
Analytically we obtain with our second approximation for $F_0$
the scaling $m^{-3}  (m^{2/3} / m^{1/4}  )^4=m^{-4/3}$.
With $|Z|^2 \approx 4 c_1^4 \left ( {2}/{m} \right )^{4/3}$
we obtain
\begin{eqnarray} 
P_m \lesssim 3.71 \times 10^{14}
     \gamma^6 m^{-3} \frac{L}{r_0}
\left ( \frac{\omega_i}{\dot \phi} \right )^4
\frac{\rm erg}{\rm s}~.
\label{power}
\end{eqnarray}

\section{Brightness Temperatures} \label{CSR:brightnesstemp}

          We consider the brightness temperatures
$T_B$ for conditions relevant to the
radio emissions of pulsars.
           Using the Rayleigh-Jeans formula
$B_\nu = 2 k_B T_B (\nu/c)^2$ for
the radiated power per unit area
per sterradian at a frequency $\nu= m \dot{\phi}/2\pi$
gives
\begin{eqnarray}
\nonumber
2k_B T_B ({\nu / c})^2
{\cal A} \Delta \Omega = 2\pi P_m/\dot{\phi}
\\
T_B \lesssim 4.5 \times 10^{21} \frac{\rm K}{\rm m} \cdot
L \gamma^6 m^{-4} \left ( \frac{\omega_i}{\dot \phi} \right )^4
\end{eqnarray}
where $k_B$ is Boltzmann's constant and
${\cal A}=2\pi r_0L$ is the area of the
E-layer.
   The solid angle of the source seen by
a distant observer has been computed in appendix A
and its value is $\Delta \Omega = {4 \pi^2 r_0}/({mL})$.
It is assumed that the angular size of the source is
small such that that radiation from the top and the bottom
emitted at an angle $\theta$ with respect to the normal
is received by the observer at the same position.
For the sample values $\gamma = 1000$,
$\zeta = 0.08$, $v_{th} = 0.04 \gamma^{-2}$,
$L = 100$ km  and  $m=m_1$ our model
predicts a maximum brightness temperature
of $T_B \approx  2 \times 10^{20} {\rm K}$.
According to our results from
previous sections there may be
degeneracy from modes with non-zero
axial wavenumbers $k_z$.
It is reasonable to assume that this will
increase the brightness
temperature by a factor in the order of
$m \tan \psi_{cr}$.
Beaming along the z-axis
may increase the brightness temperature and the observed
frequency even further.

\section{Applications in Accelerator Physics}

The next-generation linear
collider requires a beam
with very short bunches
and low emittance.
     That is,  the beam must
occupy a very small volume in phase space.
     The emittance
of the pre-accelerated
beam is reduced in a damping ring
which is operated with
longer bunches to avoid certain
instabilities.
     The bunch length has to be decreased in a
so-called bunch compressor
before the beam can be injected into
the linear collider.
      A bunch compressor consists of an
accelerating part and
an arc section.
      Since the bunch
lengths of the proposed
linear colliders are in the order of
the wavelength of the synchrotron radiation which is
being radiated in the arc section, instabilities due to
coherent synchrotron have to be taken seriously.
For a design energy of $2$ GeV
and $7 \times 10^{11}$ electrons per 100 $\mu$m our
dimensionless quantities become $\gamma=4000$ and
$\zeta=0.08$ \cite{Raubenheimer}. Our qualitative analysis
of the betatron motion suggests that CSR is suppressed
for a minimum energy spread of
$v_{th} > \zeta^{-1} \gamma^{-2} = 12.5 \gamma^{-2}$.

\section{Discussion and Conclusions}

        This work has studied the
stability of a
collisionless, relativistic,
finite-strength, cylindrical
electron (or positron) layer
by solving the Vlasov and
Maxwell equations.
         This system is of interest to
understanding the high brightness
temperature coherent synchrotron
radio emission of pulsars
and the coherent synchrotron radiation
observed in particle accelerators.
        The considered equilibrium
layers have a finite `temperature'
and therefore a finite radial
thickness.
        The electrons are considered
to move either almost perpendicular
to a uniform external magnetic
field or almost parallel to an
external toroidal magnetic field.
         A short wavelength
instability is found which
causes an exponential growth
an initial perturbation of
the charge and current densities.
         The periodicity of these
enhancements can lead to coherent
emission of synchrotron radiation.
       Neglecting betatron oscillations
we obtain an expression for
the growth rate which is similar
to the one found by Goldreich and Keeley
\cite{GoldreichKeeley1971}
if the thermal energy spread
is sufficiently small.
       The growth rate increases monotonically
approximately as $m^{1/2}$, where $m$ is the
azimuthal mode number which is proportional
to the frequency of the radiation.
       With the radial betatron oscillations included,
the growth rate varies as $m^{1/3}$ over
a significant
range before it begins to decrease.

        We argue that the growth
of the unstable perturbation saturates
when the trapping frequency of
electrons in the wave becomes
comparable to the growth rate.
        Owing to this saturation we can
predict the radiation spectrum
for a given set of parameters.
        For the realistic case including
radial betatron oscillations we
find a radiation spectrum
proportional to $m^{-5/3}$.
       This result
is in rough agreement with observations
of radio pulsars \cite{ManchesterTaylor1977,Thompson1996} (Fig.~\ref{fig:spectrum}).
The power is also
proportional to the square of
the number of particles which
indicates that the radiation is
coherent.
    Numerical simulations of electron rings
based on the fully relativistic, electromagnetic
particle-in-cell code OOPIC \cite{Schmekel:2004su}
recovers  the main scalings
found here.

\chapter{Particle in Cell Simulations$^\dagger$} \label{ch:PIC}
\symbolfootnotetext[2]{This chapter will appear as a journal article \cite{Schmekel:2004su}.
Reprinted in modified form with kind permission from the American Physical Society. 
\copyright ~ 2005 by the American Physical Society}

\section{Introduction}

Attempts to extract the nonlinear evolution of a plasma are usually unsuccessful
except in some very special cases and numerical methods have to be applied.
Seeking a straightforward numerical solution of the Vlasov equation however
is prohibitive except in lower dimensional models \cite{Venturini:2005wu} because the dimensionality of
the problem is doubled in a framework which makes use of phase space.
A possible solution to this problem is MHD which condenses the full momentum space distribution
to only a few macroscopic quantities like density and current. Numerical MHD is extremely popular 
and a vast amount of literature exists on this topic. Despite its popularity MHD has
some shortcomings. First, the results are only as good as the used closing condition.
Second, some important effects like Landau damping rely on the knowledge of the momentum
distribution. Landau damping \cite{chao1993} is a stabilization mechanism which is crucial
for the generation of stable beams in particle accelerators. Therefore, MHD is of limited
use in particle accelerator physics. Particle tracking programs avoid both problems.
The grid can be set up in position space and position and momentum can be stored for each particle. On contemporary computers this is efficient even for a large number of particles. 
Not working in the limit $N \longrightarrow \infty$ anymore the effect of changing the
number of particles being tracked needs to be investigated. Even though the number of particles
in a plasma is finite it is usually not possible to track that many and one resorts
to tracking $N_{mac}$ ``macroparticles'' with each macroparticle representing $N/N_{mac}$ particles. 
One can only hope that the results obtained with $N_{mac}$ macroparticles are sufficiently close
to convergence, i.e. $N \longrightarrow \infty$, giving a reasonably good estimate of the
behavior of the real system.

The objective in this chapter is to simulate the evolution of a particle distribution
which resembles configuration a in chapter \ref{ch:CSR}.

\section{Particle-in-Cell Simulations with OOPIC}

For the simulation the software package OOPIC \cite{verboncoeur} was used.
OOPIC is a relativistic two-dimensional particle-in-cell
code which supports both plain $(x,y)$-geometries and cylindrical
$(r,z)$-geometries.
Since the interesting dynamics takes places in the azimuthal
direction one can only simulate a thin ring (instead of a cylinder)
in the $(x,y)$-mode. Loading the initial circular particle distribution
in the $(x,y)$-mode required modifying the source code (files load.cpp, diagn.cpp
and c\_utils.c) to allow the program to handle circular particle distributions.
Some minor modifications were necessary in order to compile XOOPIC-2.5.1 with 
gcc 3.2.2 and the compiler compiler bison 1.28 under SunOS 5.9. 
The built-in function parser was extended to support elliptic integrals.

Since a thin ring of particles is simulated instead of a cylinder thereof all fields and charges
were divided by the length $L$ of the cylinder whereas the electron
mass needs to be divided by $L^2$. $L=10 r_0$ is chosen unless noted otherwise.
The electric and magnetic self-fields for a thin ring
equilibrium differ from what was used in the model.
The fields can be found in \cite{jackson}. It is ensured that OOPIC
uses these self-fields before the perturbation starts to build up.
As it turns out choosing the correct self-fields is not too crucial.
Leaving them out the system will build them up itself. Once the 
self-fields are created the system shows no difference in behavior.
The absence of the self-fields in the dispersion relation
might help to understand this feature.
As in \cite{Schmekel:2004jb} a Gaussian number density profile with RMS
width $v_{th}$ was chosen for the initial distribution. 
$5000$ macro particles were tracked on a grid
with resolution $512 \times 512$ unless noted otherwise.
Once an energy for a particle has been chosen it is placed at the equilibrium
radius $r_0 = m \gamma c (eB)^{-1}$, i.e. neglecting betatron oscillations
particles on the same orbit have the same energy. This fixes the azimuthal 
component of the canonical angular momentum. The system can
pick up transverse motion quickly.
The grid represents a rectangular region $40 {\rm m} \times 40 {\rm m}$ big where 
the ring with radius $r_0=10 {\rm m}$ is centered.

\begin{figure}
\includegraphics[width=\columnwidth]{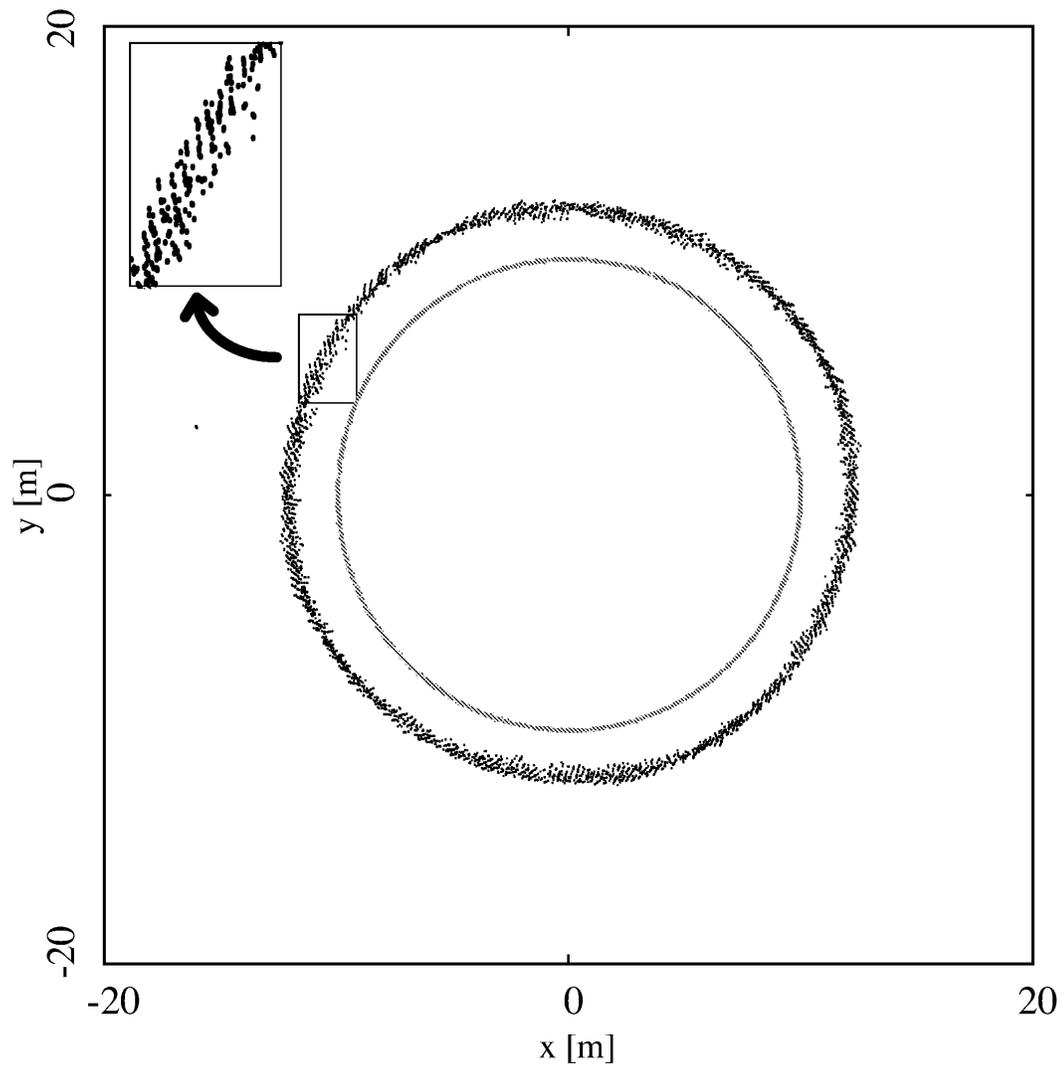}
\caption{Initial particle distribution (gray)
and the same distribution after 23ns have elapsed.
Parameters: $\zeta=0.010$, $\gamma=30$ and $v_{th}=0.002$}
\label{ely71_p23e8}
\end{figure}

\begin{figure}
\scalebox{0.35}[0.525]{\rotatebox{270}{\includegraphics[width=\columnwidth]{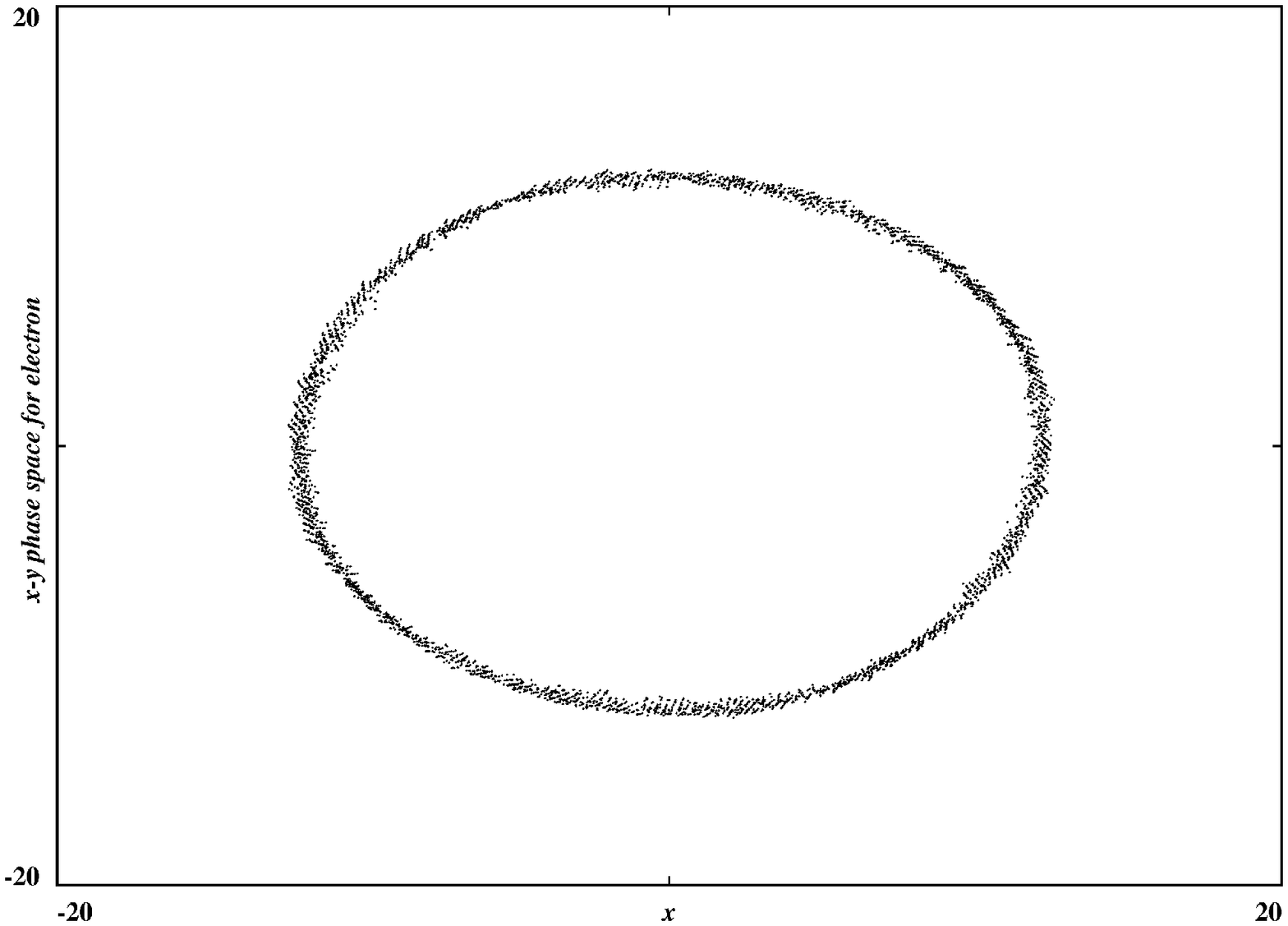}}}
\scalebox{0.35}[0.525]{\rotatebox{270}{\includegraphics[width=\columnwidth]{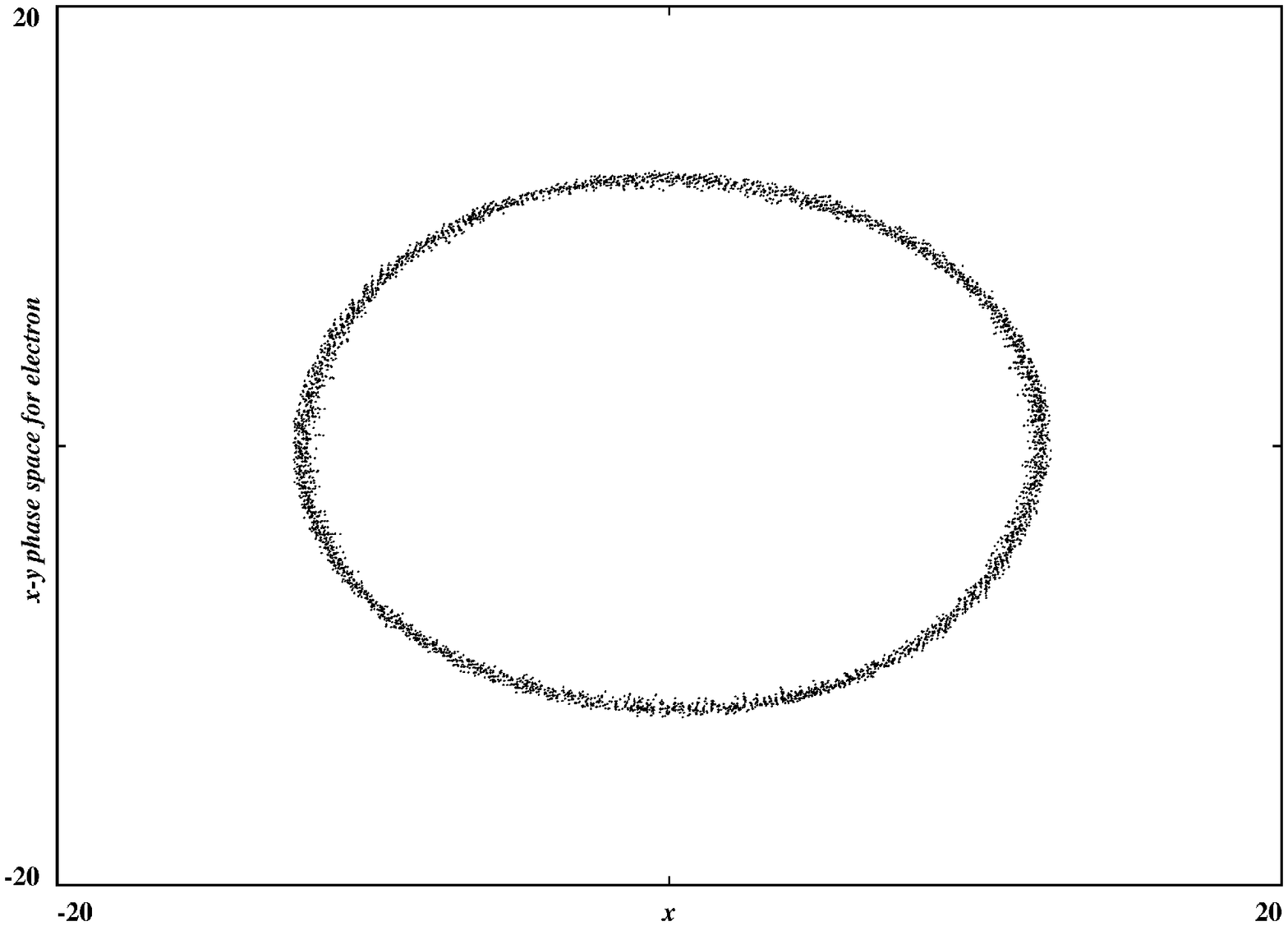}}}
\scalebox{0.35}[0.525]{\rotatebox{270}{\includegraphics[width=\columnwidth]{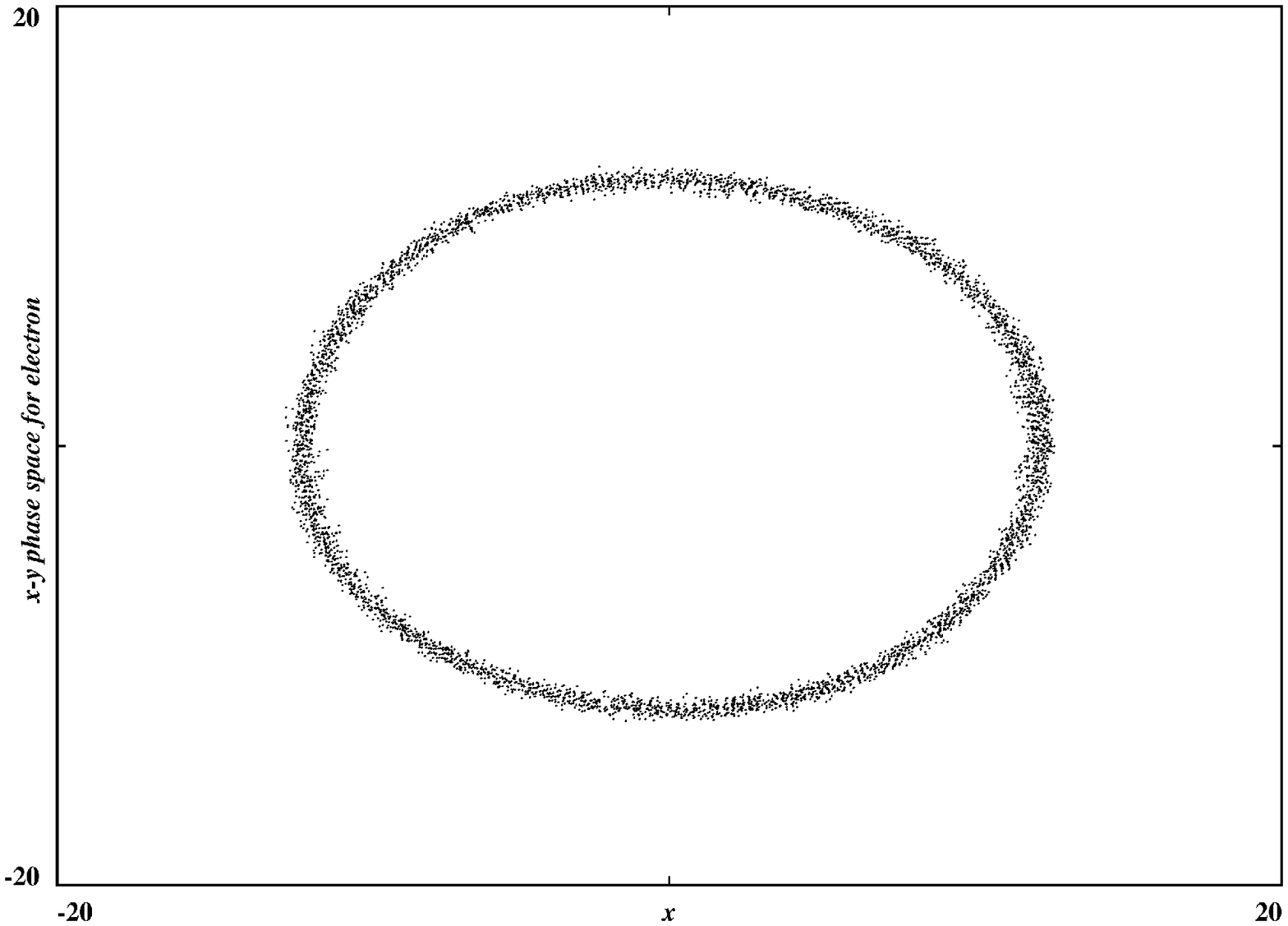}}}
\scalebox{0.35}[0.525]{\rotatebox{270}{\includegraphics[width=\columnwidth]{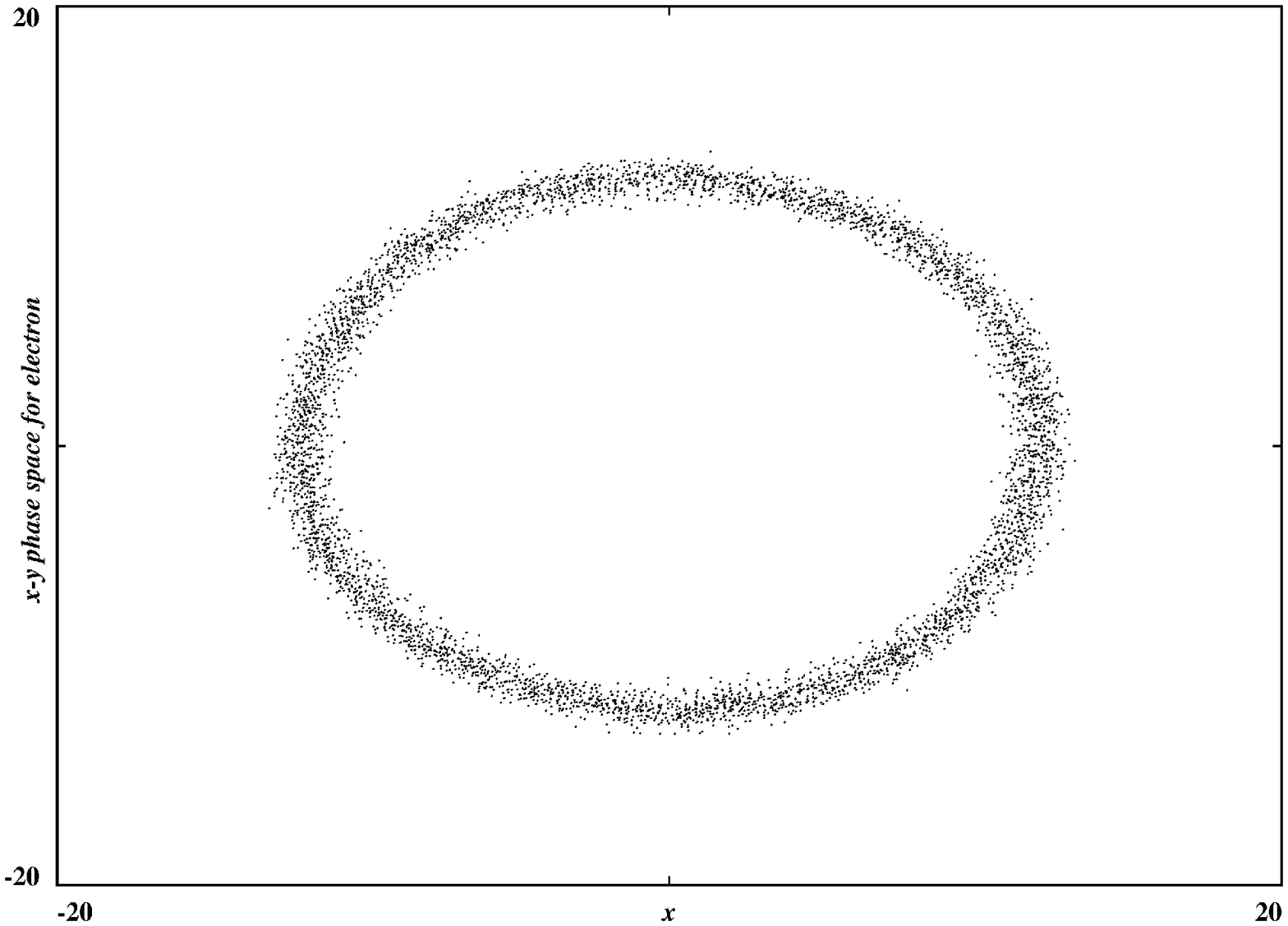}}}
\caption{Particle distribution ($\gamma=30$, $\zeta=0.01$) after 23ns
for $v_{th}=0.002$, $v_{th}=0.008$, $v_{th}=0.015$
and $v_{th}=0.033$ (from left to right, top to bottom).
All lengths are measured in units of meters. } 
\label{bunchvth}
\end{figure}

In Fig.~\ref{ely71_p23e8} the initial particle distribution (gray) and the particle distribution 
after $23ns$ are shown. The parameters are $\zeta=0.010$, $\gamma=30$, and $v_{th}=0.002$.
Qualitatively, a bunching of the particle distribution can be observed.
An enlargement of a small section of Fig.~\ref{ely71_p23e8} is also shown in the same figure.
In Fig.~\ref{bunchvth}, the bunching is shown for successively higher energy spreads.
With increasing energy spread the bunches become fuzzier and the clean gaps between bunches that
can be observed for small energy spreads are populated with ``stray particles''. This suggests
that it may be harder to achieve complete coherence for larger values of $v_{th}$. The decoherence
due to the non-zero width of the particle beam is investigated quantitatively later in the paper. 
These qualitative features are independent of $\gamma$.

\begin{figure}
\includegraphics[width=\columnwidth]{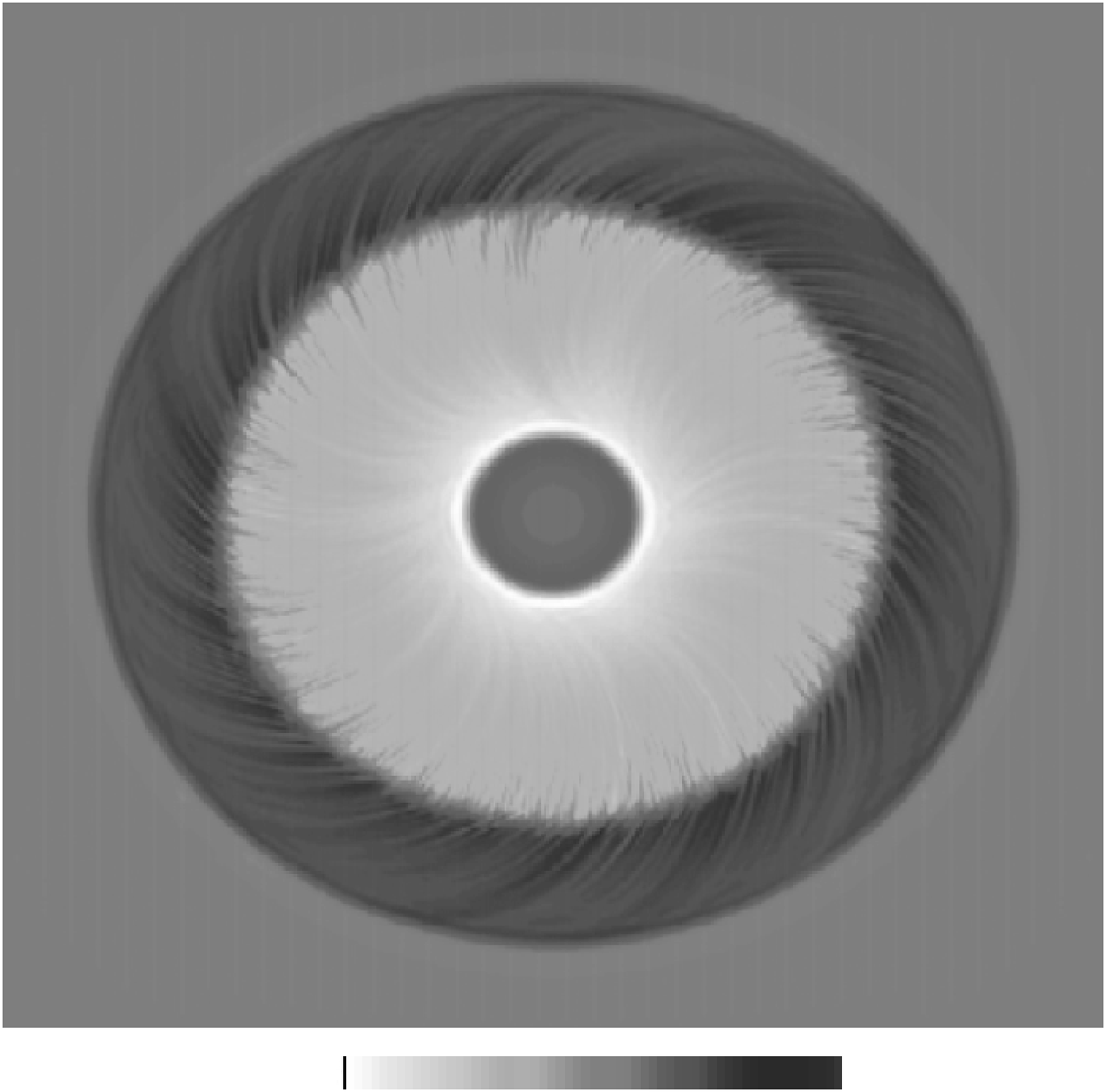}
\setlength{\unitlength}{1mm}
\flushleft\begin{picture}(0,0)(0,0)
\put(30,3){$-66.4 \mu T m^{-1}$}
\put(100,3){$158.0 \mu T m^{-1}$}
\end{picture} 
\caption{$z$-component of the magnetic field (self-field
plus perturbation without external magnetic field) after
23ns for $\zeta=0.010$, $\gamma=30$ and $v_{th}=0.025$.
The size of the area depicted is 40m $\times$ 40m.}
\label{ely84_c23e8}
\end{figure}

Also note that during the evolution of the circular charge distribution both the radius and the width
of the ring increase slightly. The former is due to a.) particles losing energy
and b.) the perturbed magnetic field changing significantly. It tends to decrease for
small energy spreads and increase for larger energy spreads.
Starting with a larger radius the radius increases even further, i.e. this is not a relaxation
from a ``false'' to a true equilibrium.
Since the non-zero mesh size imposes an upper limit on the azimuthal mode number $m$ which
can become unstable, it is expected that the distance between bunches
decreases as the resolution increases. This is indeed the case.
For larger energy spreads the bunching of the distribution becomes hardly visible, but it 
still can be observed in the $z$-component of the magnetic field (Fig.~\ref{ely84_c23e8}).

The bunches are slightly tilted and may be connected by
a very thin inner ring of particles for sufficiently high beam currents. For these
reasons it is not possible to Fourier transform the charge perturbations in order to
compute the growth rates for each value of $m$. Since the resolutions used were
low the range of $m$ values is restricted. Therefore, only the radiated power is computed which
can be obtained easily.
%
%

\begin{figure}
\scalebox{0.35}[0.525]{\rotatebox{270}{\includegraphics[width=\columnwidth]{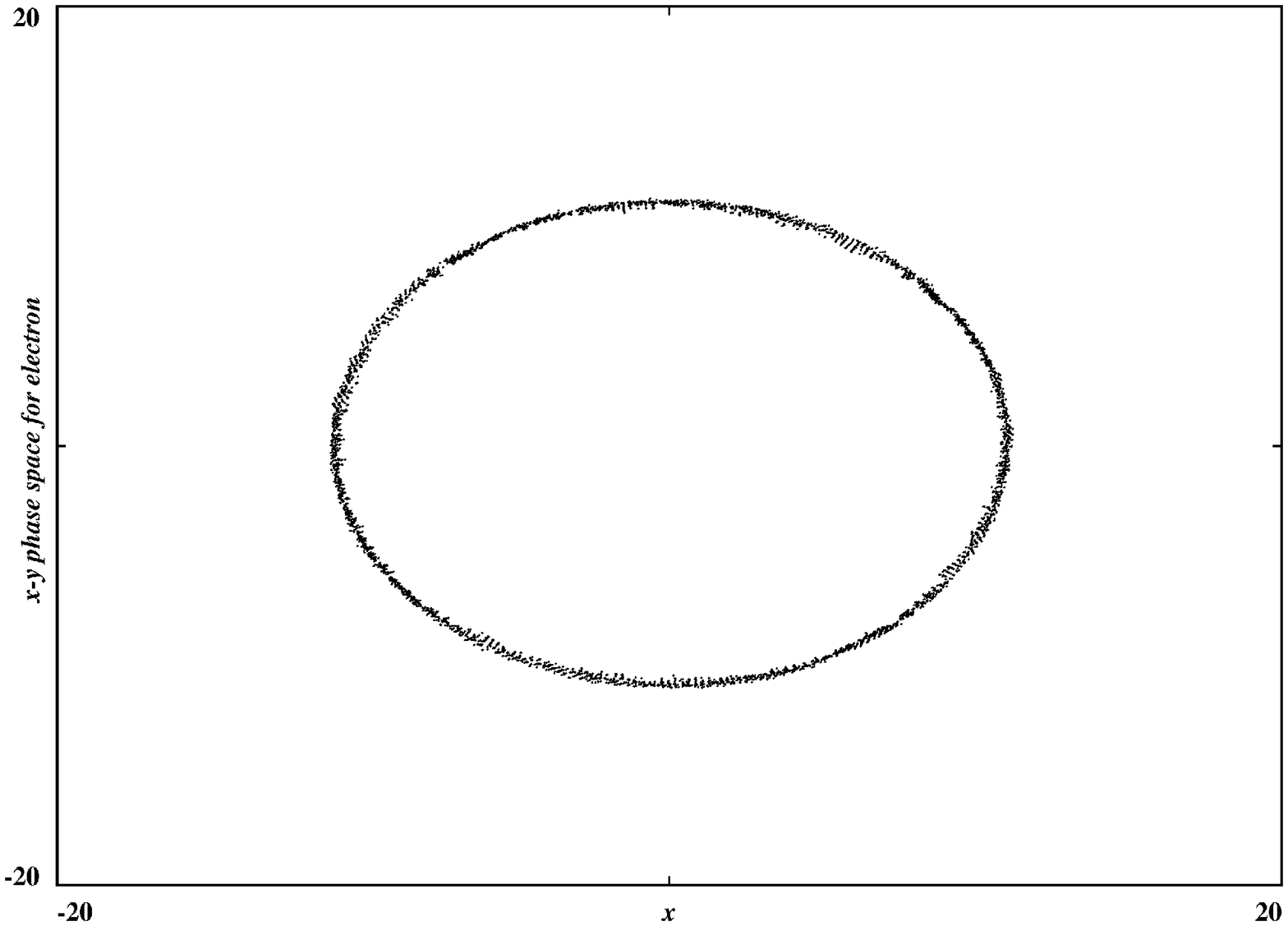}}}
\scalebox{0.35}[0.525]{\rotatebox{270}{\includegraphics[width=\columnwidth]{ly71}}}
\scalebox{0.35}[0.525]{\rotatebox{270}{\includegraphics[width=\columnwidth]{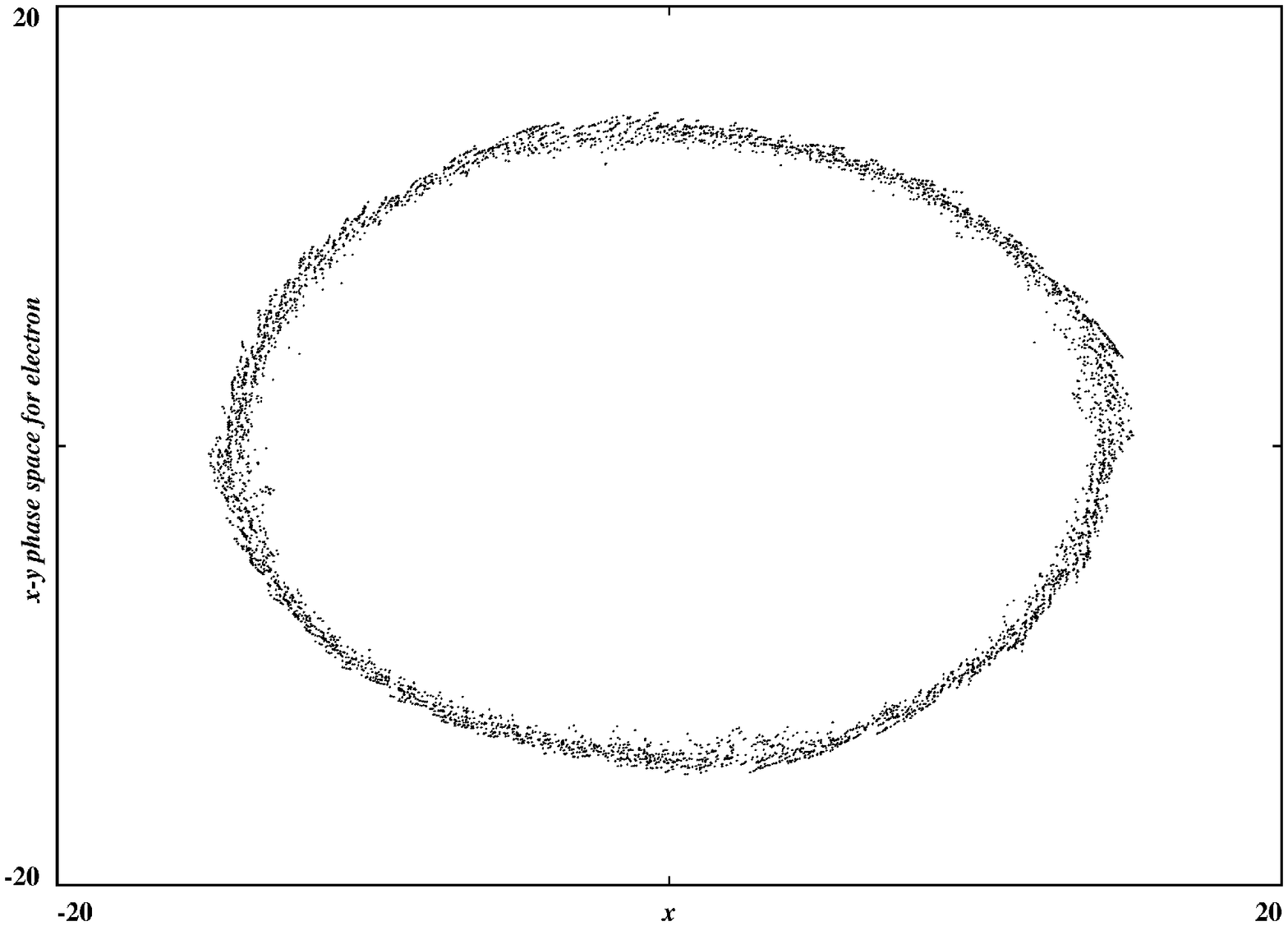}}}
\scalebox{0.35}[0.525]{\rotatebox{270}{\includegraphics[width=\columnwidth]{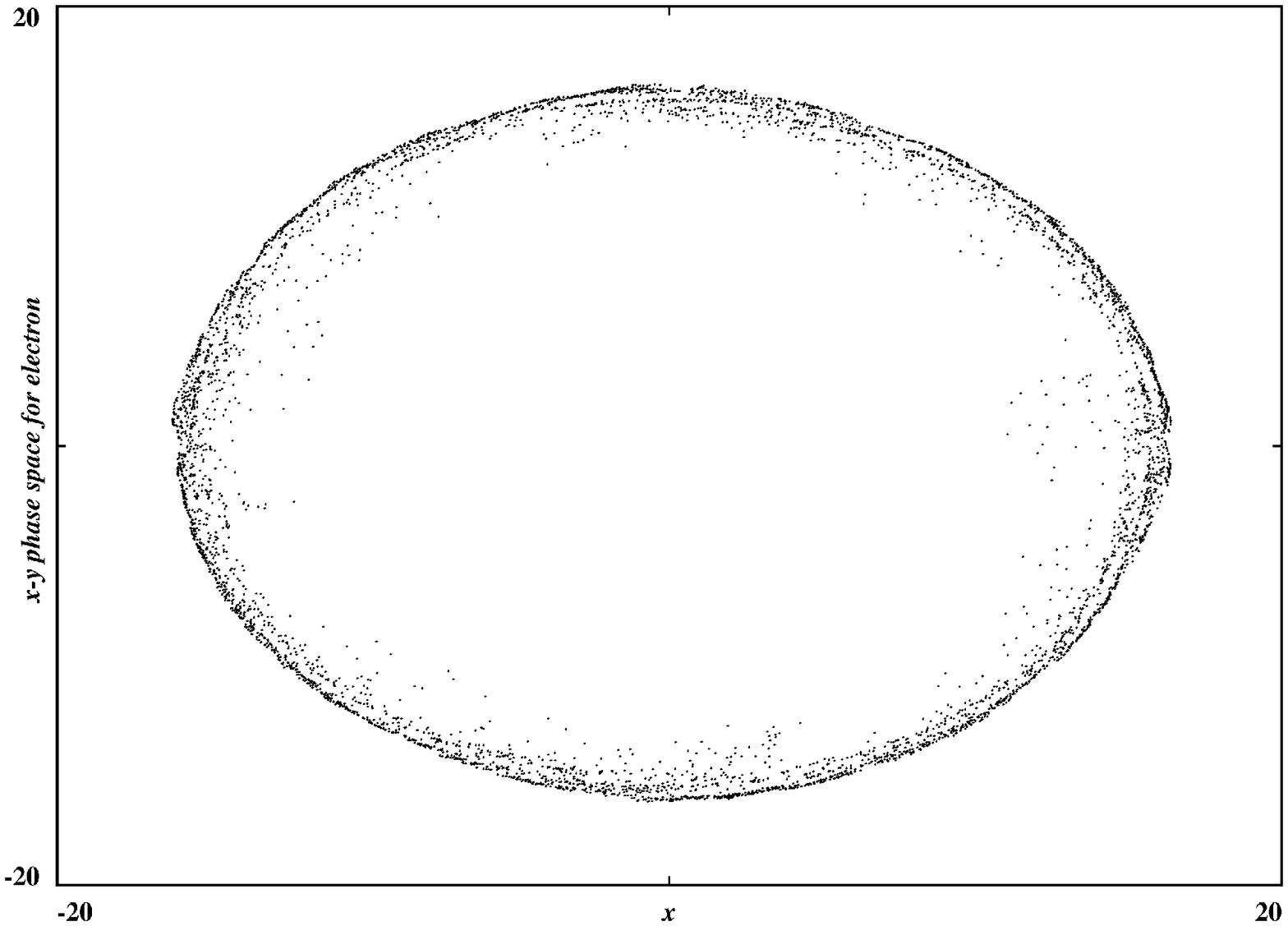}}}
\caption{Particle distribution ($\gamma=30$, $v_{th}=0.002$) after 23ns
for $\zeta=0.005$, $\zeta=0.010$, $\zeta=0.020$
and $\zeta=0.040$ (from left to right, top to bottom). 
All lengths are measured in units of meters. } 
\label{bunchvth2}
\end{figure}

\begin{figure}
\scalebox{0.35}[0.525]{\rotatebox{270}{\includegraphics[width=\columnwidth]{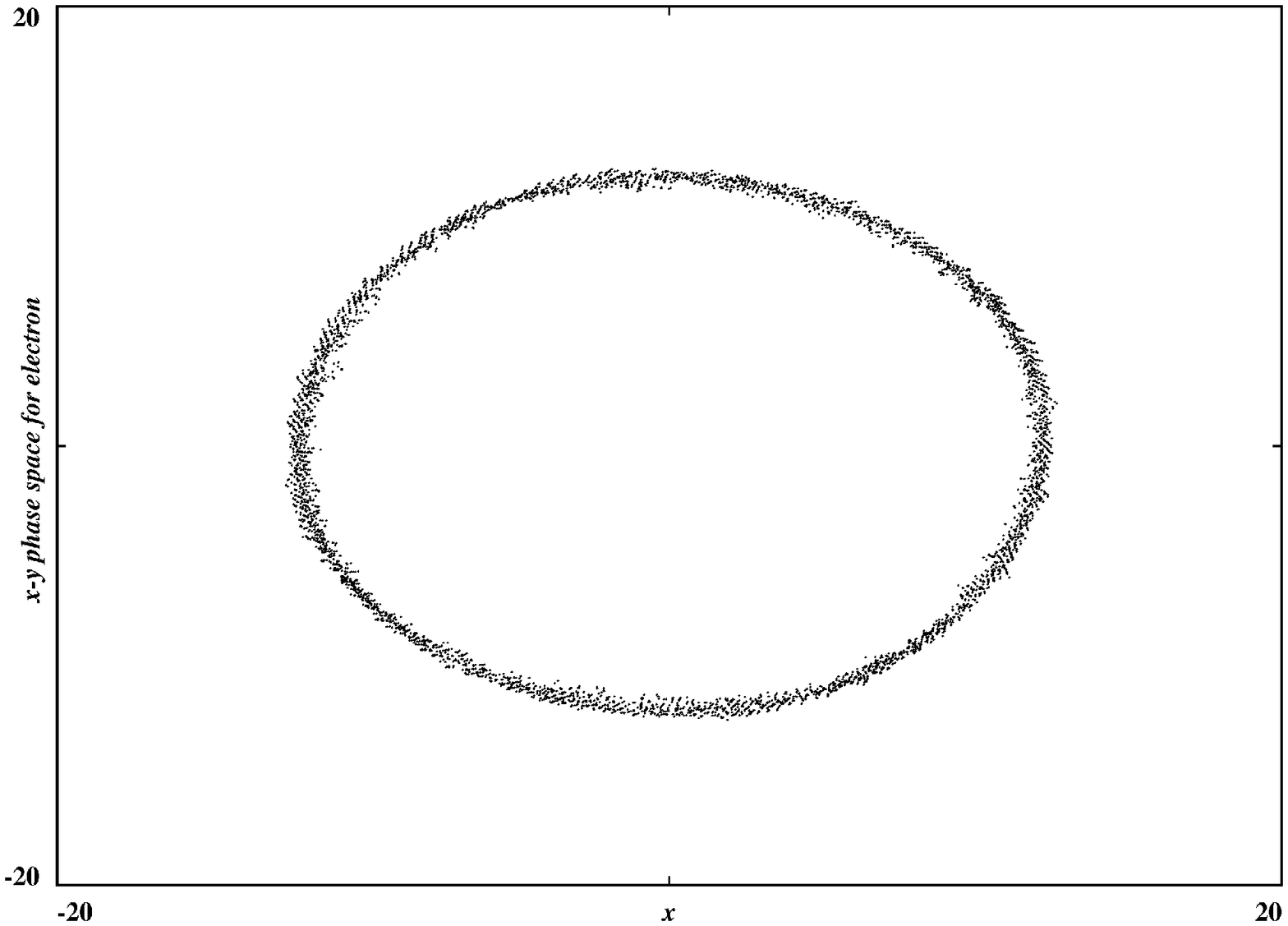}}}
\scalebox{0.35}[0.525]{\rotatebox{270}{\includegraphics[width=\columnwidth]{ly71}}}
\scalebox{0.35}[0.525]{\rotatebox{270}{\includegraphics[width=\columnwidth]{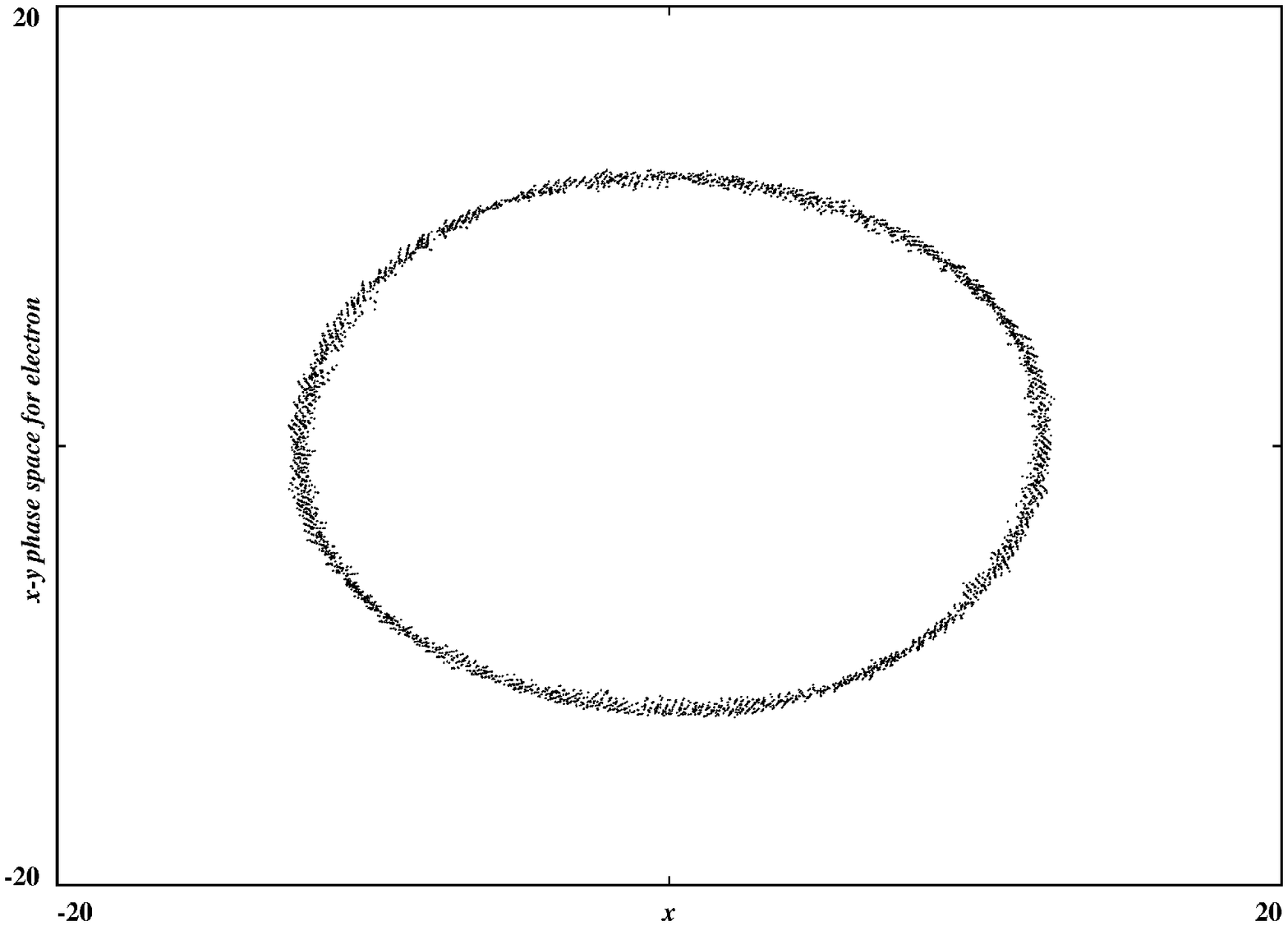}}}
\scalebox{0.35}[0.525]{\rotatebox{270}{\includegraphics[width=\columnwidth]{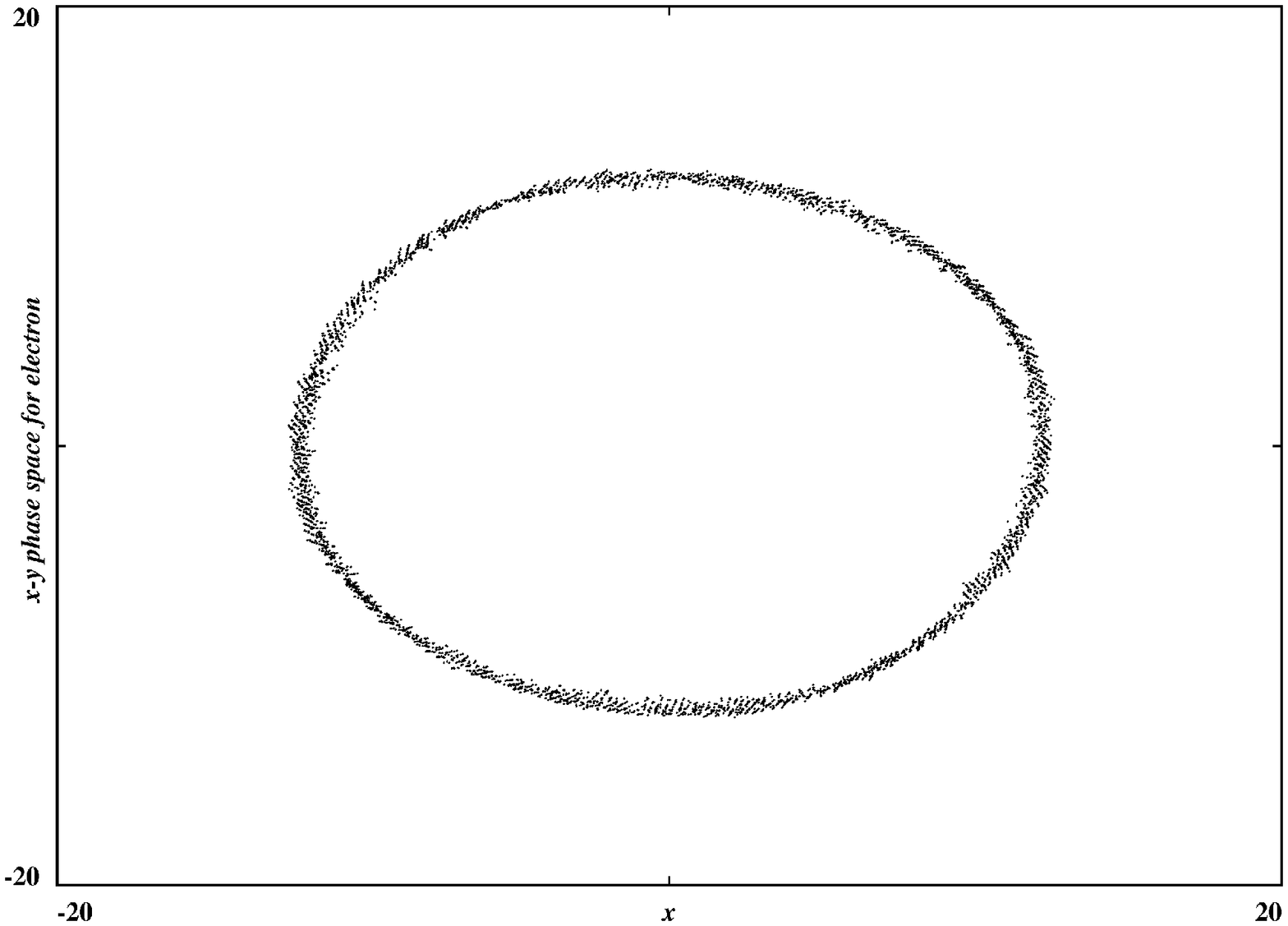}}}
\caption{Particle distribution ($\zeta=0.01$, $v_{th}=0.002$) after 23ns
for $\gamma=10$, $\gamma=30$, $\gamma=75$
and $\gamma=90$ (from left to right, top to bottom). 
All lengths are measured in units of meters. } 
\label{bunchvth3}
\end{figure}

%
%
Estimates of the growth rate are two orders of magnitude higher than what would be expected.
A possible explanation is that the ratio between the saturation amplitude and
the electric self field 
\begin{eqnarray}
\bigg | \frac{\delta E_{sat}}{E_{self}} \bigg | =
\frac{1}{m \zeta} \left({\Im(\omega)(m) \over \dot{\phi}}\right)^2
\end{eqnarray}
is typically in the order of $10^{-3}$ for the given sample cases
which is rather small. The initial perturbations due to discreteness,
numerical noise etc. are usually in the same order of magnitude.
Therefore, one cannot expect to see the regime covered by the linearized
Vlasov equation. This is another reason for focusing entirely on
the emitted power.

%
\begin{figure}
\includegraphics[width=\columnwidth]{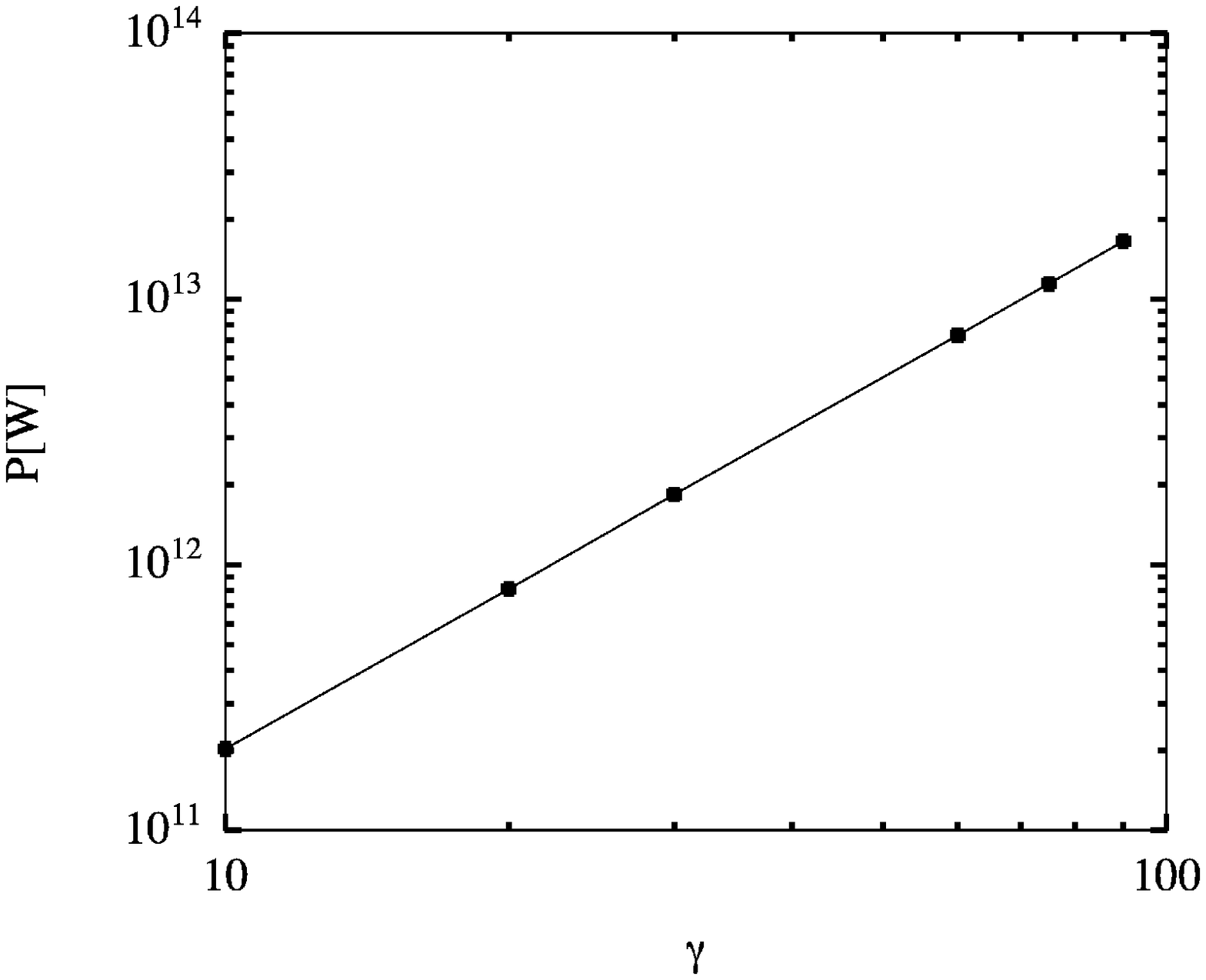}
\caption{Loss of kinetic energy in W vs.
$\gamma$ for $\zeta=0.01$ and $v_{th}=0.002$.}
\label{gamma}
\end{figure}
%
\begin{figure}
\includegraphics[width=\columnwidth]{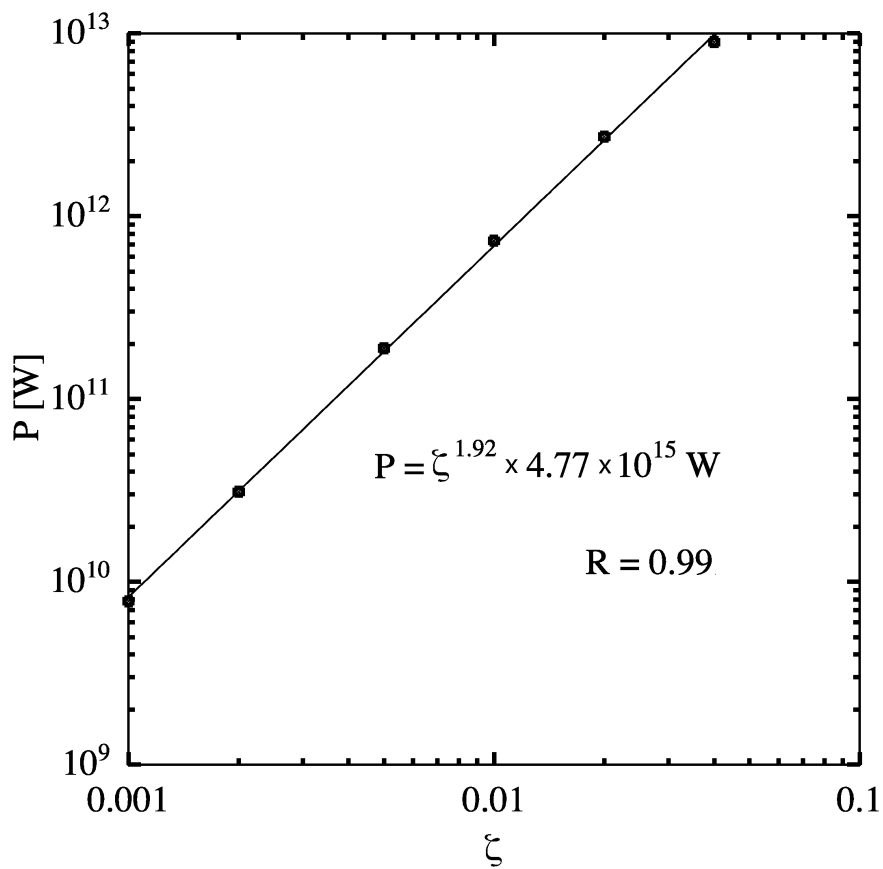}
\caption{Loss of kinetic energy in W vs.
$\zeta$ for $\gamma=30$ and $v_{th}=0.025$.
The solid line shows the best fit.}
\label{powerfit}
\end{figure}
%

\section{Radiated Power}

In Fig.~\ref{gamma} and Fig.~\ref{powerfit} the radiated power determined by measuring
the kinetic energy loss of the electron cloud after approximately 2.36 ns
is plotted as a function of $\gamma$ and $\zeta$, respectively. After 0.24 ns the
perturbations have saturated and the emitted power is fairly constant.
A quadratic dependence can be established, i.e. the first two relevant scalings 
expected from the analytical model are recovered.

The simulation has been repeated at lower (256 $\times$ 256) and higher (1024 $\times$ 1024) resolution.
No significant effect could be observed. This is consistent
with the model which predicts that most power is emitted by modes with 
low $m$. The resolution is not high enough to resolve the path length difference of orbits with different
radii at these low values of $v_{th}$. Also, decreasing the stepsize $dt$ to 2.5 ps and increasing the
number of macro particles to 50000 has a negligible effect.

\begin{figure}
\includegraphics[width=\columnwidth]{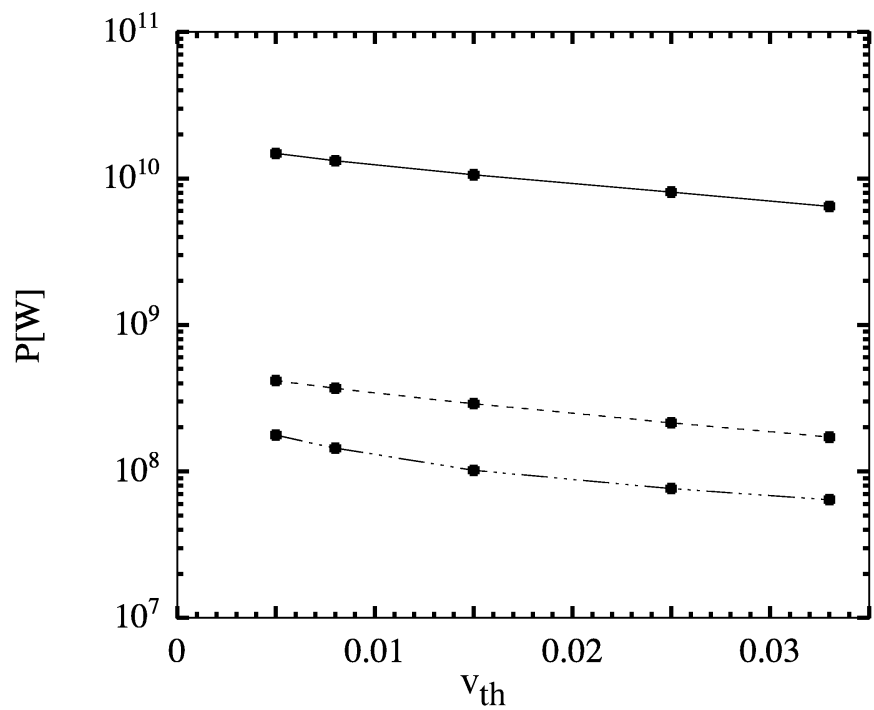}
\caption{Total power radiated as obtained from OOPIC for the parameters
$\zeta=0.01$, $\gamma=30$ (solid) and $\zeta = 0.005$, $\gamma = 10$ (dashed).
The dash-dotted line is proportional to $\sum_{m=1}^{\infty} P_m$.}
\label{compvth}
\end{figure}

Finally, the effect of the energy spread is investigated.
With increasing $v_{th}$ the power decreases
which is due to the decoherence described by the factor $F_0$
defined in Eq.~(\ref{Fnnum}). The results are plotted in
Fig.~\ref{compvth} for the parameters 
$\zeta=0.01$, $\gamma=30$ and $\zeta = 0.005$, $\gamma = 10$, respectively, and $L=r_0$.
In the former case $m_1$ is 27 and in the latter case it is 0.4.
Eq.~(\ref{power}) becomes 
\begin{eqnarray}
P \approx 3.71 \times 10^{14} \frac{\rm erg}{\rm s}  \frac{L}{r_0} \gamma^2 
\sum_m m \left [ i \pi J_m(\omega r_0) H_m(\omega r_0) \right ]^2
\end{eqnarray}
Despite $m_1$ being much larger than 1 for the first set of parameters the 
slopes in Fig.~\ref{compvth} match exactly only if the summation starts
at $m=1$.  This suggests that modes with $m < m_1$ do radiate and can
be described by the same dispersion relation. Since the power scales as $m^{-5/3}$   
these modes may actually be very important for computing the total energy loss.
Note that while the simulation suggests $P \propto L^2$ Eq.~(\ref{power}) (which was
derived under the assumption $L \gtrsim r_0$) gives
$P \propto L$. A 2D simulation cannot explain how the radiation from different
axial positions on the cylinder interacts. In the thin ring case doubling $L$
doubles the number of particles $N$ and therefore quadruples $P$.
Fortunately, as can be seen in the derivation of Eq.~(\ref{power}) in \cite{Schmekel:2004jb} 
the $\zeta$, $v_{th}$ and $\gamma$ dependent part of $P$ is independent of $L$.
In Fig.~\ref{compvth} the overall factor matches if $r_0 = 100L$ whereas the growth
rate for a perturbation of a cylinder and a thin ring coincide for $r_0 = L$ \cite{Schmekel:2004jb}.  
Also note that Fig.~\ref{ely71_p23e8} suggests $k_r v_{th} r_0 \sim 1$, whereas Eq.~(\ref{analyticm1}) was derived
under the assumption $k_r v_{th} r_0 \ll 1$.

%
%

\section{Conclusions}

The particle in cell code OOPIC was used to simulate the evolution
of density perturbations in a thin ring of charged particles which
move in relativistic almost circular motion in an external magnetic field.
The results were compared with the model in \cite{Schmekel:2004jb}.
Comparisons of the simulation with the model shows approximate agreement 
with the main predicted scaling relations.
In particular the bunching effect
could be observed very clearly and the emitted power is proportional to
the square on the number density which implies coherent radiation.
The dependence on the energy spread can be recovered exactly assuming
all modes contribute to the observed energy loss suggesting that
the model may apply even if $m < m_1$.  

\chapter{MHD Approach for a Brillouin Flow} \label{ch:MHD}

\newcommand{\Omegaa}{\dot \phi}

\section{Theory}

    We consider a laminar Brillouin
type  equilibrium of a long, non-neutral,
cylindrical relativistic
electron (or positron) layer in a uniform
external magnetic field ${\bf B}_e = B_e\hat{\bf z}$,
where we use a non-rotating cylindrical
$(r,\phi,z)$ coordinate system.
    The electron velocity is
${\bf v} = v_{\phi}(r)\hat{\rvecphi~}
=v(r)\hat{\rvecphi~}=
r\Omegaa(r)\hat{\rvecphi~}$
   The self-magnetic field is in the
$z-$direction while the self-electric
field is  in the $r-$direction.
   The radial force balance of the equilibrium
is
\begin{equation}
-\gamma \Omegaa^2 r = {q \over m_e}(E + v B)~,
\label{fbalance}
\end{equation}
where $\gamma = (1-v^2)^{-1/2}$ is the
Lorentz factor with velocities
measured in units of the speed of light, $B=B_e+B_s$
is the total (self plus external) axial magnetic
field, $E$ is the total ($=$ self) radial
electric field, and $q$ and $m_e$ are the
particle charge and rest mass.
    We have
\begin{equation}
{1\over r}{d (r E) \over dr} = 4 \pi \rho_e~,
\quad
{d B \over dr}= -4\pi \rho_e v~,
\end{equation}
where $\rho_e(r)$ is the charge density of
the electron layer.

    We consider weak layers in the sense that
the `field reversal' parameter
\begin{equation}
\zeta \equiv -~{4\pi \over B_e}
\int_{r_1}^{r_2} dr \rho_e v
\end{equation}
is small compared with unity, $\zeta^2 \ll 1$.
Under this condition Eq.~(\ref{fbalance})
gives $\Omegaa =-qB_e/(m_e \gamma)$.
Here, we have assumed that the layer exists between
$r_1$ and $r_2$.
     We also consider that the Lorentz factor
is appreciably larger than unity in the
sense that $\gamma^2 \gg 1$.

   We consider general electromagnetic
perturbations of the electron layer with
the perturbations proportional
to
\begin{equation}
f_\alpha (r) \exp(im\phi -i\omega t)~,
\end{equation}
where $\alpha =1,2,..$ for the different
scalar quantities, $m=$ integer, and $\omega$
the angular frequency of the perturbation.
   Thus the perturbations  give rise
to field components $\delta E_r$, $\delta E_\phi$,
and $\delta B_z$.
    The perturbed equation of motion is
$$
\left[{\partial \over \partial t}
+({\bf v}+\delta {\bf v})\cdot{\bf \nabla}\right]
\big(\gamma {\bf v}
+{\bf v}\delta \gamma +\gamma \delta {\bf v}\big)
$$
\begin{equation}
={q \over m_e} \big(\delta {\bf E}
+ {\bf v} \times \delta {\bf B}
+\delta {\bf v} \times {\bf B} \big)~,
\end{equation}
where the deltas indicate perturbation quantities.
     This equation can be simplified to give
$$
  ~\left [\begin {array}{cc}
{-i\gamma\Delta \omega }&
{-\gamma \Omegaa(1+\gamma^2) -{q \over m_e} B}
\\\noalign{\medskip}
{\gamma \Omegaa +(\gamma \Omegaa r)^\prime +{q\over m_e}B}&
{-i\gamma^3 \Delta \omega}
\end {array}\right ]
\left[\begin{array}{c} {\delta v_r}
\\ \noalign{\medskip}{\delta v_\phi} \end{array}\right]~
$$
\begin{eqnarray}
=~       {q\over m_e}  \left[\begin{array}{c}
{\delta E_r + v \delta B_z}
\\ \noalign{\medskip}{\delta E_\phi} \end{array}\right]~,
\label{MHDsys}
\end{eqnarray}
where the prime denotes a derivative with respect
to $r$, and
$$
\Delta \omega(r) \equiv \omega -m\Omegaa(r)
$$
is the Doppler shifted frequency seen by a
particle rotating at $\Omegaa$,

    Using the equilibrium equation (\ref{fbalance}) and
the condition $\zeta^2 \ll 1$, the matrix in
Eq.~(\ref{MHDsys}) is approximately
\begin{equation}
  {\cal D}= \left [\begin {array}{cc}
{-i\gamma\Delta \omega }&
{-\gamma^3 \Omegaa }
\\\noalign{\medskip}
{\gamma^3(\Omegaa r)^\prime }&
{-i\gamma^3 \Delta \omega}
\end {array}\right ]~,
\end{equation}
    We have used the fact that
$(\gamma\Omegaa r)^\prime =
\gamma^3(\Omegaa r)^\prime$.
For $\zeta^2 \ll 1$ we have $(\Omegaa r)^\prime
=\Omegaa /\gamma^2$ and $\Omegaa^\prime = -v^2\Omegaa/r$.
    Consequently
\begin{equation}
{\rm det}({\cal D}) =
\gamma^4(\Omegaa^2 -\Delta \omega^2)~.
\end{equation}
Inverting Eq.~(\ref{MHDsys}) gives 
\begin{equation}
\delta v_r = {q \gamma^3 \over m_e {\rm det}({\cal D})}
\bigg[-i\Delta \omega (\delta E_r +v \delta B_z)
+ \Omegaa \delta E_\phi \bigg]~,
\label{dvr}
\end{equation}
and
\begin{equation}
\delta v_\phi = {q \gamma^3\over m_e {\rm det}({\cal D})}
\bigg[-i\Delta \omega \delta E_\phi -
  \Omegaa (\delta E_r + v\delta B_z)/\gamma^2 \bigg]~,
\label{dvphi}
\end{equation}

\section{Field Sources}

     The source terms due to the perturbation are
\begin{equation}
\delta J_r = \rho_e \delta v_r~,\quad
{\rm and }\quad
\delta J_\phi = \rho_e \delta v_\phi
+\delta \rho_e v~,
\end{equation}
and from the continuity equation,
\begin{equation}
\delta \rho_e = {1 \over i \Delta \omega}
\big[D_r(\rho_e \delta v_r)
+ik_\phi \rho_e \delta v_\phi \big]~,
\end{equation}
where $D_r\equiv (1/r)[\partial/\partial r(r ...)]$
and  $k_\phi \equiv m/r$ is
the azimuthal wavenumber.
     Expanding this equation gives
$$
\delta \rho_e={1 \over i \Delta \omega}
\bigg\{ {d{\cal F}\over dr} \delta E_\phi
+{\cal F}\big[-ik_\phi(\delta E_r+v\delta B_z)
+D_r(\delta E_\phi) \big]
$$
\begin{equation}
-i\Delta \omega\bigg[D_r
\big[(\delta E_r+v\delta B_z){\cal F}/\Omegaa\big]
+ik_\phi  \delta E_\phi{\cal F}/\Omegaa\big]
\bigg]\bigg\}~.
\label{deltarhoe}
\end{equation}
Here,
\begin{equation}
{\cal F} \equiv {q\rho_e \Omegaa \gamma^3
\over m_e{\rm det}({\cal D})}~,
\end{equation}
has the role of the distribution function
of angular momentum \cite{lovelace1978}.

\section{The Limit $\Delta \omega \ll \gamma^{-1} \Omegaa$}
For
\begin{equation}
\left|{\Delta \omega \over \Omegaa} \right|^2 \ll 1
\end{equation}
the resonant term in $\delta \rho_e$ proportional to $1/\Delta \omega$
is dominant.
    
Thus we have
$$
\delta J_r =\rho_e\delta v_r
= {\cal F}\delta E_\phi+{\cal O}
\left({\Delta \omega \over\Omegaa}\right)~,
$$
$$
\rho_e \delta v_\phi=-{{\cal F}\over \gamma^2}
(\delta E_r +v \delta B_z)
+{\cal O}\left({\Delta \omega \over\Omegaa}\right)~,
$$
$$
\delta \rho_e v =
{r\Omegaa  \over i \Delta \omega}~
\bigg[{d{\cal F} \over dr} \delta E_\phi
+..\bigg]
+{\cal O}\left(\left|{\Delta \omega
\over\Omegaa}\right|^0\right)~,
$$
where the ellipsis  indicates a term equal to
  the middle line of equation (\ref{deltarhoe}).
To leading order in $|\Omegaa /\Delta \omega|$ we have
\begin{eqnarray}
\delta J_r& =&0~,
\nonumber\\
\delta J_\phi&=&\rho_e\delta v_\phi +\delta \rho_e v~,
\nonumber
\end{eqnarray}
or
\begin{equation}
\delta J_\phi={r\Omegaa  \over i \Delta
\omega}\bigg[{d{\cal F}\over dr} \delta E_\phi
+{\cal F}\big[-ik_\phi(\delta E_r+v\delta B_z)
+D_r(\delta E_\phi) \big]\bigg].
\end{equation}
In this approximation we also have
\begin{equation}
{\cal F}=- {\rho_e \over B}~.
\end{equation}
For an electron layer with $B_e >0$, we
have $\Omegaa >0$ and ${\cal F}>0$.

Thus, 
\begin{equation}
\delta v_\phi = \frac{iq}{m_e \gamma \Omegaa} \left [
-m \Delta \tilde \omega + \frac{1}{m \gamma^2 \Delta \tilde \omega} \right ] \delta E_\phi ~,
\end{equation}
where $\Delta \tilde \omega \equiv \Delta \omega / (m \Omegaa)$. Finally, 
\begin{equation}
\delta v_\phi = \frac{iq}{m_e \gamma^3 \Delta \omega} \delta E_\phi
\label{vphi2}
\end{equation}
The reason for the discrepancy from  Eq.~(\ref{drhodpot}) is the difference in the used equilibrium.
In chapter \ref{ch:CSR} the non-zero width was caused by betatron oscillations of particles with 
the same average angular velocity. In a Brillouin flow particles on different orbits have different angular velocities.
This additional source of shear is reflected in the $(\gamma \Omegaa r)'$ term. Setting this term
to zero one obtains the previous results from chapter \ref{ch:CSR} again (cf. Eq.~(\ref{constraineddvphi})~).

In the absence of radial currents the linearized continuity equation simply reads
\begin{equation}
\delta \rho = \frac{1}{\Delta \tilde \omega} \rho_0 \delta v_\phi
\end{equation}
Thus,
\begin{equation}
\delta \rho = \frac{-ie}{m_e \gamma^3 m \Omegaa} \frac{\rho_0}{(\Delta \tilde \omega)^2} \delta E_\phi
\end{equation}

\section{The Limit $\Delta \omega \gg \Omegaa$}
If we completly neglected radial motion as it was done in an earlier paragraph in chapter \ref{ch:CSR}
Eq.~(\ref{dvphi}) would read

\begin{equation} 
\delta v_\phi = {iq \over m_e \gamma \Delta \omega} \delta E_\phi ~. 
\label{constraineddvphi}
\end{equation}

In our two-dimensional MHD approach the motion is not constrained to a fixed radius
and we are wondering under which conditions the radial motion in an unconstrained model
can be neglected, i.e. when Eq.~(\ref{constraineddvphi}) and Eq.~(\ref{dvphi}) coincide.
$\delta v_r = 0$ implies that the forces due to $\delta E_\phi$, $\delta B_z$ and $\delta E_r$
have to balance. Eq.~(\ref{dvr}) gives 
\begin{equation}
\Omegaa \delta E_\phi = i \Delta \omega ( \delta E_r + v \delta B_z )
\end{equation}

Thus, Eq.~(\ref{constraineddvphi}) and Eq.~(\ref{dvphi}) coincide if
\begin{equation}
\Delta \omega \gg \Omegaa
\end{equation}
or if we choose an equilibrium distribution with zero average shear.
In both cases the growth rates are given by Eq.~(\ref{disp}).

\section{Configuration b}
In this section we are going to investigate the stability properties of an equilibrium with the same
number density and velocity profile as before, but with different external fields. Instead
of an external magnetic field in the z direction we consider an equilibrium with an azimuthal magnetic
field acting as a guiding field and a radial electric field. The latter is included in the equilibrium
condition and therefore does not enter the linearized Euler equation. $B^e_\phi$ would only enter
if we considered motion in the axial direction and non-zero axial wavenumbers. Thus, we obtain
the matrix $\mathcal{D}$ again without the $B_0$ terms, i.e. for $\gamma \gg 1$

\begin{equation}
  {\cal D}= \left [\begin {array}{cc}
{-i\gamma\Delta \omega } & {-\gamma^3 \Omegaa }
\\\noalign{\medskip}
{2 \gamma \Omegaa} & {-i\gamma^3 \Delta \omega}
\end {array}\right ]~,
\end{equation}
with
\begin{equation}
{\rm det}({\cal D}) =
\gamma^4(2 \Omegaa^2 -\Delta \omega^2)~.
\end{equation}
We obtain
\begin{equation}
\delta v_r = {q \gamma^3 \over m_e {\rm det}({\cal D})}
\bigg[-i\Delta \omega (\delta E_r +v \delta B_z)
+ \Omegaa \delta E_\phi \bigg]~,
\label{dvr2}
\end{equation}
and
\begin{equation}
\delta v_\phi = {q \gamma^3\over m_e {\rm det}({\cal D})}
\bigg[-i\Delta \omega \delta E_\phi -
  2 \Omegaa (\delta E_r + v\delta B_z)/\gamma^2 \bigg]~,
\label{dvphi2}
\end{equation}
In the limit $(\Delta \omega)^2 \gg 2 \Omegaa^2$ the azimuthal current is given by Eq.~(\ref{constraineddvphi}) again and by
Eq.~(\ref{vphi2}) for $(\Delta \omega)^2 \ll 2 \Omegaa^2$.

\section{Two Cylinder Model}
Instead of solving the full two-dimensional problem we solve it for two concentric cylinders, i.e. for the number density we have
\begin{eqnarray}
n(r) = \frac{1}{2} n_0 \sqrt{2\pi} v_{th} r_0 \left ( \delta (r-r_1) + \delta (r-r_2) \right )
\end{eqnarray}
Some interesting results can be obtained in that limit assuming zero average shear. 
The linearized continuity equation becomes
\begin{equation}
-i \omega \delta \rho + \frac{1}{r} \ddr \left ( r \rho_0 \delta v_r +
r v_{0r} \delta \rho \right ) + \frac{im}{r} \left ( \rho_0 \delta v_{\phi} + v_{0 \phi} \delta \rho \right ) = 0
\end{equation}
We drop all radial derivatives in the continuity equation which is consistent
with our previous approximation. Thus,
\begin{eqnarray}
\delta \rho = \frac{i}{2} e^2 n_0 \sqrt{2\pi} v_{th} r_0^2 
\left ( \delta (r-r_1) + \delta (r-r_2) \right ) \frac{m^{-1}}{m_e \gamma}
\frac{\delta E_\phi}{(\Delta \tilde \omega)^2} 
\end{eqnarray}

The Green function is

\begin{eqnarray}
\delta \Phi (r_1) = 2 \pi^2 i \left ( r_1 J_m(\omega r_1) H_m^{(1)}(\omega r_1) \delta \rho_1 +
r_2 J_m(\omega r_2) H_m^{(1)} (\omega r_1) \delta \rho_2 \right )
\\
\delta \Phi (r_2) = 2 \pi^2 i \left ( r_1 J_m(\omega r_2) H_m^{(1)}(\omega r_1) \delta \rho_1 +
r_2 J_m(\omega r_2) H_m^{(1)} (\omega r_2) \delta \rho_2 \right ) ~,
\end{eqnarray}
so the problem can be written as a matrix 

\begin{eqnarray}
\nonumber
A \equiv - 2 \pi^2 i e^2 \frac{1}{2} n_0 \sqrt{2\pi} v_{th} r_0^2 \frac{1}{H} \gamma^{-2}
\hspace{2in} \\ \cdot
\left ( \begin{array}{cc}
(\Domt (r_1))^{-2} \frac{r_1}{r_0} J_m(\omega r_1) H_m(\omega r_1) & (\Domt (r_2))^{-2} \frac{r_2}{r_0} J_m(\omega r_1) H_m(\omega r_2)
\\
(\Domt (r_1))^{-2} \frac{r_1}{r_0} J_m(\omega r_1) H_m(\omega r_2) & (\Domt (r_2))^{-2} \frac{r_2}{r_0} J_m(\omega r_2) H_m(\omega r_2)
\end{array} \right )
\end{eqnarray}

acting on the column vector $(\delta \Phi(r_1)),\delta \Phi(r_2))^{T}$
which returns the same column vector. Nontrivial solutions can only exist if 
\begin{eqnarray}
\det(A-\mathbbm{1})=0~.
\end{eqnarray}
Finally, this equation can be solved for $\Delta \tilde \omega$.

\begin{figure}
\includegraphics[width=\columnwidth]{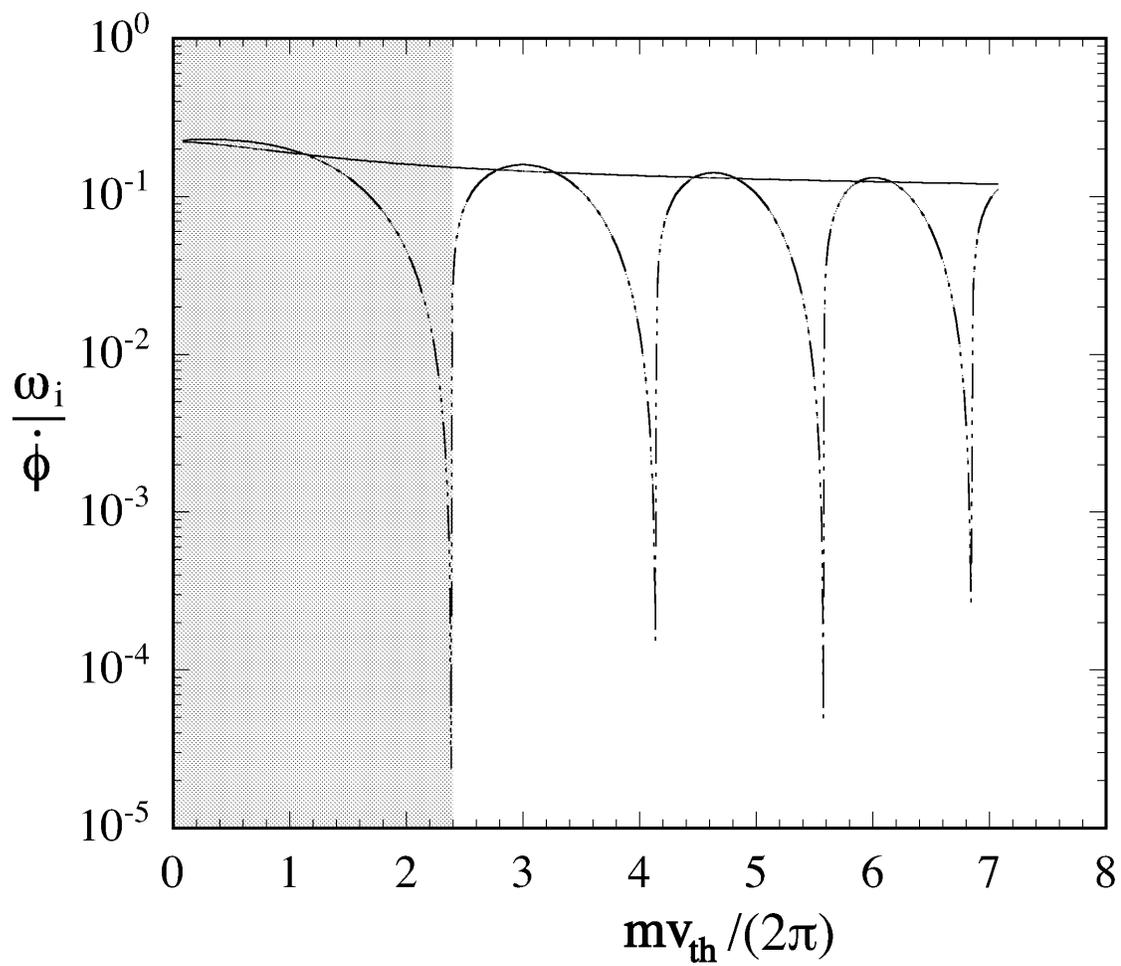}
\caption{Growth rates as a function of energy spread. The Airy fuctions were retained.
Parameters: $\zeta=0.02$, $\gamma=30$, $m=500$ }
\label{twoAiry}
\end{figure}

The coefficient in front of the matrix is equal to $-\frac{\pi}{2} i \zeta \gamma^{-2}$.
We introduce the notation $\Domt \equiv \Domt (r_0)$, $\Domt(r) = (1+ \Domt) \frac{r}{r_0} - 1$
and solve the dispersion relation for $\Domt$ approximating the Bessel functions by Airy functions
\begin{eqnarray}
J_m(z) = \left ( \frac{2}{m} \right )^{1/3} \Ai \left ( -(2/m)^{1/3} (z-m) \right )
\\
Y_m(z) = - \left ( \frac{2}{m} \right )^{1/3} \Bi \left ( -(2/m)^{1/3} (z-m) \right )
\end{eqnarray}
and neglecting terms proportional to $\Domt$ which are assumed to be small compared with $\gamma^{-2}$.
The growth rates for the parameters $\zeta=0.02$, $\gamma=30$, $m=500$ are plotted in Fig.~\ref{twoAiry}.

Numerically, we find that the drop occurs very roughly for $v_{th}=j\pi/m$ if $j$ is an even integer,
i.e. when the thickness of the layer is an integer multiple of 4 times the wavelength of the radiation.
This periodicity is due to the Bessel functions. The drop in the growth rate is
caused by the Coulomb term as can be seen by using the following
approximation for the Airy functions
\begin{eqnarray}
\Ai(w) \approx c_1 - c_2 w
\\
\Bi(w) \approx \sqrt{3}[c_1 + c_2 w]
\end{eqnarray}
where $c_1 = 1/[3^{2/3} \Gamma(2/3)]$ and $c_2 = 1/[3^{1/3} \Gamma(1/3)]$ which is justified for $|w|^2 \lesssim 0.5$. 
This gives $Z_m(w) \equiv i J_m(w) H_m(w) \approx (2/m)^{2/3} [\sqrt{3}(c_1^2 - c_2^2 w) +
i (c_1 - c_2 w)^2]$. 
Evaluating the Green function at $r_0$ for $w=0$ we just obtain
the radiation term quoted by Goldreich and Keeley \cite{GoldreichKeeley1971} and the two unstable modes shown in
Fig.~\ref{twoAiry} become degenerate. Keeping the first order term in $w$
we recover the Coulomb term as well. This approximation is valid in the shaded area of Fig.~\ref{twoAiry}.

The dependence of the larger mode in Fig.~\ref{twoAiry} on $v_{th}$ seems to be in rough agreement with the dependence
obtained from including the decoherence due to betatron oscillations in chapter \ref{ch:CSR}. We speculate that the betatron
oscillations themselves are not too important. What is important is the fact that particles move on different orbits
with different angular velocities $\Omegaa(r)$.

\chapter{Summary}

What have we learnt in the last three chapters?
We analyzed the stability properties and the
power spectra of charged particles executing
circular motion at relativistic speeds using
three different techniques (Vlasov, PIC, MHD). 
Relativistic plasmas are capable of self-bunching
and emitting coherent synchrotron radiation.
This has been established by all three methods.
The particular field geometry of the external guiding
fields played only a minor role (MHD). In particular
there were only minor differences in the stability 
properties of configuration a (Fig.~\ref{configa})
and configuration b (Fig.~\ref{configb}). The 
self-fields do not enter the dispersion relation
either as long as $\zeta \ll 1$ (Vlasov, PIC).
For the Vlasov treatment we selected a rather special
distribution function with no spread in the
azimuthal component of the canonical angular momentum
and a small energy spread. Other choices would
have been possible. However, since we are able to
recover the main effects with MHD the only important 
kinetic effect seems to be the Landau damping observed
for non-zero axial wavenumbers (Vlasov). We put a lot 
of effort in understanding the effect of a small energy 
spread. In the absence of such an energy spread the
results by Goldreich and Keeley were recovered 
\cite{GoldreichKeeley1971}. Once the energy spread
exceeds a critical threshold (which is a function
of the azimuthal mode number $m$) the growth rate
obtained by Goldreich and Keeley is attenuated by
a factor which is due to the decoherence caused by
the betatron oscillations. Once this threshold has
been exceeded there is no further dependence of
this factor on the energy spread as long as the
spread remains small. Fig.~\ref{ratioF0} summarizes 
these findings. The characteristic power spectrum 
which scales as $\nu^{-5/3}$ is due to this decoherence
and it is in good agreement with observations from
actual radio pulsars. 
In a Brillouin flow the CSR instability may be almost
completely suppressed by the shear intrinsic to
such an equilibrium. The amount of shear is very small
and the resolution of the PIC simulations were not high enough
to resolve the path length differences of orbits with different
radii at the low energy spreads that were used.
In the Vlasov approach we used a different equilibrium where 
the non-zero thickness was due to betatron oscillations. At 
least on average all particles were moving at the same angular
velocity. Still, both the naive two cylinder model and the
Vlasov approach for a thin layer with betatron oscillations give the
same dependence of the growth rate on the energy spread, but further
investigation is needed to establish this result.
Whether the negative mass instability is present in the system
under investigation is a tough call. As Fig.~\ref{twoAiry}
and the accompanying paragraph suggest the Coulomb term tends
to stabilize the second unstable mode. This conclusion is in agreement
with earlier findings by Goldreich and Keeley \cite{GoldreichKeeley1971}.
However, the very presence of a second unstable mode might
be an artifact of the negative mass instability. 
The treatment in \cite{GoldreichKeeley1971} and chapter \ref{ch:CSR} differ
in the following way: In the former an increase in energy leads to higher
angular velocity whereas in the latter the angular velocity decreases because
of the negative effective mass. Still the results resemble each other.
In some sense the CSR instability in chapter \ref{ch:CSR} is a 
``negative mass instability'' with a dominating radiation term instead of
a Coulomb term. The effect of the Coulomb term can only be seen for non-zero
energy spreads.
In Fig.~\ref{twoAiry} the second mode does drop down to zero for larger energy spreads
which is a characteristic of the negative mass instability, but a slight
increase in energy spread leads to a recovery of that mode again. 
Both the negative mass instability and the CSR instability manifest themselves
as a bunching of the distribution. It is conceivable that the two instabilities
interfere with each other in the sense that the weak negative mass instability
caused by the Coulomb term can only disturb the CSR instability caused by
the much stronger radiation term. In this case it would be very hard to tell
them apart.
 
As was pointed out in chapter \ref{intro:neutron} there are a couple of competing 
theories in the literature trying to explain the radio 
emission of pulsars. With the exception of the CSR instability
none of them seems viable. The biggest obstacle to applying
the CSR instability to pulsars has been a lack of more detailed
models. We hope that we bridged that gap. It is certainly far
from easy to figure out how a system evolves by only looking
at its saturated state. However, our model is not arbitrary.
It is well motivated and rather simple. Only very few dimensionless
parameters are needed and its results are strikingly generic.
We certainly do not claim that our model is the final word,
but dismissing it as a coincidence without further studies 
would be too easy. The fact that many of the presented
results only depend on dimensionless parameters implies scale
invariance. This opens up the possibility of testing pulsar radiation
mechanisms in the lab. Particle accelerators suitable for such
experiments already exist and the similarity of instabilities
found in accelerators and astrophysical objects is hard to overlook.
Proof of principle experiments  for different astrophysical problems 
have already been performed \cite{Lebedev:2004cs}.

\appendix

\chapter{Green's Function}

The Green's function for the potentials give
\begin{eqnarray}
\delta \Phi({\bf r},t)=
\int dt^\prime d^3 r^\prime ~
G({\bf r}-{\bf r^\prime},t-t^\prime) ~
\delta \rho({\bf r}^\prime,t^\prime)~.
\nonumber\\
\delta {\bf A}({\bf r},t)=
\int dt^\prime d^3 r^\prime ~
G({\bf r}-{\bf r^\prime},t-t^\prime) ~
\delta {\bf J}({\bf r}^\prime,t^\prime)~,
\end{eqnarray}
where
$$
\left(\nabla^2 - {\partial^2 \over \partial t^2}\right)
G({\bf r},t)= -4\pi \delta(t)\delta({\bf r})~,
~~ \tilde{G}({\bf k},\omega)=
{4\pi \over  {\bf k}^2-\omega^2  }~,
$$
\begin{eqnarray}
G({\bf r},t)=\frac{4\pi}{(2 \pi)^4} \int_C d\omega \int d^3k~
{\exp(i{\bf k\cdot r}-i\omega t) \over  {\bf k}^2-\omega^2 }~,
\end{eqnarray}
where $\tilde{G}$ is the Fourier transform of the Green's
function. The ``C'' on the integral indicates an
$\omega-$integration parallel to but above the real axis, ${\rm
Im}(\omega)>0$, so as to give the retarded Green's function.

           Because of the assumed dependences
of Eq.~(\ref{ansatz}),
we  have for the electric potential,
$$
\delta \Phi_{\omega m k_z}(r) = 2
\int_0^\infty r^\prime dr^\prime
\int_0^\infty \kappa d\kappa \int_0^{2\pi} d\alpha~
~ \delta
\rho_{\omega m k_z}(r^\prime)\big[.. \big]
$$
\begin{equation}
=4\pi \int_0^\infty r^\prime dr^\prime
\int_0^\infty \kappa d\kappa~
{J_m(\kappa r) J_m(\kappa r^\prime)
\over \kappa^2- (\omega^2 - k_z^2) }~
\delta\rho_{\omega m k_z}(r^\prime)~,
\end{equation}
where
$$
\big[.. \big]\equiv{\exp(im\alpha)
J_0\{\kappa [r^2+ (r^\prime)^2-2r
r^\prime
\cos\alpha]^{1/2}\}
\over \kappa^2-(\omega^2 - k_z^2) }~,
$$
where $\kappa^2 \equiv k_x^2+k_y^2$.
Because $\omega$ has a positive imaginary part,
this solution corresponds to the retarded field.
         Also because ${\rm Im}(\omega)>0$, the
$\kappa-$integration can be done by
a contour integration as discussed in \cite{Watson1966}
which gives
\begin{equation}
\delta \Phi_{\omega m k_z}(r)\!=\!
2\pi^2 i \int_0^\infty \!\!\!r^\prime dr^\prime
J_m(kr_<)H_m^{(1)}(kr_>)\delta\rho_{\omega m k_z}(r^\prime),
\label{dpot1}
\end{equation}
where $k\equiv  (\omega^2-k_z^2)^{1/2}$, where $r_<$ ($r_>$)
is the lesser (greater) of $(r,r^\prime)$, and where
$H_m^{(1)}(x)=J_m(x)+iY_m(x)$ is the  Hankel
function of the first kind.
From the Lorentz gauge condition
\begin{equation}
\delta \Psi^{\omega m k_z}(r)= r\delta A_\phi^{\omega m k_z}=
r_0 v_{\phi} \left ( 1 + \Delta \tilde \omega \right )
\delta \Phi^{\omega m k_z}(r)
\label{dpot2}
\end{equation}
Eqs.~(\ref{dpot1}) and (\ref{dpot2}) are useful in subsequent
calculations.

          To determine the total synchrotron radiation
from the E-layer it is sufficient to calculate
$\delta {\bf A}$ at a large distance from the
E-layer.
          We assume that the E-layer has a finite
axial length and exists between $-L/2 \leq z \leq L/2$.
          Thus we evaluate $\delta {\bf A}$ in a spherical
coordinate system ${\bf R}=(R,\theta, \phi)$ at a distance
$R \gg L$.
          The retarded solution is
$$
\delta {\bf A}({\bf R})= {1 \over R}
\int d^3 r^\prime~\delta {\bf J}\bigg({\bf r}^\prime,
t - |{\bf R} -{\bf r}^\prime|\bigg)
=
$$
\begin{equation}
{\exp(i\omega R)\over R}\int\!\!\! d^3 r^\prime~ \delta {\bf
J}(r^\prime)
\exp \bigg[im \phi^\prime +ik_z z^\prime
-i\omega (t +\hat{\bf R}\cdot {\bf r}^\prime)\bigg],
\end{equation}
(see, e.g. ch. 9 of \cite{LandauLifshitz1962}).
          The source point is at
$(x^\prime=r^\prime\cos\phi^\prime,~ y^\prime=r^\prime \sin
\phi^\prime,~ z^\prime)$.
          The observation point is taken to be at
$(x=0,~y=R\sin \theta,~ z=R\cos\theta)$.
          Consequently, $\hat{\bf R}\cdot {\bf r}^\prime=
r^\prime \sin\theta \sin\phi^\prime +z^\prime \cos\theta$.
         The phase factor $\exp(i\omega R)$ does
not affect the radiated power and is henceforth dropped.

           For the cases where $\delta J_\phi$ is the dominant
component of the current-density perturbation we have
$$
\bigg[\delta A_x^\omega,~\delta A_y^\omega\bigg]=
{S(\theta)\over R}\int r^\prime d r^\prime d\phi^\prime~
\bigg[-\sin\phi^\prime, \cos \phi^\prime\bigg] \times
$$
\begin{equation}
\delta J_\phi(r^\prime)~
\exp(im\phi^\prime-i\omega r^\prime \sin \theta
\sin\phi^\prime)~,
\label{dAomega1}
\end{equation}
where
\begin{eqnarray}
S(\theta) \equiv L~ {\sin[(k_z-\omega \cos \theta)L/2]
\over (k_z-\omega \cos \theta)L/2}
\end{eqnarray}
is a structure function accounting for the finite
axial length of the E-layer, and $\omega$ superscript
indicates $\omega=m\dot{\phi}$.
          Carrying out the $\phi^\prime$ integration in
Eq.~(\ref{dAomega1}) gives
\begin{equation}
\bigg[\delta A_x^\omega,~\delta A_y^\omega \bigg]
={S(\theta)\over R}\int r^\prime d r^\prime~
\delta J_\phi(r^\prime)\big[..\big]
\label{dAomega}
\end{equation}
where
$$
\big[..\big]\equiv\left[ i J_m^\prime
(\omega r^\prime \sin \theta),~
{m \over \omega r^\prime \sin \theta}
J_m(\omega r^\prime \sin \theta)\right]
$$
and where the prime on the Bessel function
indicates its derivative with respect to its
argument.
           The radiated power per unit solid angle is
\begin{eqnarray}
{d P_\omega \over d \Omega}
={R^2 \over 8\pi}|\delta {\bf B}^\omega|^2
= {R^2 \over 8\pi}|{\bf k} \times \delta {\bf A}^\omega |^2 =
        \nonumber\\
{R^2 \omega^2 \over 8\pi} \left( |\delta A_x^\omega|^2  +
\cos^2 \theta |\delta A_y^\omega|^2\right)~,
\label{dIdOmega1}
\end{eqnarray}
where ${\bf k} \equiv \omega \hat{\bf R}$ is the
far field  wavevector.

          For a radially thin E-layer, $(\Delta r/r_0)^2 \ll 1$,
Eqs.~(\ref{dAomega}) and (\ref{dIdOmega1}) give
\begin{eqnarray}
\nonumber
{d P_\omega \over d \Omega}
={S^2(\theta) \over 8\pi}\left|
\int r^\prime dr^\prime~\delta J_\phi(r^\prime)
\omega J_m^\prime(\omega r_0\sin\theta) \right |^2 +
\\
{S^2(\theta) \over 8\pi} \left | \int r^\prime dr^\prime~\delta
J_\phi(r^\prime) {m~{\rm ctn}~\theta \over r_0} J_m(\omega
r_0\sin\theta)\right |^2~.
\label{dIdOmega}
\end{eqnarray}
The factor within the curly brackets is the same as
that for the radiation pattern of a single charged particle
(see ch. 9 of \cite{LandauLifshitz1962}).

          The factor $S^2(\theta)$
in Eq.~(\ref{dIdOmega})  tightly
constrains
the radiation to be in the direction
$\theta_* = \cos^{-1}(k_z/\omega)$ if the  angular width of
$S^2(\theta)$, the half-power half-width
$\Delta \theta_{1/2} \approx \pi /(\omega L)$, is
small compared with the angular spread of the single
particle synchrotron radiation, $1/\gamma$,
which is the angular width due to the Bessel
function terms in Eq.~(\ref{dIdOmega}).
         This corresponds to E-layers with $L \gg \pi \gamma/\omega
=\pi r_0 \gamma/m$.  For $L \sim r_0$, we
need $m  \gg \pi \gamma$, which is satisfied by
the spectra discussed later in \S \ref{CSR:thick}.
In this case, Eq.~(\ref{dIdOmega})
can be integrated over
the solid angle to give
$$
P_\omega = {\pi  L\sin\theta_* \over 2 \omega} \left \{ \left|
\int r^\prime dr^\prime~\delta J_\phi(r^\prime)
\omega J_m^\prime( \omega r_0\sin\theta_*) \right |^2 + \right .
$$
\begin{equation}
\left . \left| \int r^\prime dr^\prime~\delta J_\phi(r^\prime)
{m~{\rm ctn}~\theta_* \over r_0} J_m(\omega r_0
\sin\theta_*) \right |^2 \right \}~.
\label{Iomega}
\end{equation}

One limit of  interest of Eq.~(\ref{Iomega}) is that where
$k_z=0$ so that $\theta_*=\pi/2$ and
\begin{eqnarray}
P_m
={\pi m v_\phi L  \over 2r_0 }\left|
\int r^\prime dr^\prime~\delta J_\phi(r^\prime)
J_m^\prime(\omega r_0) \right|^2~,
\label{Imeq}
\end{eqnarray}
where we have set $\omega \rightarrow m \dot{\phi}$.
         The total radiated power is $P = \sum_m P_m$.

\chapter{Bessel Function Approximations}

The Bessel functions $J_m (\omega r)$ and $Y_m (\omega r)$ are the
two linear independent solutions of the differential equation

\begin{eqnarray}
\left ( \frac{1}{r} \frac{\partial}{\partial r} r \frac{\partial}{\partial r}
- \frac{m^2}{r} + \omega^2 \right ) \delta E_{\phi} (r) = 0
\end{eqnarray}

Computing these functions is very cumbersome for high values of $m$ - even
on modern computers. If analytical results are desired approximating those
functions may be a necessity. The following approximations are extremely useful
\cite{Abramowitz1965,Watson1966}
\begin{eqnarray}
\nonumber
J_m \left ( m + z m^{1/3} \right ) = \left ( \frac{2}{m} \right )^{1/3}
\Ai \left ( - 2^{1/3} z \right ) + O(m^{-1})
\\
Y_m \left ( m + z m^{1/3} \right ) = - \left ( \frac{2}{m} \right )^{1/3}
\Bi \left ( - 2^{1/3} z \right ) + O(m^{-1})
\label{BessAiry}
\end{eqnarray}
The Airy functions can be evaluated very quickly on modern computers. For
analytical results further approximations may be required.

For $|w|^2 \gg 1$ 
\begin{eqnarray}
\nonumber
\Ai(w) \approx (2 \sqrt{\pi})^{-1} w^{-1/4} \exp(-2w^{3/2}/3)
\\
\Bi(w) \approx (\sqrt{\pi})^{-1} w^{-1/4} \exp(2w^{3/2}/3)
\label{expAiry}
\end{eqnarray}
and for $|w|^2 \lesssim 0.5$ 
\begin{eqnarray}
\nonumber
\Ai(w) = c_1 - c_2 w + O(w^3)
\\
\Bi(w) = \sqrt{3} [c_1 + c_2 w + O(w^3) ]
\label{linAiry}
\end{eqnarray}
where $c_1 = 1/[3^{2/3} \Gamma(2/3)] \approx 0.355$
and $c_2 = 1/[3^{1/3} \Gamma(1/3)] \approx 0.259$.

\chapter{Source Code Listings (Maple Worksheets)}
The following worksheets were generated using Maple 6 for IRIX 6.5.
\section{Solver for Eq.~(\ref{disp})}
\begin{verbatim}
> eqn:=-1+Pi*zeta*Z*(Domt-1/gamma0^2)/Domt^2;
             
> Z:=(2/m)^(2/3)*(AiryAi(w)*AiryBi(w)+I*AiryAi(w)^2);

> w:=(m/2)^(2/3)*(1/gamma0^2-2*Domt);

> gamma0:=30: zeta:=0.02: 

> m:=1000;

> fsolve(eqn,Domt=1e-6+1e-6*I);
             -0.00001518689566 + 0.0005418589616 I
> Im(%)*m;
                          0.5418589616

\end{verbatim}

\section{Solver for Eq.~(\ref{epsu})}
\begin{verbatim}
> gamma0:=30: zeta:=0.02: vth:=1/gamma0: 
  psi:=arctan(0.005): m:=1e3: eps:=1e-4:
> Z:=(2/m)^(2/3)*(AiryAi(w)*AiryBi(w)+I*AiryAi(w)^2):
> w:=(m/2)^(2/3)*(1/gamma0^2+tan(psi)^2-2*u*tan(psi)):
> F:=unapply(Heaviside(-Im(z))*I*sqrt(2*Pi)*exp(-z^2/2)+1/sqrt(2*Pi)
     *Int(exp(-x^2/2)/(x-z),x=-infinity..infinity,
          digits=5,method=_NCrule),z):
> C:=unapply(Pi/vth^2*zeta*Z/tan(psi)*(u-1/gamma0^2/tan(psi)),u):
> B:=unapply(Pi/vth^3*zeta*Z*(u-1/gamma0^2/tan(psi))*
            (0*1-u/tan(psi)),u):
> u:=(0.1+0.3*I)/m/tan(psi);
              u := 0.02000000000 + 0.06000000000 I
> evalf(1-B(u)*F(u/vth)+C(u));
                 -0.278474072 - 0.9518667283 I
> for i from 0 to 8 do
> eps1:=evalf(1-B(u+eps)*F((u+eps)/vth)+C(u+eps));
> eps0:=evalf(1-B(u)*F(u/vth)+C(u));
> un:=u-eps/(eps1-eps0)*eps0:
> u:=un; printf("%g %g %g %g \n",Re(u),Im(u),Re(eps0),Im(eps0));
> end:

> Domt:=u*tan(psi)/(1-1/2/gamma0^2);
         Domt := -0.00003303461986 + 0.0004692459188 I
> Im(Domt)*m;
                          0.4692459188
> Re(Domt)*m;
                         -0.03303461986

\end{verbatim}
\section{Evaluator for Eq.~(\ref{soln0})}
\begin{verbatim}
> Ff:=unapply(I^n*exp(-I*n*arctan(krb/m))/2/Pi*
      Int(exp(-I*n*theta-Chi^2/2*(m/sqrt(m^2+krb^2)-sin(theta))^2),
         theta=-Pi..Pi),n,Chi);

> gamma0:=30: zeta:=0.02: vth:=1/gamma0: 
> m:=5e4; krb:=10:

>  
> F[0]:=evalf(Ff(0,m*vth));
>  
    
> Domt:=(2/Pi)^(1/4)*sqrt(-zeta*F[0])/sqrt(vth*(m^2+gamma0^2*krb^2));
        
> evalf(m*Im(Domt));
                         0.07543389577




\end{verbatim}

\section{Solver for Eq.~(\ref{dispF0})}
\begin{verbatim}
> eqn:=1=Pi*zeta*Z*(-1/gamma0^2*sum(F[n]/(Domt-n/l)^2,n=-N..N));
 
> Ff:=unapply(I^n/2/Pi*Int(exp(-I*n*theta-Chi^2/2*sin(theta)^2 + 
      Chi^2*sin(theta)-Chi^2/2),theta=-Pi..Pi),n,Chi);
 
> Z:=(2/l)^(2/3)*(AiryAi(w)*AiryBi(w)+I*AiryAi(w)^2);
     
> w:=(l/2)^(2/3)*(1/gamma0^2);
                 
> N:=0:
> gamma0:=4000: zeta:=0.08: vth:=0.04: 
> l:=1e3; 
                      
> for j from -N to N do 
> F[j]:=evalf(Ff(j,l*vth));
> od; 
        
> solve(eqn,Domt);
           
> l*Im(%[2]);
                         0.002124396570
                         

\end{verbatim}
\chapter{Source Code Listings (XOOPIC)}
The results presented in chapter \ref{ch:PIC} were obtained using
a modified version of XOOPIC-2.5.1. In the following sections we
present the output of UNIX {\it diff} command applied to the original
and the modified source files. Together with the original source files
it is possible to recover the modified version of XOOPIC.

\section{File c\_utils.c}

\begin{verbatim}
> diff /tmp/oopic/otools/c_utils.c ~/oopic/otools/c_utils.c
61a62
> /* wrappers for atan2 and erf 
69a71
> float ATAN2W(double y, double x) { return atan2(y, x); }
70a73,74
> float ERFW(double x) {return erf(x); }
> 
92a97,115
> /* some useful functions 
> 
> float EllipticE(float x) {
>    float m1,res;
>    m1 = 1-x;
>    res = 1+.46301*m1+.10778*m1*m1 + 
           (.24527*m1+.04124*m1*m1)*log(1/m1);
>    if (x<0.999) return res;
>    return 1.;
> }
> 
> float EllipticK(float x) {
>    float m1,res;
>    m1 = 1-x;
>    res = 1.38629+.11197*m1+.07252*m1*m1 + 
           (.5+.12134*m1+.02887*m1*m1)*log(1/m1);
>    if (x<0.999) return res;
>    return 4.5;
> 
> }
> 
125d147
< 
\end{verbatim}

\section{File dump.cpp}
\begin{verbatim}
> diff /tmp/oopic/otools/dump.cpp ~/oopic/otools/dump.cpp
0a1
> extern "C++" void write_validation();
58a60,65
> 
> 
> // 
> 
> 
> /*
72a80,81
> 
> */
73a83,86
> // 
> // 
>  
>         write_validation(); 
75d87
< 
\end{verbatim}

\section{File diagn.cpp}
\begin{verbatim}
> diff /tmp/oopic/otools/diagn.cpp ~/oopic/otools/diagn.cpp
879c879,881
< #ifdef BENCHMARK
---
> //  #ifdef BENCHMARK
> //  
> //
885c887
< void write_validation() {
---
> void write_validation2() {
888a891,900
>   int j,k;
>   double Bval[1024],Btr[1024];
>   double Eval[1024],Etr[1024];
>   double rhoval[1024],rhotr[1024];
>   double a_sum,b_sum;
>   double R,max,count,countmax,maxR;
>   
> 
> 
>   
890,892d901
<   for(int j=0;j<J;j++) 
<     for(int k=0;k<K;k++) 
<               fprintf(trace_file,"%10.4g\n",E[j][k].e1());
893a903,975
>   for (R=110;R<=135;R+=5)
>   {
>   max = 2*3.1415*R;
>   for(j=0;j<max;j++) 
>     {
>     Bval[j] = (theSpace->getBNodeDynamic())[(int) 
                 (J/2+R*sin(j/R))][(int) (J/2+R*cos(j/R))].e3();
>     fprintf(trace_file,"%d %d %d %10.4g\n",0,(int) R,j,Bval[j]);
>     }
>     
>    for(j=0;j<max;j++)
>          { 
>          a_sum = 0;
>          b_sum = 0;
>          for(k=0;k<max;k++)
>              {
>              a_sum += Bval[k]*cos(j*2*3.1415*k/max);
>              b_sum += Bval[k]*sin(j*2*3.1415*k/max);
>              }
>          a_sum = 2*a_sum/max;
>          b_sum = 2*b_sum/max;
>          Btr[j] = sqrt(a_sum*a_sum+b_sum*b_sum);
>          fprintf(trace_file,"%d %d %d %10.4g\n",1,(int) R,j,Btr[j]);
>          }
> 
>    
>    for(j=0;j<max;j++) 
>     {
> /*    Eval[j] = (theSpace->getENode())[(int) (J/2+R*sin(j/R))]
                 [(int) (J/2+R*cos(j/R))].e1()*(-1)*cos(j/R)+
>                 (theSpace->getENode())[(int) (J/2+R*sin(j/R))]
                 [(int) (J/2+R*cos(j/R))].e2()*sin(j/R);          */
>       Eval[j] = sqrt(pow((theSpace->getENode())
                      [(int) (J/2+R*sin(j/R))]
                      [(int) (J/2+R*cos(j/R))].e1(),2)+
>                      pow((theSpace->getENode())
                      [(int) (J/2+R*sin(j/R))]
                      [(int) (J/2+R*cos(j/R))].e2(),2));
>     fprintf(trace_file,"%d %d %d %10.4g\n",2,(int) R,j,Eval[j]);
>     }
>     
>    for(j=0;j<max;j++)
>          { 
>          a_sum = 0;
>          b_sum = 0;
>          for(k=0;k<max;k++)
>              {
>              a_sum += Eval[k]*cos(j*2*3.1415*k/max);
>              b_sum += Eval[k]*sin(j*2*3.1415*k/max);
>              }
>          a_sum = 2*a_sum/max;
>          b_sum = 2*b_sum/max;
>          Etr[j] = sqrt(a_sum*a_sum+b_sum*b_sum);
>          fprintf(trace_file,"%d %d %d %10.4g\n",3,(int) R,j,Etr[j]);
>          }
>          
>    for(j=0;j<max;j++) 
>     {
>     rhoval[j] = (theSpace->getRho())
                  [(int) (J/2+R*sin(j/R))][(int) (J/2+R*cos(j/R))];
>     fprintf(trace_file,"%d %d %d %10.4g\n",4,(int) R,j,rhoval[j]);
>     }
>     
>    for(j=0;j<max;j++)
>          { 
>          a_sum = 0;
>          b_sum = 0;
>          for(k=0;k<max;k++)
>              {
>              a_sum += rhoval[k]*cos(j*2*3.1415*k/max);
>              b_sum += rhoval[k]*sin(j*2*3.1415*k/max);
>              }
>          a_sum = 2*a_sum/max;
>          b_sum = 2*b_sum/max;
>          rhotr[j] = sqrt(a_sum*a_sum+b_sum*b_sum);
>          fprintf(trace_file,"%d %d %d %10.4g\n",5,
                                (int) R,j,rhotr[j]);
>          }
> 
>   }
>    
>    
896c978,1012
< #endif
---
> 
> void write_validation() {
>   FILE *trace_file;
>   int J = theSpace->getJ();
>   int K = theSpace->getK();
>   int j,k;
>   double minB,maxB,minE,maxE,temp;
>   
>   
>   if((trace_file=fopen("trace.dat","a"))==NULL) exit(1);
> 
>   
>   minB = (theSpace->getBNodeDynamic())[0][0].e3();
>   maxB = (theSpace->getBNodeDynamic())[0][0].e3();
>   
>   minE = sqrt(pow((theSpace->getENode())[0][0].e1(),2) + 
                pow((theSpace->getENode())[0][0].e2(),2));
>   maxE = sqrt(pow((theSpace->getENode())[0][0].e1(),2) + 
                pow((theSpace->getENode())[0][0].e2(),2));
>  
>   
>   for(j=0;j<J;j++)
>    for(k=0;k<K;k++) 
>     {
>     if ((theSpace->getBNodeDynamic())[j][k].e3() < minB) 
           minB = (theSpace->getBNodeDynamic())[j][k].e3();
>     if ((theSpace->getBNodeDynamic())[j][k].e3() > maxB) 
           maxB = (theSpace->getBNodeDynamic())[j][k].e3();
>     temp = sqrt(pow((theSpace->getENode())[j][k].e1(),2) + 
                  pow((theSpace->getENode())[j][k].e2(),2));
>     if (temp < minE) minE = temp;
>     if (temp > maxE) maxE = temp;
>     }
>     fprintf(trace_file,"%10.4g %10.4g %10.4g %10.4g\n",
              minB, maxB, minE, maxE);
>     
>     fclose(trace_file);
>   
> }
> 
> // #endif
\end{verbatim}

\section{File evaluator.y}
\begin{verbatim}
> diff /tmp/oopic/otools/evaluator.y ~/oopic/otools/evaluator.y
83a84
> ATAN2W();
87a89,90
> EllipticE();
> EllipticK();
91a95
> ERFW();
97a102
>            "atan2",ATAN2W,
104a110,111
>            "EllipticE",EllipticE,
>            "EllipticK",EllipticK,
105a113
>            "erf",  ERFW, 
\end{verbatim}

\section{File evaluator.h}
\begin{verbatim}
> diff /tmp/oopic/otools/evaluator.h ~/oopic/otools/evaluator.h
42a43
>       float ATAN2W(float y,float x) {return (float)atan2(y,x); }
47a49
>   //  float ATAN2W(float );
\end{verbatim}

\section{File load.cpp}
\begin{verbatim}
> diff /tmp/oopic/physics/load.cpp ~/oopic/physics/load.cpp
146a147
>   Scalar Jmx,Kmx;
211a213
> 
213c215,222
<    u = maxwellian->get_U();
---
>    Jmx=grid->getJ();
>    Kmx=grid->getK();
>    u = maxwellian->get_v0();
>    Vector3 beta = iSPEED_OF_LIGHT*u;
>    Scalar gamma0 =1/sqrt(1-beta*beta)*sqrt((x-Vector2(Jmx/2,Kmx/2))
                         *(x-Vector2(Jmx/2,Kmx/2)))/(Jmx/4); 
>    Scalar phi = atan2(x*Vector2(1,0)-Kmx/2,x*Vector2(0,1)-Jmx/2);
>    u = -gamma0*SPEED_OF_LIGHT*(1-1/2/gamma0/gamma0)
                *cos(phi)*Vector3(1,0,0) +
>         gamma0*SPEED_OF_LIGHT*(1-1/2/gamma0/gamma0)
                *sin(phi)*Vector3(0,1,0);
> 
\end{verbatim}

\section{Copyright}

Copyright (C) 1994-2002 The Regents of the University of California
(Regents). All Rights Reserved.

The code XOOPIC is referred to herein as the Software.

Permission to use, copy, modify, and distribute the Software and its
documentation for educational and research purposes, without fee and
without a signed licensing agreement, is hereby granted, provided: (1)
that the above copyright notice, this paragraph and the following two
paragraphs appear in all copies, modifications, and distributions; (2)
you will not charge more than the cost of duplication for copies of
the original or derivative versions of the Software; (3) any export of
the Software must be in compliance with U. S. export control
regulations.  For a license permitting for-profit distribution of the 
Software or its derivatives, contact The Office of Technology Licensing, 
UC Berkeley, 2150 Shattuck Avenue, Suite 510, Berkeley, CA 94720-1620,
(510) 643-7201.

IN NO EVENT SHALL REGENTS BE LIABLE TO ANY PARTY FOR DIRECT, INDIRECT,
SPECIAL, INCIDENTAL, OR CONSEQUENTIAL DAMAGES, INCLUDING LOST PROFITS,
ARISING OUT OF THE USE OF THIS SOFTWARE AND ITS DOCUMENTATION, EVEN IF
REGENTS HAS BEEN ADVISED OF THE POSSIBILITY OF SUCH DAMAGE. 
  
REGENTS SPECIFICALLY DISCLAIMS ANY WARRANTIES, INCLUDING, BUT NOT LIMITED
TO, THE IMPLIED WARRANTIES OF MERCHANT-ABILITY AND FITNESS FOR A
PARTICULAR PURPOSE. THE SOFTWARE AND ACCOMPANYING DOCUMENTATION, IF
ANY, PROVIDED HEREUNDER IS PROVIDED "AS IS". REGENTS HAS NO OBLIGATION TO
PROVIDE MAINTENANCE, SUPPORT, UPDATES, ENHANCEMENTS, OR MODIFICATIONS.

For custom modification and support of the Software, contact
Prof. J. P. Verboncoeur (johnv@eecs.berkeley.edu).  If you use the
Software for publication or other form of communication, as a
courtesy, please use the following or similar citation:

J. P. Verboncoeur, A. B. Langdon and N. T. Gladd, Comp. Phys. Comm. 87, 
199 (1995). Code available via http://ptsg.eecs.berkeley.edu.

\chapter{OOPIC Input Files}
\section{A Sample Input File}
In this section a valid OOPIC input file is shown which can be used to recover
the results presented in chapter \ref{ch:PIC}. The values for $\zeta$, $v_{th}$ and
$\gamma$ can be changed easily.
\begin{verbatim}
E-Layer
{
Simulation of E-Layer
}

Variables
{
        c  = 2.99792e8
        re = 2.82e-15
        e  = -1.6e-19
        me = 9.11e-31
        mu0 = 1.256e-6
        eps0 = 8.8542e-12
        gamma = 30
        zeta = 0.020
        vth = 0.002
        r0 = 10.0
        L = 10*r0
        range = 20.0
        sigma = vth*r0/sqrt(1-2*zeta)
        beta = 1 - 1/(2*gamma^2)
        Bext = 9.11e-31*c*beta*gamma/1.609e-19/r0*(1+2*zeta)
        N0 = gamma*L*zeta/re
        dN = 0.0
        NMacPart = 5000
        norm = sqrt(2*PI)*sigma*2*PI*r0
        KK = N0/norm
        Q = N0*e
        I = Q*c/(2*PI*r0)
}

Region
{
Grid
{
        J = 512     
        x1s = -range   
        x1f =  range   
        
        K = 512     
        x2s = -range   
        x2f =  range  
        Geometry = 1   
}
Control
{
        dt = 5.0e-12
        B3init = mu0*I/L/2/PI*(
            EllipticE(2*sqrt(sqrt(x1^2+x2^2)*r0)*(sqrt(x1^2+x2^2)+r0)/
             ((sqrt(x1^2+x2^2)+r0)^2+r0^2*vth^2))*(sqrt(x1^2+x2^2)-r0)/
             ((sqrt(x1^2+x2^2)-r0)^2+r0^2*vth^2)-
            EllipticK(2*sqrt(sqrt(x1^2+x2^2)*r0)*(sqrt(x1^2+x2^2)+r0)/
             ((sqrt(x1^2+x2^2)+r0)^2+r0^2*vth^2))/
             (sqrt(x1^2+x2^2)+r0))
        B03    = Bext/L
        ElectrostaticFlag=0
}
Species
{
        name = electron
        m = me/L^2
        q = e/L
        collisionModel = 1
}

Load
{
        speciesName = electron
        LoadMethodFlag = 0
        analyticF = KK*exp(-(x1^2+x2^2-2*r0*sqrt(x1^2+x2^2)+r0^2)/
                           2.0/sigma^2)*(1+dN*sin(50*atan2(x1,x2)))
        np2c = N0/NMacPart
        v1drift = c*(1-1/(2*gamma^2))
        v2drift = 0.0
        v3drift = 0.0
        x1MinMKS = -range
        x1MaxMKS =  range
        x2MinMKS = -range
        x2MaxMKS =  range
}

}

\end{verbatim}

\end{document}